\documentclass{aa}  

\usepackage{graphicx}
\usepackage{txfonts}
\usepackage{bm}
\usepackage{lipsum}
\usepackage{hyperref}
\usepackage[switch]{lineno}
\modulolinenumbers[5]   
\nolinenumbers

\usepackage{subcaption}         
\usepackage{lscape}             
\usepackage{placeins}           
\usepackage{gensymb}                                
\newcommand{\MA}{CH$_3$CCH}
\newcommand{\CCH}{C$_2$H}
\newcommand{\TCC}{$^{13}$CCH}
\newcommand{\CTC}{C$^{13}$CH}

\newcommand{\TC}{$^{13}$C}
\newcommand{\NE}{RCW122-NE}
\newcommand{\SW}{RCW122-SW}

\newcommand{\TE}{$T_{\rm ex}$}
\newcommand{\NC}{$N_{\rm col}$}

\newcommand{\radec}{\hbox{RA, Dec.(J2000)}}
\newcommand{\gra}{$^{\circ}$}

\begin{document}

   \title{Cold \CCH\ and warm \MA:\ complementary tracers in the massive star-forming complex RCW~122 (NE \& SW)}


%
%
%

\author{Edgar Mendoza\inst{1,2}\fnmsep\thanks{edgar@xao.ac.cn}        
        \and Nicolas Duronea\inst{3} 
        \and Leonardo Bronfman\inst{4}
        \and Xiaohu Li\inst{1,2}
        \and Mart\'in Ortega\inst{5}
        \and Xia Zhang\inst{1,2}
        \and Ricardo Finger\inst{4}
        }

   \institute{State Key Laboratory of Radio Astronomy and Technology, Xinjiang Astronomical Observatory, CAS, 150 Science 1-Street, Urumqi, Xinjiang, 830011, China.
   \and Xinjiang Key Laboratory of Radio Astrophysics, 150 Science 1-Street, Urumqi, Xinjiang 830011, China.
   \and Instituto de Astrof\'{\i}sica de La Plata (UNLP - CONICET), La Plata, Argentina.
   \and Departamento de Astronomía, Universidad de Chile, Camino El Observatorio 1515, Las Condes, Santiago de Chile, Chile
   \and Instituto de Astronomía y Física del Espacio (UBA, CONICET), CC 67, Suc. 28, 1428 Buenos Aires, Argentina}

   \date{Received 20XX, Accepted 20XX}

 
  \abstract
   {In the study of star-forming environments, \CCH\ and \MA\ serve as distinct, complementary tracers of the low-complexity hydrocarbon network. Their rich rotational spectra constrain the gas temperature and density, while their distinct chemistries diagnose the cold-to-warm chemical stratification surrounding embedded high-mass  young stellar objects (YSOs).}
   {By combining \MA\ and \CCH, we present a first multi-transition analysis of the \NE\ and \SW\ regions in the massive star-forming complex RCW 122. The aim is to investigate the cold and warm gas layers that favour carbon-chain chemistry and to constrain the nature, chemistry, and evolutionary stage of each region.}
   {The observations were performed with the Atacama Pathfinder Experiment (APEX). The analysis is based on 39 \MA\ lines with $J=10$--9 to 20--19 (170.8--341.8~GHz) and four hyperfine components of the \CCH\ $N=4$--3 transition (349.1--349.6~GHz), observed independently toward the NE and SW positions.   Excitation conditions were estimated by applying both local thermodynamic equilibrium (LTE) and non-LTE methods, using rotational-diagram analysis and Markov chain Monte Carlo (MCMC) calculations. Gas--grain chemical models were performed with the  \texttt{Nautilus} code to reproduce the observed abundances over timescales characteristic of high-mass  YSO evolution.}
   {The excitation analysis of \MA\ indicates two physical components suggestive of a thermal gradient. The NE region yields $T\thickapprox 41$--61\,K and $N_{\rm tot}\thickapprox 1\times 10^{16}$\,cm$^{-2}$, while SW exhibits $T\thickapprox 33$--48\,K and $N_{\rm tot}\thickapprox 7\times 10^{15}$\,cm$^{-2}$. In both regions, \CCH\ traces cold gas below 15\,K with an H$_2$ density of $0.5$--$4\times 10^6$\,cm$^{-3}$.  The detection of the $^{13}$C isotopologues of \CCH \ indicates that \CTC\ is more abundant than \TCC. Chemical models at various evolutionary stages, with cosmic rays as a key parameter, reproduce the observed \CCH\ and \MA\ abundances, although the two species form through chemically distinct routes.}
   {Observational analyses and chemical modelling provide a coherent scenario in which \CCH\ traces colder gas at an early evolutionary stage, whereas \MA\ traces a region characterized by a warmer temperature gradient, associated with chemically more evolved gas at a later stage. These results constrain the conditions governing the origin and survival of interstellar species in a thermally and chemically stratified medium.}

\keywords{ISM: Molecules -- Molecular data -- Methods: data analysis -- Astrochemistry -- Methods: miscellaneous}

\titlerunning{\CCH\ and \MA}
\authorrunning{RCW 122}

   \maketitle


\section{Introduction}\label{sec1}

Massive stars ($M>8\,M_{\odot}$) are believed to have a critical effect on the physical and chemical evolution of galaxies through their energetic feedback, stellar winds, outflows, HII regions, and, eventually, supernova explosions. Although there remains controversy regarding the formation of massive stars, theories agree that they begin their lives deeply embedded in dense condensations called molecular clumps,  with their earliest observable stage corresponding to deeply embedded high-mass young stellar objects (YSOs) (see \citealt{Z&Y2007,Tan2014,Motte2018,Beuther2025} for a complete review). Within an evolved molecular clump, it is possible to identify the so-called hot molecular core phase, which is important because these objects are believed to be the birthplace of the next generation of massive stars prior to the onset of ionising radiation. Hot molecular cores (HMCs) are small ($\lesssim 0.1$ pc), hot ($>100$ K), dense ($>10^6$ cm$^{-3}$), and luminous ($>10^4$ L$_{\odot}$) (e.g., \citealt{vD&B1998,kurtz2000}). Once the HMC phase begins, temperatures above $\sim 100$ K lead to the sublimation of icy dust mantles that accumulated in cold clumps during the earliest stages of star formation; this, together with the high temperatures, leads to the emergence of a wide variety of molecular species, namely O-, N-, and S-bearing molecules, carbon chains, and complex organic molecules (COMs; \citealt{H&vD2009}). 

Single-dish observations of HMCs often reveal a dense forest of rotational lines at millimetre and sub-millimetre wavelengths, which can be used to constrain key physical properties (e.g., \citealt{font2007,bell2013,suz2016,wid2017,osp2018,bous2022,bous2024}). These lines also provide the observables needed for the rotational-diagram method, which, under LTE assumptions, derives excitation temperatures and molecular column densities from multiple transitions spanning a range of upper-state energies \citep{G&L1999}. Complementary interferometric observations are often necessary to resolve the true spatial extent of an astronomical source, allowing for a correction by the beam dilution factor, and ensuring that the rotational temperature and column density are more accurate (e.g. \citealt{Hervias2019,dur2019,canelo2021,sant2022,Mendoza2023}).

In this paper, we derive key physical properties of two well-identified HMCs embedded in the massive star-forming complex RCW122, hereafter \NE\ (NE) and \SW (SW), through a wide-frequency-coverage, multi-transition analysis of the ethynyl radical (\CCH) and methyl acetylene (\MA), using the Atacama Pathfinder Experiment (APEX) telescope. We also use complementary interferometric data available through the Atacama Large Millimeter/submillimeter Array (ALMA) archive. A comprehensive survey of \MA\ and \CCH\ lines toward both clumps provides a powerful means to assess how different components of the molecular gas are associated with ongoing star-formation activity. In particular, the complementary nature of these tracers raises the question of whether \MA\ identifies relatively warm and chemically processed gas surrounding the  high-mass YSOs \citep{Hickson2016}, while \CCH\ traces a more extended, colder, and less evolved component \citep{beu2008}. Their common origin within the same molecular complex further enables a direct comparison of the two HMCs, allowing us to investigate whether differences in their physical conditions, chemistry, and kinematics are linked to distinct formation and evolutionary pathways.

\CCH\ is a linear molecule ($^{\bullet}$C$\equiv$C$-$H) with an unpaired electron in its valence shell, i.e. a radical. In the vibrational ground state of \CCH\, spin--rotation interaction produces fine-structure splitting (FS), while coupling with the hydrogen nuclear spin gives rise to hyperfine structure (HFS). The corresponding angular-momentum coupling scheme is $J = N + S$ (FS) and $F = J + I$ (HFS), where $N$ is the rotational angular momentum, $S=1/2$ is the electron spin, $I=1/2$ is the hydrogen nuclear spin, $J$ is the total angular momentum excluding nuclear spin, and $F$ is the total angular momentum including nuclear spin. \CCH\ is one of the first radicals observed in the ISM \citep{tuc1974}. It is one of the most abundant ISM molecules and has been detected toward a wide variety of objects and regions, including massive star-forming regions (MSFRs), photodissociation regions (PDRs), infrared dark clouds (IRDCs), and protoplanetary discs (PPDs)  \citep{prat1997,dut1997,tey2004,beu2008}.

\MA\ is a symmetric-top molecule that exhibits groups of closely spaced rotational transitions, commonly referred to as $K$-ladders, allowing multiple lines to be observed simultaneously within the same spectral bandwidth. The rotational spectrum of \MA\ is characterized by the quantum numbers $J$, which represent the total rotational angular momentum, and $K$, corresponding to its projection along the molecular symmetry axis (\citealt{tow1975,mul2000}). The molecule has a relatively small electric dipole moment ($\mu$ = 0.75 D; \citealt{dub1978}), which enables its rotational transitions to thermalise at densities as low as $\sim$ 10$^4$ cm$^{-3}$ (\citealt{berg1994,font2002}). Since radiative transitions with $\Delta K \ne 0$ are forbidden, the populations of the different $K$-ladders are primarily determined by collisions. Consequently, the relative intensities among the ladders are highly sensitive to the gas kinetic temperature, making \MA\ an excellent tracer of physical conditions and a reliable molecular thermometer (\citealt{moli2016,gian2017}). Because of these properties, \MA\ has been extensively used to investigate Galactic star-forming regions and dense cores (\citealt{holl1981,kuip1984,guzm2018,boge2019,sant2022}), as well as extragalactic objects (\citealt{maue1991,qiu2020}). It has also been reported in planetary nebulae (\citealt{schm2019}).

\subsection*{Targeted sources}\label{sec1.2}

\begin{figure*}
\centering
\includegraphics[width=0.9\linewidth]{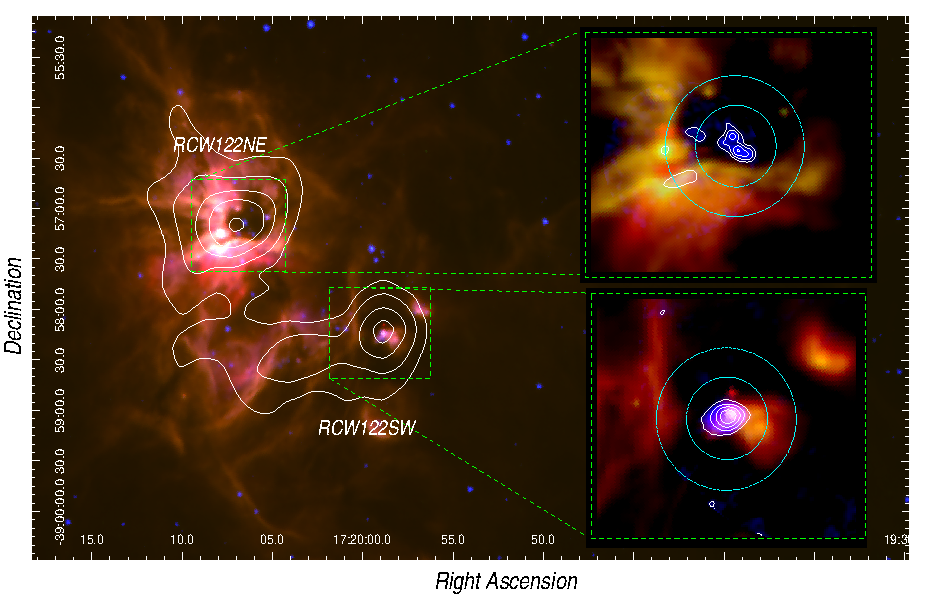}
\caption{Main panel: Studied region of RCW 122. Composite images in red,  green, and blue colours show emission at 8.0,  4.5, and 3.6 $\mu$m  ({\it Spitzer}), respectively. White contours showcase the emission distribution of C$_2$H (3,7/2,4-2,5/2,3) averaged in the velocity range from $-14.0$ to $-12.6$ km s$^{-1}$ obtained with APEX data (see \citealt{dur2024} for a description). Insets: Three colour composite images showing a zoom-in of the targeted sources. Red and green tonalities show the Spitzer 8.0 and 4.5 $\mu$m emission, respectively. In contrast, blue tonalities along with white contours depict the ALMA continuum emission at 220 and 100 GHz for \NE\ and \SW, respectively. The cyan dashed circles depict the beams of the SEPIA-180 and SEPIA-345 APEX observations. }
\label{fig-presentacion}
\end{figure*}

In Fig.~\ref{fig-presentacion}, we show the central region of the star-forming region RCW\,122, a giant molecular cloud/HII region located in the fourth quadrant of the Galactic plane [($l,b$)=(348\gra.24,--1\gra.04)] at a distance of 3.38 kpc \citep{Wu2012}. This complex is associated with a large molecular cloud of $2.2 \times 10^5~M_{\odot}$ (see \citealt{arn2008}). The central region contains two main sources identified by \citet{mcbre1985} as RCW\,122A and RCW\,122B; these sources were later studied by \citet{soll2004}, who dubbed them \NE\ and \SW, respectively, and detected strong HC$_3$N emission (a typical HMC molecule) in both objects. These sources are spatially coincident with regions bright in the four {\it Spitzer} bands (see Fig.~\ref{fig-presentacion}) and with two $870~\mu$m emission clumps, classified in the ATLASGAL catalog of \citet{con13} as AGAL348.726-01.039 and AGAL348.698-01.027. Furthermore, \NE\ and \SW\ are strong maser emitters (\citealt{sca80,bra83,for89,chen11,yang17}).

A complete millimetre and sub-millimetre study of \NE\ and \SW\ (designated in that work as clumps B and D, respectively), as well as other compact molecular clumps in RCW\,122 with AGAL counterparts, was performed by \citet{dur2024}. In addition to typical molecular gas tracers such as CO and HCO$^+$ (and their corresponding isotopologues), they reported a preliminary analysis of SO, H$_2$CO, CH$_3$OH, \CCH,\ and \MA\ in \NE\ and \SW\ using APEX data. The sources also exhibit high temperatures, densities, and luminosity-to-mass ratios, as well as strong and very compact radio continuum emission, which the authors interpreted as indicating the HMC phase, or a very early stage of the subsequent ultracompact HII region (UCHII) phase. In Table~\ref{tab:propiedades}, we summarize some of the key physical properties derived by \citet{dur2024} for the molecular clumps associated with \NE\ and \SW.  We outline the column densities obtained from the C$^{18}$O(2-1) line, $N$(C$^{18}$O)$_{\rm LTE}$, assuming LTE and excitation temperatures obtained from the optically thick emission of CO(2-1), $\boldsymbol{T_{\rm ex}}$; the total masses, $M({\rm H_2})_{\rm LTE}$, were obtained from the total hydrogen column densities, $N$(H$_2$)$_{\rm LTE}$, derived assuming a standard abundance of C$^{18}$O. We also summarize the temperatures and bolometric luminosities, $T_{\rm c}$ and $L_{\rm bol}$, respectively, derived by fitting spectral energy distributions (SEDs) in the submillimeter range, as well as the masses derived from optically thin emission at 870 microns, $M(\rm H_2)_{\rm 870}$.

\begin{table}
\begin{center}
\caption{Main physical properties derived for the molecular clumps associated with \NE\ and \SW\ (extracted from \citet{dur2024})}
\label{tab:propiedades}
{\small
\begin{tabular}{lcc}
\hline
\hline
                                              &  \NE       & \SW  \\
\hline                                    
$\boldsymbol{T_{\rm ex}}$ [K]                             &    57            &  45        \\
$N$(C$^{18}$O)$_{\rm LTE}$ [10$^{16}$cm$^{-2}$]           &   5.2            &  2.7       \\
$N$(H$_2$)$_{\rm LTE}$ [10$^{22}$cm$^{-2}$]              &   26.1            &  13.4       \\   
$M({\rm H_2})_{\rm LTE}$  [10$^3$$M_{\odot}$] &   4.4            &  1.6       \\
$T_{\rm c}$[K]                                &   33.9           & 32.7       \\
$L_{\rm bol}$[10$^3$L$_{\odot}$]              &   238            & 93         \\
$M(\rm H_2)_{\rm 870}$ [$10^3$$M_{\odot}$]    &   3.8            & 1.5            \\
\hline
\hline
\end{tabular}
}
\end{center}
\end{table}

In this work, we provide a more detailed view of \NE\ and \SW\ in the insets of Fig.~\ref{fig-presentacion}, using recently acquired archival ALMA data. The images show millimeter continuum emission at 100 and 220 GHz, revealing a clearer picture of the clumps identified by \citet{dur2024} (Ortega et al., in prep.). As a first estimate of the source sizes, we estimated their diameters from their effective and deconvolved radii measured in the ALMA continuum emission.\footnote{$R_{\rm D}=\left(R_{\rm eff}^{2}-({\rm hpbw}/2)^{2}\right)^{1/2}$, where $R_{\rm eff}=(A/\pi)^{1/2}$, $A$ is the area of the source, and hpbw is the synthesized beam size.} For \SW, the 100 GHz continuum emission is concentrated into a spot with a diameter of $\sim9.8$\arcsec \ ($\sim0.16$ pc at a distance of $\sim3.38$ kpc).  For \NE, the bulk of the 220 GHz continuum emission appears concentrated into an irregular, clumpy structure with a diameter of $\sim8.2$\arcsec \ ($\sim0.13$ pc), nearly coincident with the peak of the integrated C$_2$H emission. 

The paper is organised as follows. Section~\ref{sec2} describes the methodology adopted in this work. Section~\ref{sec3} presents the identification and analysis of the molecular lines, and Section~\ref{sec4} focuses on the excitation analysis, including both LTE and non-LTE approaches. The astrochemical implications of the results are discussed in Section~\ref{sec5}. Section~\ref{sec6} summarises the main conclusions. Appendices~\ref{app1}~and~\ref{app2}\ present supplementary material.

\section{Methodology}\label{sec2}

\subsection{Observations}\label{sec2.1}

\subsection*{APEX dataset}

The molecular single-dish observations presented in this paper were obtained with the 12-m APEX telescope  \citep{gu06} at Llano de Chajnantor (Chilean Andes) between October 2020 and November 2021. As a front-end for the observations, we used the SEPIA-180 and SEPIA-345 receivers \citep{beli2018,beli2018b}. The SEPIA180 channel is a dual-polarisation (2SB) receiver with two IF outputs per polarisation,  the upper and lower sidebands (USB and LSB, respectively), each covering 4-8 GHz, adding up to a total of 16 GHz instantaneous IF bandwidth. The central frequencies of the two sidebands are separated by 12 GHz. The sideband rejection ratio is by design $>$10 dB and 18.5 dB on average. The single-sideband noise temperature is below 55 K at all frequencies within the RF band. The  half-power beam width  (HPBW) of APEX at the SEPIA-180 frequencies is in the range 30 - 39\arcsec.\  The total frequency range coverage of the instrument is 159 -  211 GHz.  The SEPIA345 channel is a dual polarisation, 2SB, receiver with two IF outputs per polarisation, USB and LSB, each covering 4 - 12 GHz, adding up to a total of 32 GHz instantaneous IF bandwidth. The central frequencies of the two sidebands are separated by 16 GHz. Each sideband (and polarisation) is recorded by two FFTS spectrometer units, each of them sampling 4 GHz, in the following configuration:FFTS1: 4.17 – 8.17 GHz IF bandwidth, and FFTS2: 8.07 – 12.07 GHz IF bandwidth;  both units overlap in the middle for about 100 MHz, and the full coverage is 7.9 GHz. The sideband rejection ratio is  $\>$ 10 dB over 90 $\%$ of the band. The HPBW of APEX at the SEPIA-345 frequencies is in the range 17 - 23\arcsec.\  The total frequency range coverage of the instrument is 272 -  376 GHz. 

The spectra were obtained in the single-point mode at the coordinates  \radec = 17$^h$20$^m$06.1$^s$, $-$38\gra57\arcmin15.4\arcsec\ and  \radec = 17$^h$19$^m$58.9$^s$, $-$38\gra58\arcmin14.1\arcsec\ (the estimated centers of \NE\ and \SW, respectively; see Fig.~\ref{fig-presentacion}). For SEPIA-180, we observed in seven frequency setups to achieve a total frequency coverage of 162.630 - 210.770 GHz. For the SEPIA-345 observations, we used nine frequency setups to cover the frequency range of 272.400 - 369.100 GHz. The data reduction was performed using the CLASS package within the GILDAS software suite. The spectra were retrieved from the APEX real-time calibration pipeline, already stored in the CLASS data format and calibrated to the corrected antenna temperature scale ($T_A^*$) \citep{Ulich1976,Pety2005}.\footnote{\url{https://www.apex-telescope.org/telescope/efficiency/index.php}}  Throughout this work, all details pertaining to the line identification and the characterization of the spectra are provided in units of $T_A^*$ as a function of velocity relative to the local standard of rest ($V_{\rm LSR}$), adopting systemic velocities of $V_{\rm LSR}$ = -11.3 km s$^{-1}$ and -12.3 km s$^{-1}$ for \NE\ and \SW,\ respectively. All spectra were smoothed to a common velocity resolution of 1 km s$^{-1}$ to improve uniformity across the dataset. 

Spectral line identification was performed by cross-referencing databases such as the Cologne Database for Molecular Spectroscopy (CDMS, \citealt{Muller2001,Endres2016})\footnote{\url{https://cdms.astro.uni-koeln.de/}} and the JPL Molecular Spectroscopy Catalog \citep{pick98}.\footnote{\url{https://spec.jpl.nasa.gov/}} Spectral line modeling and the derivation of physical conditions were carried out using the CASSIS software \citep{Vastel2015}.\footnote{\url{http://cassis.irap.omp.eu/}}  For the radiative transfer and quantitative analyses, the line intensities and integrated areas were obtained from the spectra converted to the main-beam temperature scale ($T_{\rm MB}$) by applying Ruze's equation and using typical main-beam efficiencies ($\eta_{\rm mb}$ = 0.80--0.85) for the SEPIA180 and SEPIA345 receivers, respectively, and a surface accuracy (rms) of 17$\mu$m. To account for beam dilution, source angular sizes of $\theta_s \simeq$ 8.2\arcsec \ and 9.8\arcsec \ were adopted for \NE\ and \SW, respectively.  A
larger, nominal size of $20\arcsec$ was adopted for emission components
tracing more extended gas. A calibration uncertainty equivalent to 15\% has been adopted as part of the data treatment and analysis procedures \citep{dum2010}.

For \MA, we performed an Local Thermodynamic Equilibrium (LTE) analysis using the population diagram method \citep{G&L1999} to constrain the excitation conditions. Thus, the so-called rotational diagrams  were constructed  using CASSIS. For each transition, the integrated-intensity uncertainty ($\Delta W$) combined the calibration uncertainty and the rms noise:
\begin{equation}
\Delta W = \sqrt{(\mathrm{cal}/100 \times W)^2 + \left(\mathrm{rms}\sqrt{2\,\mathrm{FWHM}\,\Delta v}\right)^2},
\end{equation}
where $W$ is the integrated area, cal is the calibration uncertainty ($\%$), FWHM is the full width at half maximum, and $\Delta v$ is the velocity bin size (e.g., \citealt{Vastel2015,dur2019}). This uncertainty is propagated to the rotational diagram, and a weighted least-squares fit to all detected transitions returned the excitation temperature and total column density ($T_{\rm ex}$ and $N_{\rm tot}$, respectively), after applying beam dilution and opacity corrections to each transition. For the $^{13}$C isotopologues of \CCH,\ whose hyperfine components are unresolved and blended, column densities were also estimated  assuming LTE. This assumption is justified by the optically thin regime of the emission.  We used the Markov chain Monte Carlo (MCMC) sampler \citep{Foreman2013} to infer the joint posterior distribution of $N_{\rm col}$ and $\boldsymbol{T_{\rm ex}}$ conditioned on the observed line intensities under the LTE model.

For \CCH,\ non-LTE radiative transfer calculations were performed with the RADEX code \citep{van2007}, using collisional rate coefficients retrieved from the LAMDA database (e.g., \citealt{Schoier2005,Spielfiedel2012,Spielfiedel2013}.\footnote{\url{https://home.strw.leidenuniv.nl/~moldata/}} Parameter estimation was carried out using the MCMC algorithm, running 10$^5$ steps and discarding the first 30 \% as burn-in, with a mean acceptance rate of $<$0.5. The fitting was performed independently for each source using all available transitions and comparing the 4 most intense components of the \CCH\ $N=4-3$ hyperfine structure.

\subsection*{ALMA complementary data}

ALMA continuum data for both targets, \NE \ and \SW,\ were obtained from the ALMA Science Archive\footnote{\url{http://almascience.eso.org/aq/}}. For \NE, we used data from Project 2019.1.00195 (PI: S. Molinari) observed with the 12~m array. For this configuration, the L5BL and L80BL of the baseline distribution were 26.2~m and 139.4~m, respectively. The observed frequency range goes from 217.012 to 220.947~GHz (Band 6). The primary beam field of view (FoV) is $\sim$26\farcs59. The final beam size and beam position angle (B.P.A.) are 1\farcs30 $\times$ 1\farcs12 and 82.1$^\circ$, respectively. The rms noise level  (1 sigma) is $\sim$0.1507~mJy~beam$^{-1}$ for the continuum emission, and the maximum resolvable scale is 10\farcs58. At the assumed distance of 3.38 kpc, this beam size corresponds to a spatial resolution of approximately 0.020~pc ($\sim$4100~au). For \SW, we used data from Project 2015.1.01209 (PI: L. Sjouwerman), also taken with the 12~m array. For this configuration,  the L5BL and L80BL of the baseline distribution were 27.75~m and 158.06~m, respectively. The observed frequency range covers 85.406 to 98.45~GHz (Band 3). The primary beam FoV is $\sim$63\farcs34. The resulting beam size and B.P.A. are 2\farcs8 $\times$ 2\farcs3 and 75.8$^\circ$, respectively, with a continuum rms noise level of $\sim$0.13~mJy~beam$^{-1}$ and a maximum resolvable scale of 23\farcs82. This setup provides a spatial resolution of approximately 0.042~pc ($\sim$8670~au) at the source distance. The archived datasets for both projects successfully passed the QA2 quality level, ensuring reliable calibration for science-ready data. 

\subsection{Chemical modelling}\label{sec2.2}

Gas-grain chemical models were computed to predict the molecular abundances of \CCH\ and \MA\ with the  \texttt{ Nautilus} astrochemical code \citep{Ruaud2016}, which computes the time-dependent abundances of chemical species in the gas phase and on interstellar dust grains, including grain surfaces and mantles. It solves the kinetic rate equations for a prescribed set of physical conditions, such as gas density ($n$, cm$^{-3}$), temperature ($T$, K), visual extinction ($A_{\rm v}$, mag), and the  cosmic-ray ionisation rate of H$_2$ ($\zeta$, s$^{-1}$), for which the code adopts a fiducial value of $\simeq$ 10$^{-17}$ s$^{-1}$ \citep{Wakelam2024}. Earlier work provided a preliminary investigation of \MA\ formation in hot molecular cores \citep{sant2022}. Here we extend that analysis by examining the formation and evolution of \CCH\ and \MA\ using the Kinetic Database for Astrochemistry (KIDA) \citep{Wakelam2015,Wakelam2024}. We consider multiple evolutionary stages, with physical conditions guided by the tracers observed in the NE and SW regions.

The modelling is designed (i) to identify the ranges of physical conditions and chemical ages that reproduce abundances consistent with those derived from the observations, and (ii) to determine the dominant formation and destruction routes to assess whether the chemistry supports a direct link between \CCH\ and \MA.\ The simulations are framed within a high-mass YSO evolutionary sequence. In this context, \CCH\ predominantly traces extended, weakly processed gas at earlier stages \citep{beu2008,Fontani2025}, whereas \MA\ is more representative of chemically evolved material in warmer layers surrounding the high-mass YSOs \citep{miettinen2006}.

All models start from a common molecular-cloud stage with standard elemental abundances. We first evolve the system for $1$~Myr at a gas number density of $2 \times 10^4$~cm$^{-3}$ and a temperature of $10$~K. For \CCH,\ the resulting abundances are used as initial conditions for a second stage, where we explore temperatures in the range $10$--$20$~K and enhanced cosmic-ray ionisation rates, $\zeta = \zeta_0(1$--$100)$, where $\zeta_0$ is the standard cosmic-ray ionisation rate, equal to $1.3 \times 10^{-17}$~s$^{-1}$  (e.g., \citealt{vanderTak2000}), as a function of time up to 0.1~Myr. For \MA, a substantial fraction of the abundance is expected to be produced on grains \citep{miettinen2006,Hickson2016,Andron2018,sant2022}. Accordingly, we compute a second stage starting from the abundances at the end of the molecular-cloud phase. This stage represents the peak-temperature phase, in which the temperature and density are increased up to $200$~K and $10^{6}$~cm$^{-3}$, respectively, and the model is run for $0.01$~Myr. We then compute a cooling stage for \MA\ using the end-of-stage abundances, allowing the gas to cool to $T =(30-60)$~K and decreasing the density to $10^{5}$~cm$^{-3}$. In these final runs, we tested lower ionisation rates, $\zeta = (0.1$--$1)\zeta_{0}$, adopting $\zeta = 0.1\zeta_{0}$ to avoid rapid destruction of \MA.

This multi-stage framework is intended to capture the chemical evolution of \CCH\ and \MA\ and to reproduce their observed abundances under quasi-steady-state conditions. We therefore compare with the observations using the final evolutionary stage, for which the modelled abundances vary by less than $20\%$ and do not show rapid excursions (peaks, oscillations, or sharp transients) on timescales of $\sim 0.01$~Myr under the adopted physical conditions \citep{Lee1998,Pilleri2013,Roueff2020,Dufour2023}. The resulting abundance profiles and their chemical implications are analysed in Sects.~\ref{sec5.1}--\ref{sec5.2}, and supplementary modelling details are provided in Appendix~\ref{app1}.

\section{Spectral line analysis}\label{sec3}

\subsection{Detection of \MA}\label{sec3.1}

Here we present an extensive survey of \MA\ toward \NE\ and \SW. Across the frequency range of 170.8--341.8 GHz, 39 spectral lines were identified in each region and fitted using Gaussian profiles. 
 The spectroscopic and line-fit parameters of   \MA\ in the NE and SW regions are summarised in Tables~\ref{tab:ma-ne} and~\ref{tab:ma-sw}, respectively. The listed rest frequencies, upper-state energies ($E_{\rm u}$), and Einstein coefficients ($A_{\rm ul}$) were taken from spectroscopic databases (Sect.~\ref{sec2.1}). The beamwidth corresponds to the APEX HPBW at the line frequency. The kinematic quantities (systemic velocity, $V_{\rm LSR}$), FWHM and $W$ were derived from single-component Gaussian fits to the $T_{\rm A}^{*}$ spectra. Figure~\ref{ma-lines} presents all analysed lines of \MA\ in the NE and SW regions.
There have been covered all the rotational levels from $J=10$--$9$ to $J=20$--$19$, with the exception of three transitions ($J_{\rm u}$=13, 14, and 15) due to missing data. Regarding the $K$-ladders associated with each rotational level of \MA, all ladders corresponding to $K$ = 0 through 4 have been analysed. Among these, transitions from the $K$ = 0 and $K$ = 1 ladders--despite their close spectral separation--are resolved and represent the most intense features in the spectra. Furthermore, the $K$ = 4 ladder, corresponding to the faintest lines across the entire spectral range, has also been detected and analysed in all but a few cases (e.g., the $J$=19$_4$--18$_4$ transition). It is worth noting that a search for all the \TC\ and D-substituted isotopologues of \MA\ was conducted; however, no clear identifications were made.

\begin{table*}[h!]
\centering
\caption{Spectroscopic and line parameters of the \MA\ and \CCH\ transitions identified toward \NE. Numbers in parentheses represent the 
1$\sigma$ uncertainty in the last digits.}\label{tab:lines-ne} 
\begin{tabular}{lccccccc}
\hline\hline
Transition & Frequency  & Beamwidth  & $E_{\rm u}$ & $A_{\rm ul}$  & $V_{\rm LSR} $  & FWHM  & $\int T_{\rm A}dV_{\rm lsr}$   \\
 & MHz & \arcsec & K &  10$^{-5}$ s$^{-1}$ & km s$^{-1}$ & km s$^{-1}$ & K km s$^{-1}$  \\
\hline
\multicolumn{8}{c}{\MA} \\
 $10_4$--$9_4$	&	170853.576	&	36.5 & 160.70	&	1.43	&	-11.312	(58)	&	3.11	(25)	&	0.373	(01)	\\
 $10_3$--$9_3$	&	170876.41	&	36.5 & 110.14	&	1.55	&	-11.503	(12)	&	3.20	(06)	&	2.307	(01)	\\
 $10_2$--$9_2$	&	170892.726	&	36.5 & 74.01	&	1.63	&	-11.561	(12)	&	3.28	(07)	&	2.782	(01)	\\
 $10_1$--$9_1$	&	170902.518	&	36.5 & 52.33	&	1.68	&	-11.620	(13)	&	3.29	(06)	&	4.686	(03)	\\
 $10_0$--$9_0$	&	170905.783	&	36.5 & 45.11	&	1.70	&	-11.605	(07)	&	3.30	(03)	&	5.425	(02)	\\
\hline
 $11_4$--$10_4$	&	187936.23	&	33.2 & 169.72	&	1.97	&	-11.401	(71)	&	3.48	(31)	&	0.419	(01)	\\
 $11_3$--$10_3$	&	187961.342	&	33.2 & 119.16	&	2.10	&	-11.416	(15)	&	3.31	(07)	&	2.037	(01)	\\
 $11_2$--$10_2$	&	187979.286	&	33.2 & 83.04	&	2.20	&	-11.492	(12)	&	3.08	(06)	&	2.216	(01)	\\
 $11_1$--$10_1$	&	187990.055	&	33.2 & 61.36	&	2.25	&	-11.520	(10)	&	3.31	(06)	&	3.910	(02)	\\
 $11_0$--$10_0$	&	187993.645	&	33.2 & 54.13	&	2.27	&	-11.505	(07)	&	3.25	(03)	&	4.348	(01)	\\
\hline
 $12_4$--$11_4$	&	205018.114	&	30.4 & 179.56	&	2.63	&	-11.103	(12)	&	3.32	(23)	&	0.634	(02)	\\
 $12_3$--$11_3$	&	205045.501	&	30.4 & 129.01	&	2.77	&	-11.473	(11)	&	3.21	(10)	&	2.806	(01)	\\
 $12_2$--$11_2$	&	205065.07	&	30.4 & 92.88	&	2.88	&	-11.548	(29)	&	3.14	(68)	&	3.020	(03)	\\
 $12_1$--$11_1$	&	205076.816	&	30.4 & 71.20	&	2.94	&	-11.599	(17)	&	3.14	(04)	&	4.869	(03)	\\
 $12_0$--$11_0$	&	205080.732	&	30.4 & 63.97	&	2.96	&	-11.562	(13)	&	3.17	(02)	&	5.506	(03)	\\
\hline
 $16_4$--$15_4$	&	273336.519	&	22.8 & 227.11	&	6.64	&	-10.893	(12)	&	3.59	(40)	&	0.704	(05)	\\
 $16_3$--$15_3$	&	273372.99	&	22.8 & 176.56	&	6.84	&	-11.047	(45)	&	3.62	(65)	&	2.757	(04)	\\
 $16_2$--$15_2$	&	273399.051	&	22.8 & 140.44	&	6.98	&	-11.103	(33)	&	3.12	(49)	&	2.305	(03)	\\
 $16_1$--$15_1$	&	273414.692	&	22.8 & 118.77	&	7.06	&	-11.135	(30)	&	3.51	(41)	&	3.997	(04)	\\
 $16_0$--$15_0$	&	273419.906	&	22.8 & 111.54	&	7.09	&	-11.107	(39)	&	3.50	(20)	&	4.437	(08)	\\
\hline
 $17_4$--$16_4$	&	290413.488	&	21.5 & 241.05	&	8.04	&	-10.708	(19)	&	2.78	(67)	&	0.460	(06)	\\
 $17_3$--$16_3$	&	290452.224	&	21.5 & 190.50	&	8.25	&	-10.930	(63)	&	3.43	(43)	&	2.350	(05)	\\
 $17_2$--$16_2$	&	290479.904	&	21.5 & 154.38	&	8.40	&	-11.029	(81)	&	3.21	(54)	&	2.060	(07)	\\
 $17_1$--$16_1$	&	290496.516	&	21.5 & 132.71	&	8.49	&	-11.112	(35)	&	3.46	(15)	&	3.540	(06)	\\
 $17_0$--$16_0$	&	290502.054	&	21.5 & 125.48	&	8.52	&	-11.068	(24)	&	3.29	(09)	&	3.748	(04)	
\\
\hline
 $18_4$--$17_4$	&	307489.264	&	20.3 & 255.81	&	9.62	&	-10.426	(29)	&	3.88	(11)	&	0.543	(07)	\\
 $18_3$--$17_3$	&	307530.263	&	20.3 & 205.26	&	9.84	&	-11.044	(20)	&	3.90	(93)	&	2.426	(02)	\\
 $18_2$--$17_2$	&	307559.559	&	20.3 & 169.14	&	10.00	&	-11.018	(52)	&	3.38	(24)	&	1.982	(05)	\\
 $18_1$--$17_1$	&	307577.141	&	20.3 & 147.47	&	10.10	&	-11.096	(30)	&	3.55	(12)	&	3.111	(05)	\\
 $18_0$--$17_0$	&	307583.003	&	20.3 & 140.25	&	10.10	&	-11.019	(35)	&	3.52	(17)	&	3.470	(06)	
\\
\hline
 $19_3$--$18_3$	&	324607.035	&	19.2 & 220.84	&	11.60	&	-11.131	(29)	&	4.19	(21)	&	1.689	(14)	\\
 $19_2$--$18_2$	&	324637.946	&	19.2 & 184.72	&	11.80	&	-11.069	(19)	&	3.95	(66)	&	1.169	(14)	\\
 $19_1$--$18_1$	&	324656.497	&	19.2 & 163.05	&	11.90	&	-11.075	(23)	&	3.31	(52)	&	1.929	(09)	\\
 $19_0$--$18_0$	&	324662.682	&	19.2 & 155.83	&	11.90	&	-11.009	(10)	&	3.58	(21)	&	2.374	(10)	
\\
\hline
 $20_4$--$19_4$	&	341636.954	&	18.3 & 287.78	&	13.40	&	-10.698	(13)	&	4.12	(26)	&	0.625	(02)	\\
 $20_3$--$19_3$	&	341682.47	&	18.3 & 237.23	&	13.60	&	-10.899	(78)	&	3.33	(25)	&	1.254	(04)	\\
 $20_2$--$19_2$	&	341714.994	&	18.3 & 201.12	&	13.80	&	-10.859	(77)	&	3.54	(53)	&	1.429	(04)	\\
 $20_1$--$19_1$	&	341734.513	&	18.3 & 179.45	&	13.90	&	-10.858	(55)	&	4.40	(71)	&	3.018	(04)	\\
 $20_0$--$19_0$	&	341741.021	&	18.3 & 172.23	&	13.90	&	-10.935	(28)	&	3.52	(13)	&	2.261	(03)	\\
\hline
\multicolumn{8}{c}{\CCH} \\
4 9/2 4--3	5/2	3	&	349108.494	&	17.9 & 41.91	&	0.0058	&	-9.12 (06)		&	7.86 (18)		&	2.741 (50)		\\
4 9/2 5--3	7/2	4	&	349337.706	&	17.9 &41.91	&	13.10	&	-11.58 (05)		&	4.74 (13)		&	17.27 (37)		\\
4 7/2 4--3	5/2	3	&	349399.276	&	17.9 &41.93	&	12.50	&	-11.45 (07)		&	4.58 (18)		&	12.85 (39)		\\
4 7/2 4--3	7/2	4	&	349603.614	&	17.8 &41.93	&	0.52	&	-10.87 (06)		&	3.23 (16)		&	0.437 (18)		\\
\hline
\end{tabular}
\label{tab:ma-ne}
\tablefoot{Column definitions: 
Transition: upper to lower quantum numbers ($J_K-(J{-}1)_K$ for symmetric-top \MA; $N_{J,F}\to N'_{J',F'}$ for \CCH\ hyperfine;
Frequency: rest frequency;
Beamwidth: $\theta=1.22\,\lambda/D$, $D=12$~m \citep{gu06};
$E_{\rm u}$: upper-level energy above the ground state;
$A_{\rm ul}$: Einstein $A$ coefficient;
$V_{\rm LSR}$: velocity relative to the local standard of rest;
FWHM: full width at half maximum of the Gaussian fit;
$\int T_{\rm A}^{*}\,{\rm d}V_{\rm LSR}$: integrated line intensity (Sect~\ref{sec2}).}
\end{table*}

\begin{table*}[h!]
\centering
\caption{Same description as in Table~\ref{tab:lines-ne}, but for \SW.}\label{tab:lines-sw}
\begin{tabular}{lccccccc}
\hline\hline
Transition & Frequency  & Beamwidth  & $E_{\rm u}$ & $A_{\rm ul}$  & $V_{\rm LSR} $  & FWHM  & $\int T_{\rm A}dV_{\rm lsr}$   \\
 & MHz & \arcsec & K &  10$^{-5}$ s$^{-1}$ & km s$^{-1}$ & km s$^{-1}$ & K km s$^{-1}$  \\
\hline
\multicolumn{8}{c}{\MA} \\
$10_4$--$9_4$	&	170853.576	&	36.5 &	160.70	&	1.43	&	-12.24	(20)	&	2.66	(33)	&	0.193	(31)	\\
$10_3$--$9_3$	&	170876.41	&	36.5  &	110.14	&	1.55	&	-12.33	(03)	&	2.76	(04)	&	1.612	(37)	\\
$10_2$--$9_2$	&	170892.726	&	36.5 & 74.01	&	1.63	&	-12.27	(02)	&	2.75	(79)	&	2.197	(33)	\\
$10_1$--$9_1$	&	170902.518	&  36.5  &	52.33	&	1.68	&	-12.348	(11)	&	2.82	(41)	&	4.044	(36)	\\
$10_0$--$9_0$	&	170905.783	&	36.5  &	45.11	&	1.70	&	-12.331	(09)	&	2.83	(08)	&	4.922	(39)	\\
\hline
$11_4$--$10_4$	&	187936.23	&	33.2  &	169.72	&	1.97	&	-12.16	(07)	&	2.95	(11)	&	0.250	(14)	\\
$11_3$--$10_3$	&	187961.342	&	33.2  &	119.16	&	2.10	&	-12.24	(01)	&	2.84	(02)	&	1.585	(14)	\\
$11_2$--$10_2$	&	187979.286	&	33.2  &	83.04	&	2.20	&	-12.303	(12)	&	2.78	(87)	&	2.159	(19)	\\
$11_1$--$10_1$	&	187990.055	&	33.2  &	61.36	&	2.25	&	-12.357	(08)	&	2.79	(12)	&	3.706	(22)	\\
$11_0$--$10_0$	&	187993.645	&	33.2  &	54.13	&	2.27	&	-12.329	(06)	&	2.84	(01)	&	4.410	(22)	\\
\hline
$12_4$--$11_4$	&	205018.114	&	30.4 &	179.56	&	2.63	&	-12.58	(17)	&	2.67	(29)	&	0.245	(34)	\\
$12_3$--$11_3$	&	205045.501	&	30.4 &	129.01	&	2.77	&	-12.23	(03)	&	2.81	(04)	&	1.554	(33)	\\
$12_2$--$11_2$	&	205065.07	&	30.4 &	92.88	&	2.88	&	-12.302	(17)	&	2.88	(71)	&	1.859	(26)	\\
$12_1$--$11_1$	&	205076.816	&	30.4 &	71.20	&	2.94	&	-12.283	(12)	&	2.79	(15)	&	3.316	(33)	\\
$12_0$--$11_0$	&	205080.732	&	30.4 &	63.97	&	2.96	&	-12.279	(10)	&	2.88	(02)	&	3.996	(32)	\\
\hline
$16_4$--$15_4$	&	273336.519	&	22.8  &	227.11	&	6.64	&	-12.62	(27)	&	3.70	(46)	&	0.286	(62)	\\
$16_3$--$15_3$	&	273372.99	&	22.8  &	176.56	&	6.84	&	-12.268	(35)	&	3.07	(58)	&	1.228	(36)	\\
$16_2$--$15_2$	&	273399.051	&	22.8  &	140.44	&	6.98	&	-12.301	(44)	&	3.07	(06)	&	1.311	(47)	\\
$16_1$--$15_1$	&	273414.692	&	22.8  &	118.77	&	7.06	&	-12.278	(22)	&	2.93	(03)	&	2.057	(37)	\\
$16_0$--$15_0$	&	273419.906	&	22.8  &	111.54	&	7.09	&	-12.294	(28)	&	3.12	(03)	&	2.534	(58)	\\
\hline
$17_4$--$16_4$	&	290413.488 &	21.5 	&	241.05	&	8.04	&	-12.31	(17)	&	3.25	(26)	&	0.218	(30)	\\
$17_3$--$16_3$	&	290452.224 &	21.5 	&	190.50	&	8.25	&	-12.289	(38)	&	3.22	(06)	&	1.113	(35)	\\
$17_2$--$16_2$	&	290479.904 &	21.5 	&	154.38	&	8.40	&	-12.236	(31)	&	2.94	(04)	&	1.064	(27)	\\
$17_1$--$16_1$	&	290496.516 &	21.5 	&	132.71	&	8.49	&	-12.238	(23)	&	3.14	(04)	&	1.806	(34)	\\
$17_0$--$16_0$	&	290502.054 &	21.5 	&	125.48	&	8.52	&	-12.241	(19)	&	3.34	(04)	&	2.124	(32)	\\
\hline
$18_4$--$17_4$	&	307489.264	&	20.3 &	255.81	&	9.62	&	-12.25	(11)	&	3.31	(39)	&	0.195	(17)	\\
$18_3$--$17_3$	&	307530.263	&	20.3 & 205.26	&	9.84	&	-12.286	(25)	&	3.25	(53)	&	0.949	(20)	\\
$18_2$--$17_2$	&	307559.559	&	20.3 & 169.14	&	10.00	&	-12.301	(27)	&	3.12	(05)	&	0.983	(22)	\\
$18_1$--$17_1$	&	307577.141	&	20.3 & 147.47	&	10.10	&	-12.275	(16)	&	3.22	(03)	&	1.514	(20)	\\
$18_0$--$17_0$	&	307583.003	&	20.3 & 140.25	&	10.10	&	-12.235	(16)	&	3.36	(02)	&	1.775	(23)	\\
\hline
$19_3$--$18_3$	&	324607.035	&	19.2 & 220.84	&	11.60	&	-12.31	(21)	&	3.19	(40)	&	0.698	(119)	\\
$19_2$--$18_2$	&	324637.946	&	19.2 & 184.72	&	11.80	&	-13.06	(18)	&	4.47	(98)	&	1.041	(145)	\\
$19_1$--$18_1$	&	324656.497	&	19.2 & 163.05	&	11.90	&	-12.32	(18)	&	3.40	(26)	&	0.962	(143)	\\
$19_0$--$18_0$	&	324662.682	&	19.2 & 155.83	&	11.90	&	-11.63	(25)	&	3.72	(30)	&	1.295	(276)	\\
\hline
$20_4$--$19_4$	&	341636.95	&	18.3 & 287.78	&	13.40	&	-12.60 (28)     &	2.57 (50)           	&	0.083 (17)          	\\
$20_3$--$19_3$	&	341682.47	&	18.3 & 237.23	&	13.60	&	-12.41	(07)	&	3.19	(16)	&	0.499	(27)	\\
$20_2$--$19_2$	&	341714.994	&	18.3 & 201.12	&	13.80	&	-12.48	(16)	&	3.29	(10)	&	0.606	(79)	\\
$20_1$--$19_1$	&	341734.513	&	18.3 & 179.45	&	13.90	&	-12.181	(39)	&	3.43	(09)	&	0.797	(26)	\\
$20_0$--$19_0$	&	341741.021	&	18.3 & 172.23	&	13.90	&	-12.287	(47)	&	3.79	(10)	&	0.954	(36)	\\
\hline
\multicolumn{8}{c}{\CCH} \\
4 9/2	4--	3	5/2	3	&	349108.494	&	17.9 & 41.91	&	0.0058	&	-12.66 (29)		&	3.90 (59)		&	0.142 (20)		\\
4	9/2	5--	3	7/2	4	&	349337.706	&	17.9 & 41.91	&	13.10	&	-12.776 (14)		&	4.287 (36)		&	12.938 (90)		\\
4	7/2	4--	3	5/2	3	&	349399.276	&	17.9 & 41.93	&	12.50	&	-12.780 (23)		&	4.262 (59)		&	10.126 (12)		\\
4	7/2	4--	3	7/2	4	&	349603.614	&	17.8 & 41.93	&	0.52	&	-12.156 (76)		&	3.08 (19)		&	0.279 (14)		\\
\hline
\end{tabular}
\label{tab:ma-sw}
\tablefoot{Column definitions: Same as Table~\ref{tab:lines-ne}}.
\end{table*}

The \MA\ lines are predominantly unblended and free of detectable contamination across the observed spectral range, with the most robust detections occurring in the lower rotational transitions ($J < 12$).  For instance, the integrated-intensity ratio between the $J$=10$_0$--9$_0$ and $J$=20$_0$--19$_0$ lines is $\sim$2.4 in the NE and $\sim$5.2 in the SW, consistent with the lowest-$J$ lines being the brightest in both regions. The spectral baseline is generally flat and stable in both the NE and SW regions. However, baseline anomalies are observed toward NE: a localized deviation is evident in the $J = 12-11$ transition, with additional, milder anomalies detected in the $J = 18-17$ and $J = 20-19$ lines. These features are absent in the corresponding SW spectra, suggesting a region-specific environmental influence--such as continuum structure, kinematic complexity, or warm dust emission. Yet, further examination is needed to better constrain such features. Regarding the $J = 19-18$ transition, it was detected and analysed up to $K = 3$ for both regions.

\begin{figure*}[!h]
\centering
\includegraphics[width=0.42\linewidth]{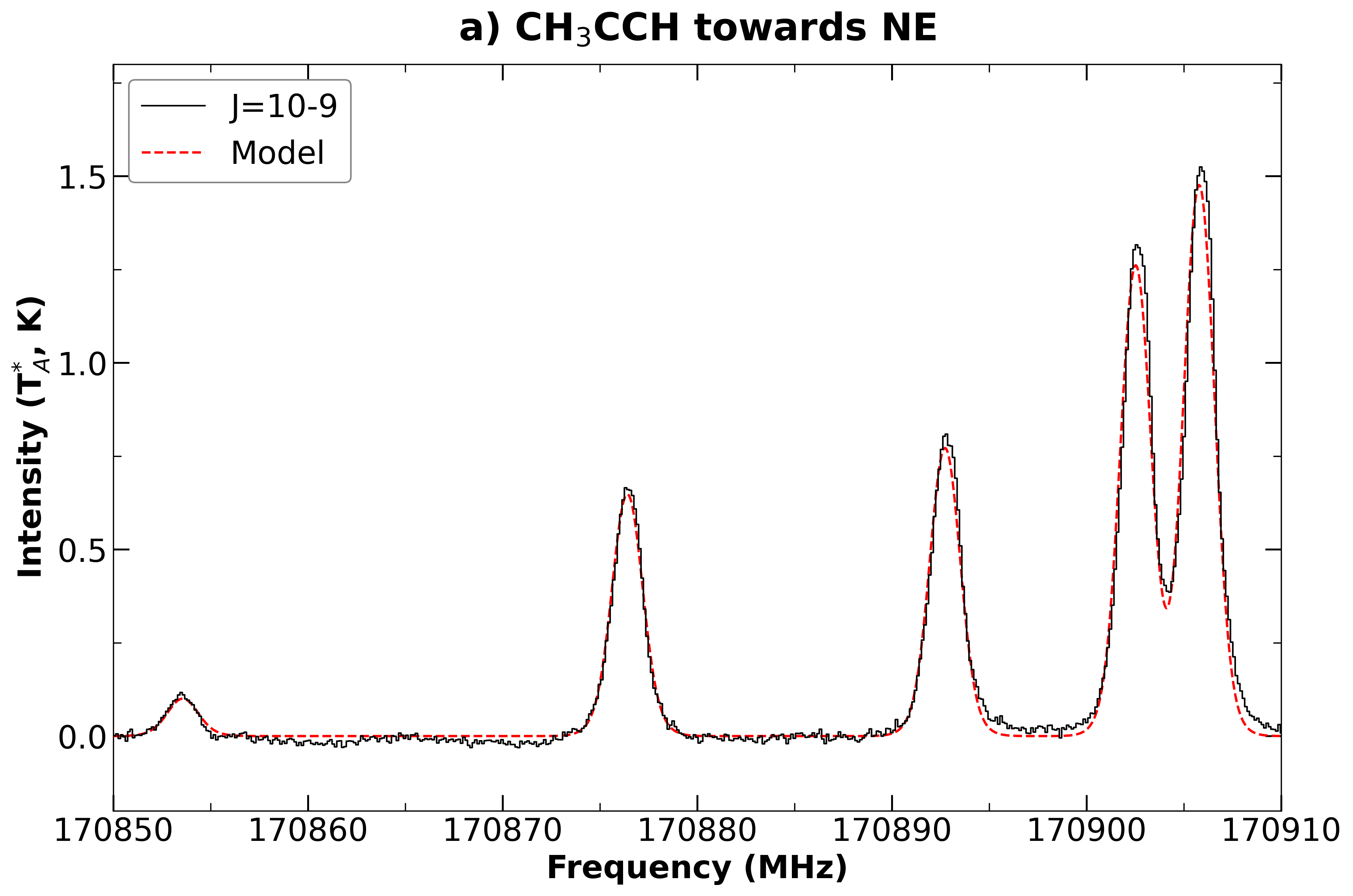}\hspace{0.5cm}
\includegraphics[width=0.42\linewidth]{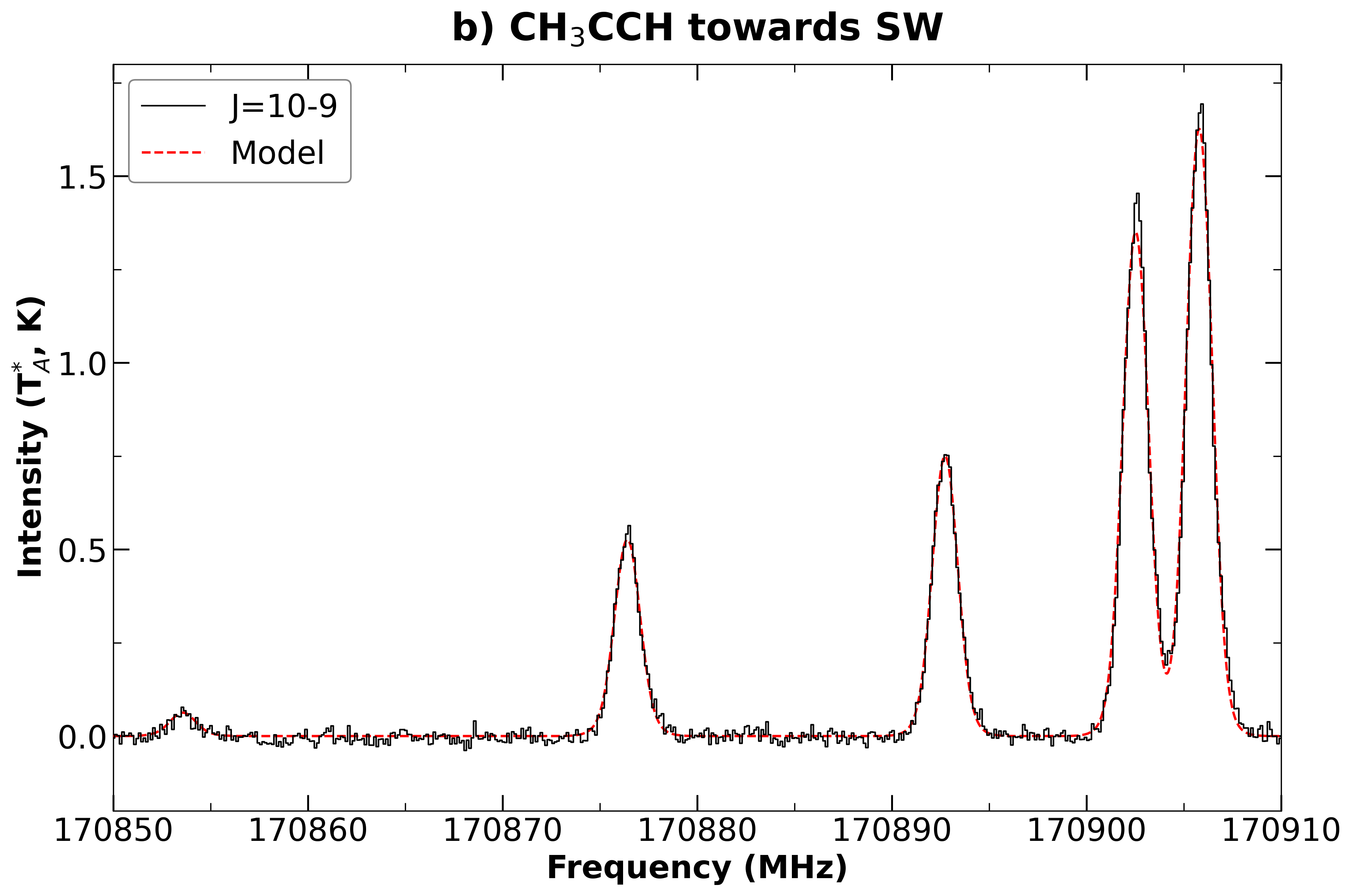}
\includegraphics[width=0.42\linewidth]{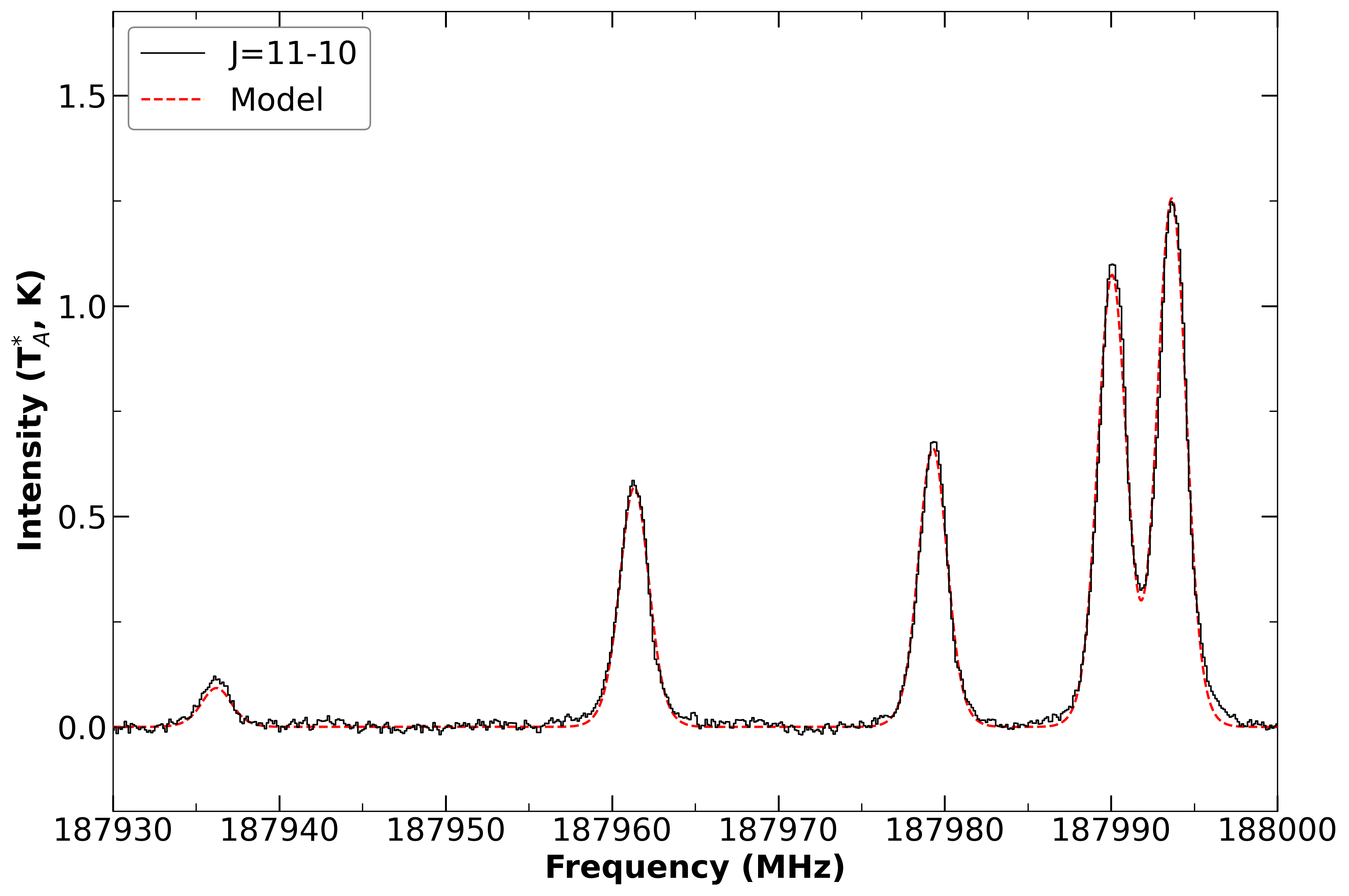}\hspace{0.5cm}
\includegraphics[width=0.42\linewidth]{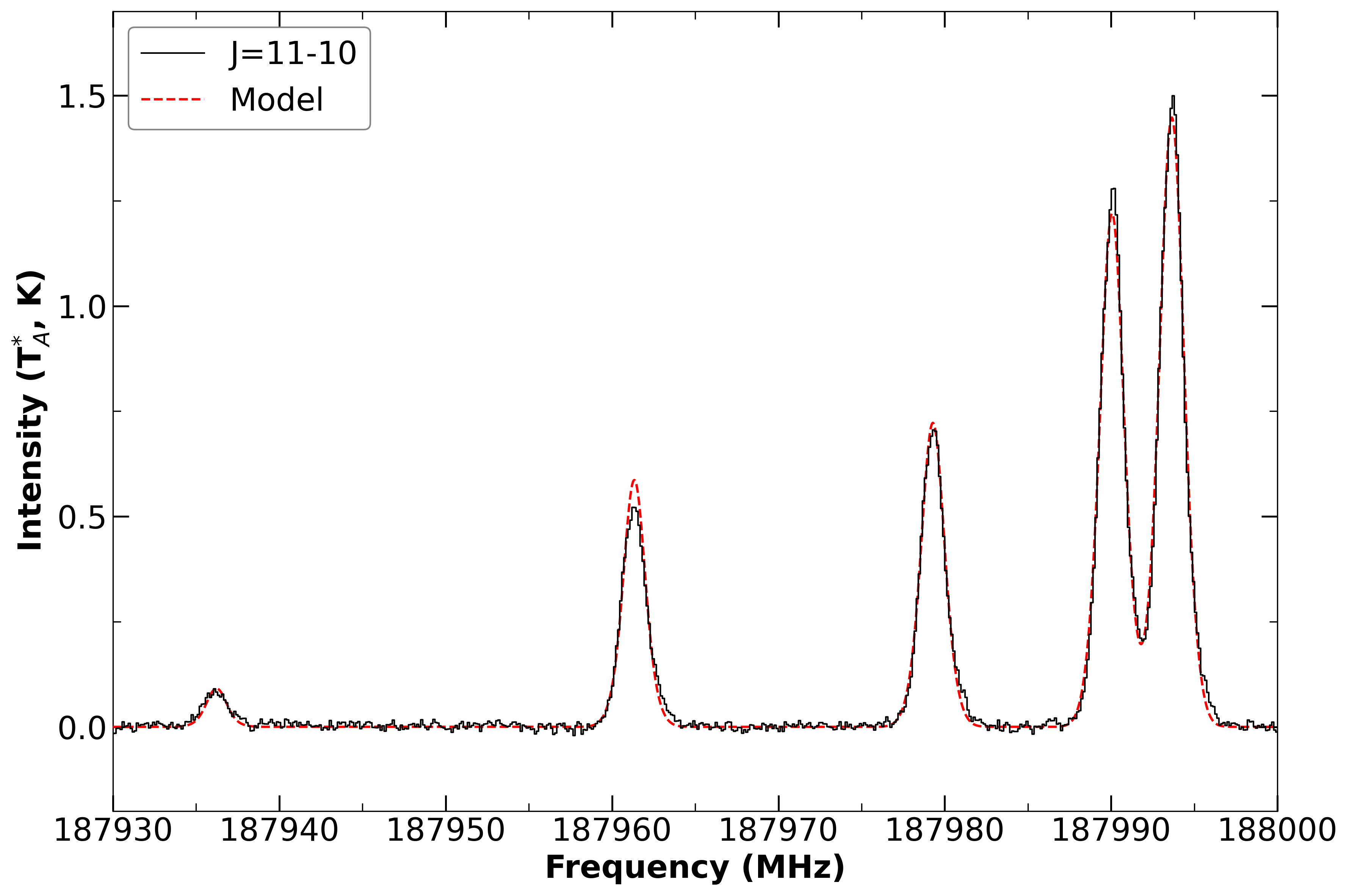}
\includegraphics[width=0.42\linewidth]{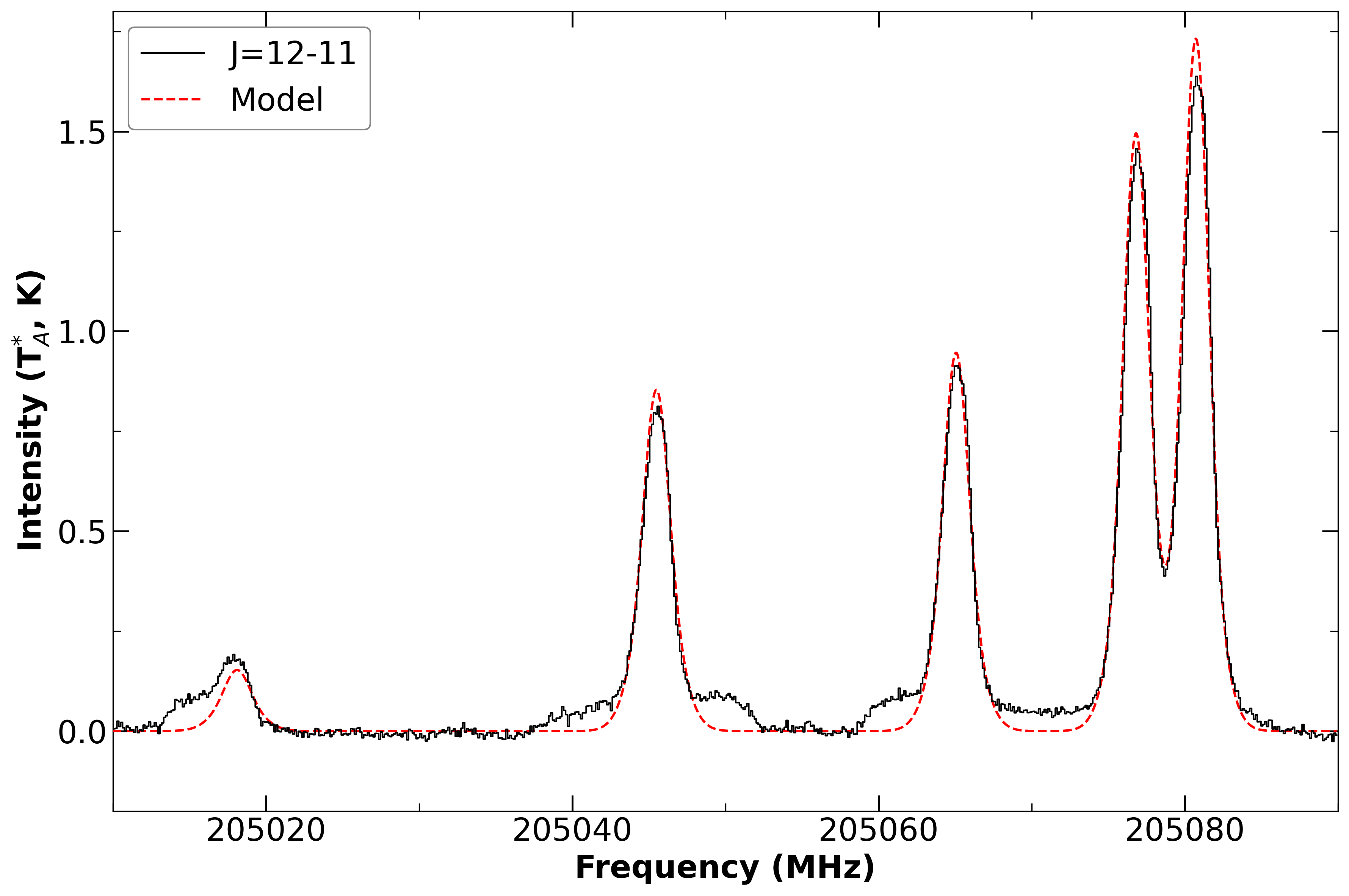}\hspace{0.5cm}
\includegraphics[width=0.42\linewidth]{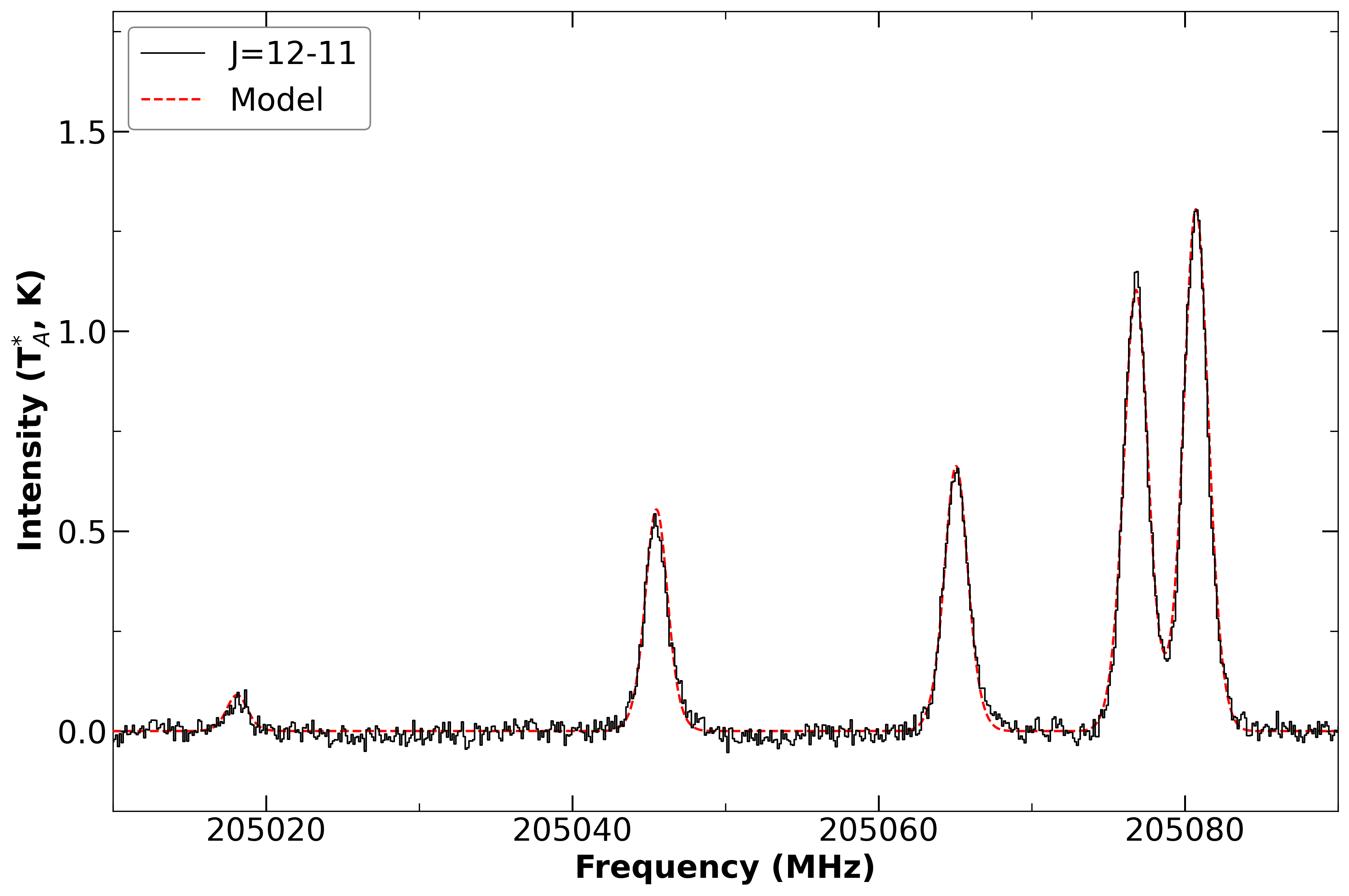}
\includegraphics[width=0.42\linewidth]{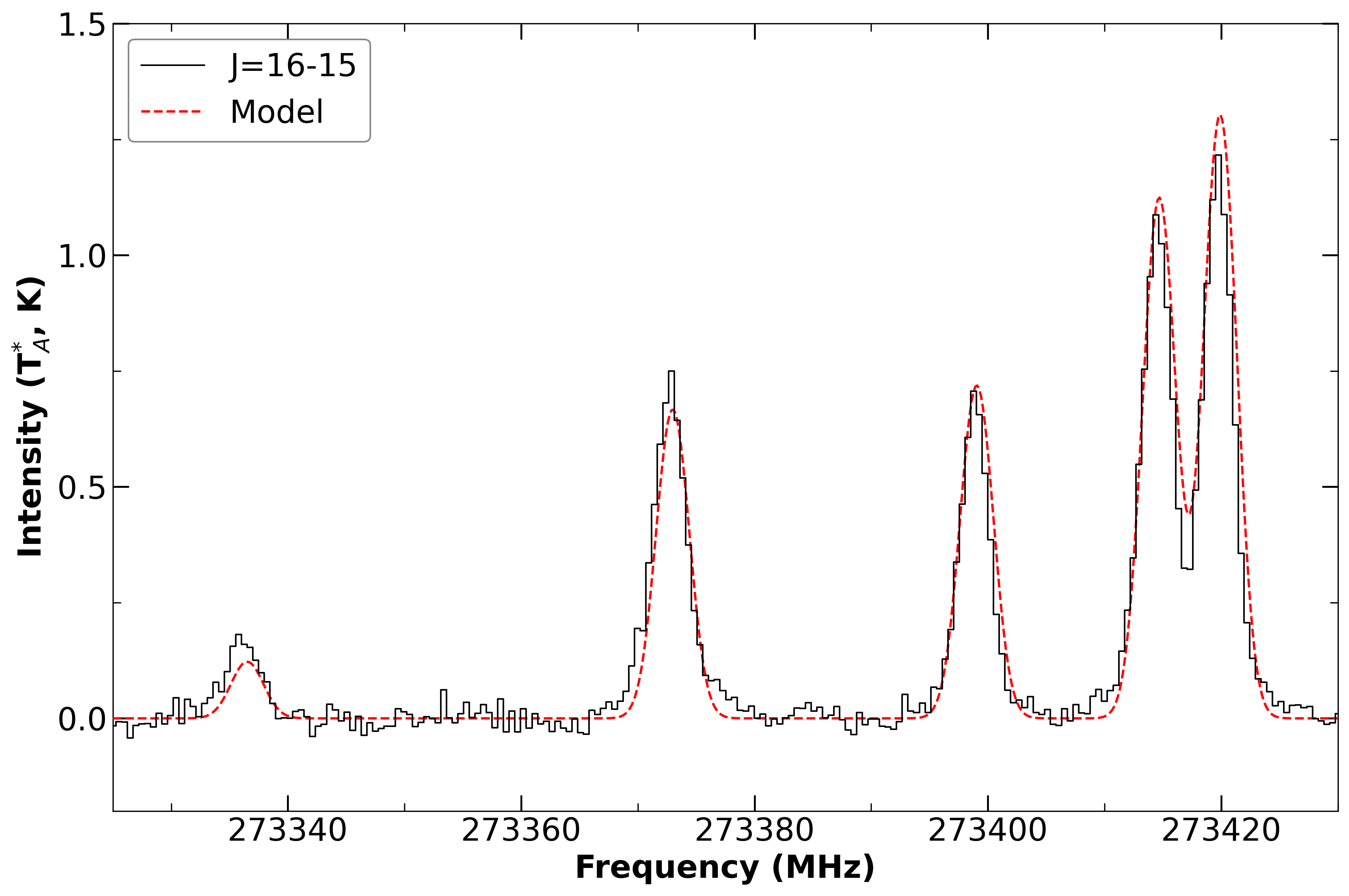}\hspace{0.5cm}
\includegraphics[width=0.42\linewidth]{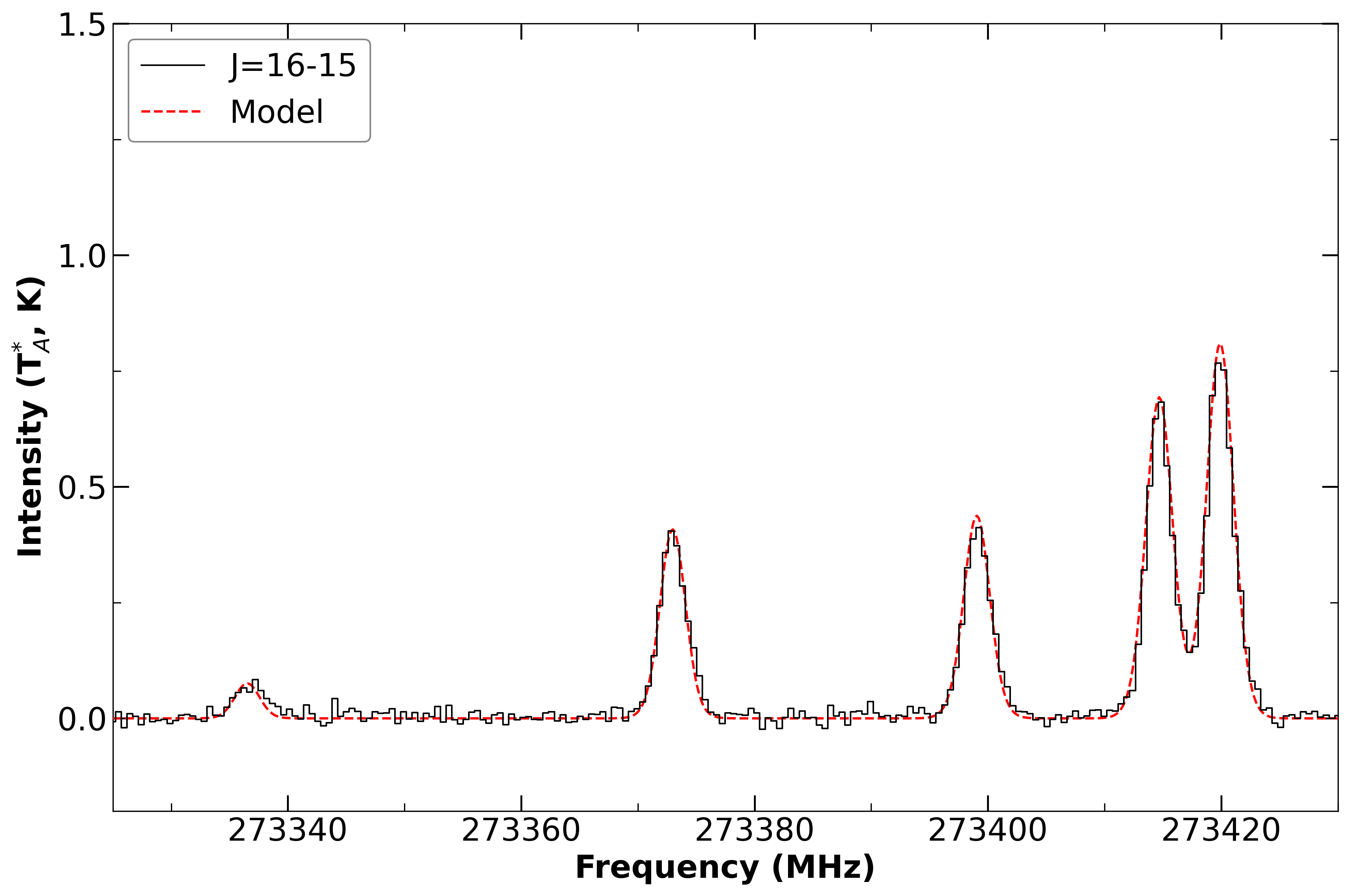}
\caption{Comparison of \MA\ spectral lines observed toward (a) \NE\ and (b) \SW. The observed spectra are shown as black histograms, with the composite LTE model fit overlaid as red dashed lines ({\it Continued on next page}).}
\end{figure*}

\begin{figure*}[!h]
\ContinuedFloat
\centering
\includegraphics[width=0.42\linewidth]{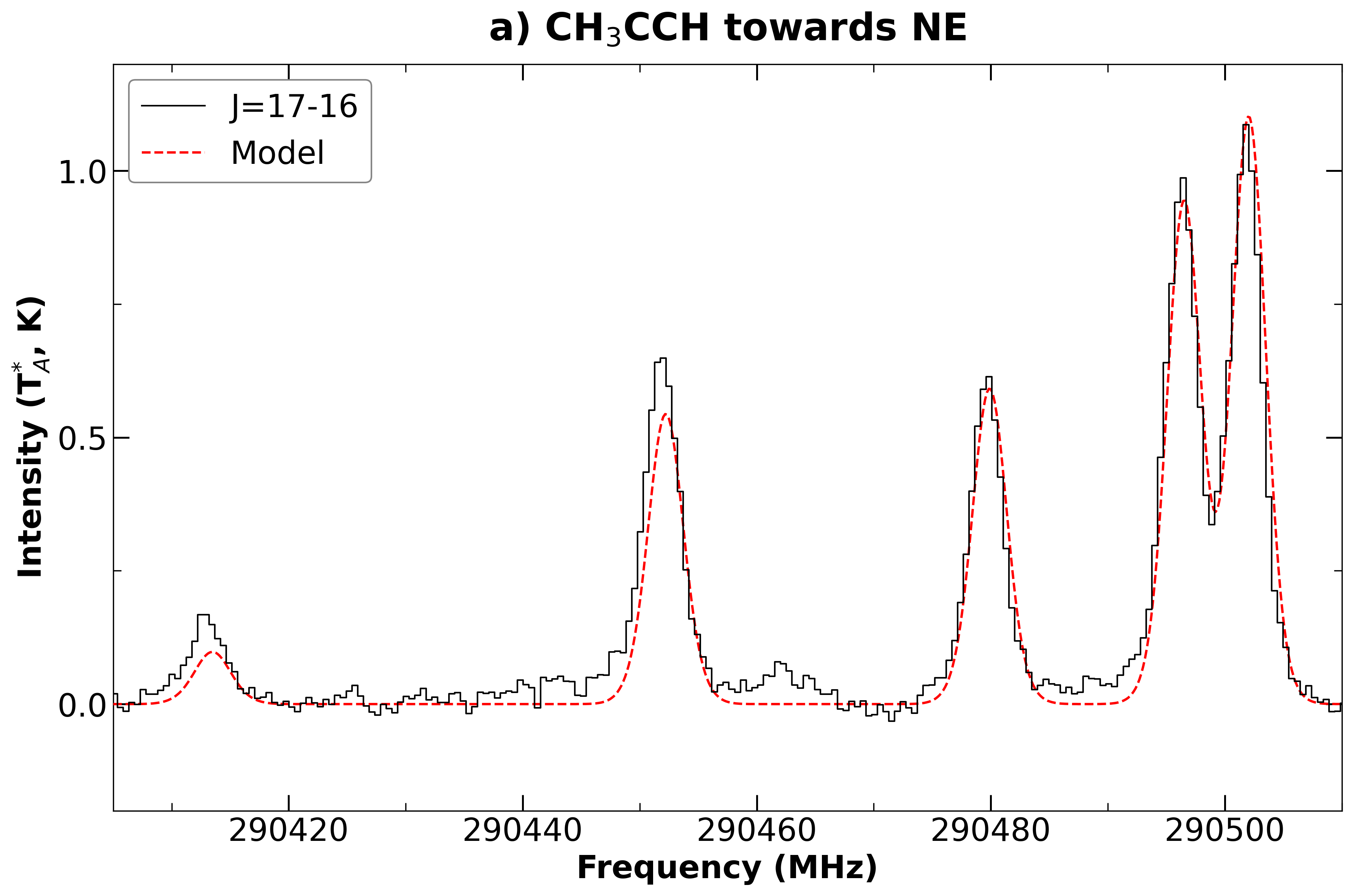}\hspace{0.5cm}
\includegraphics[width=0.42\linewidth]{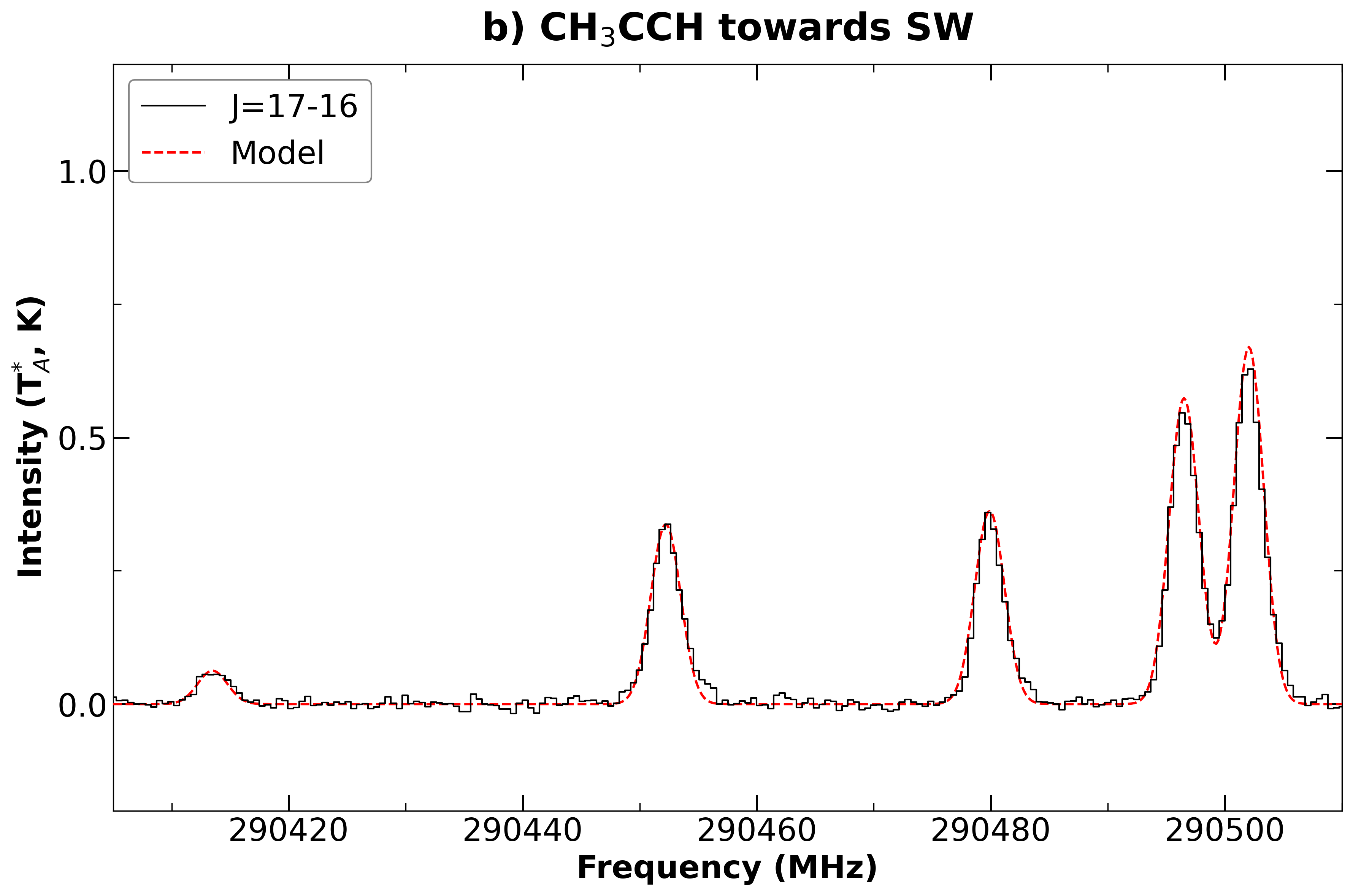}
\includegraphics[width=0.42\linewidth]{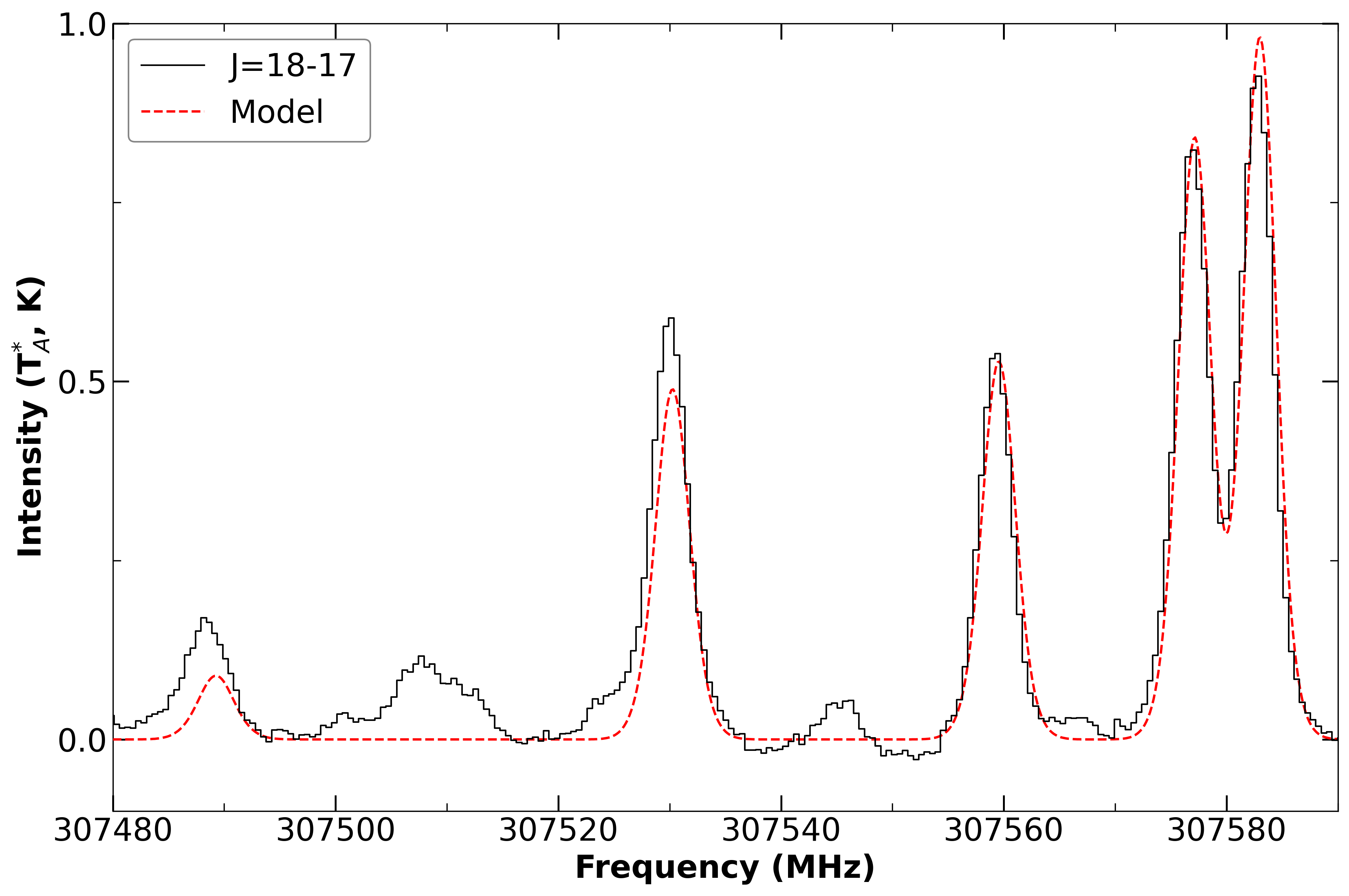}\hspace{0.5cm}
\includegraphics[width=0.42\linewidth]{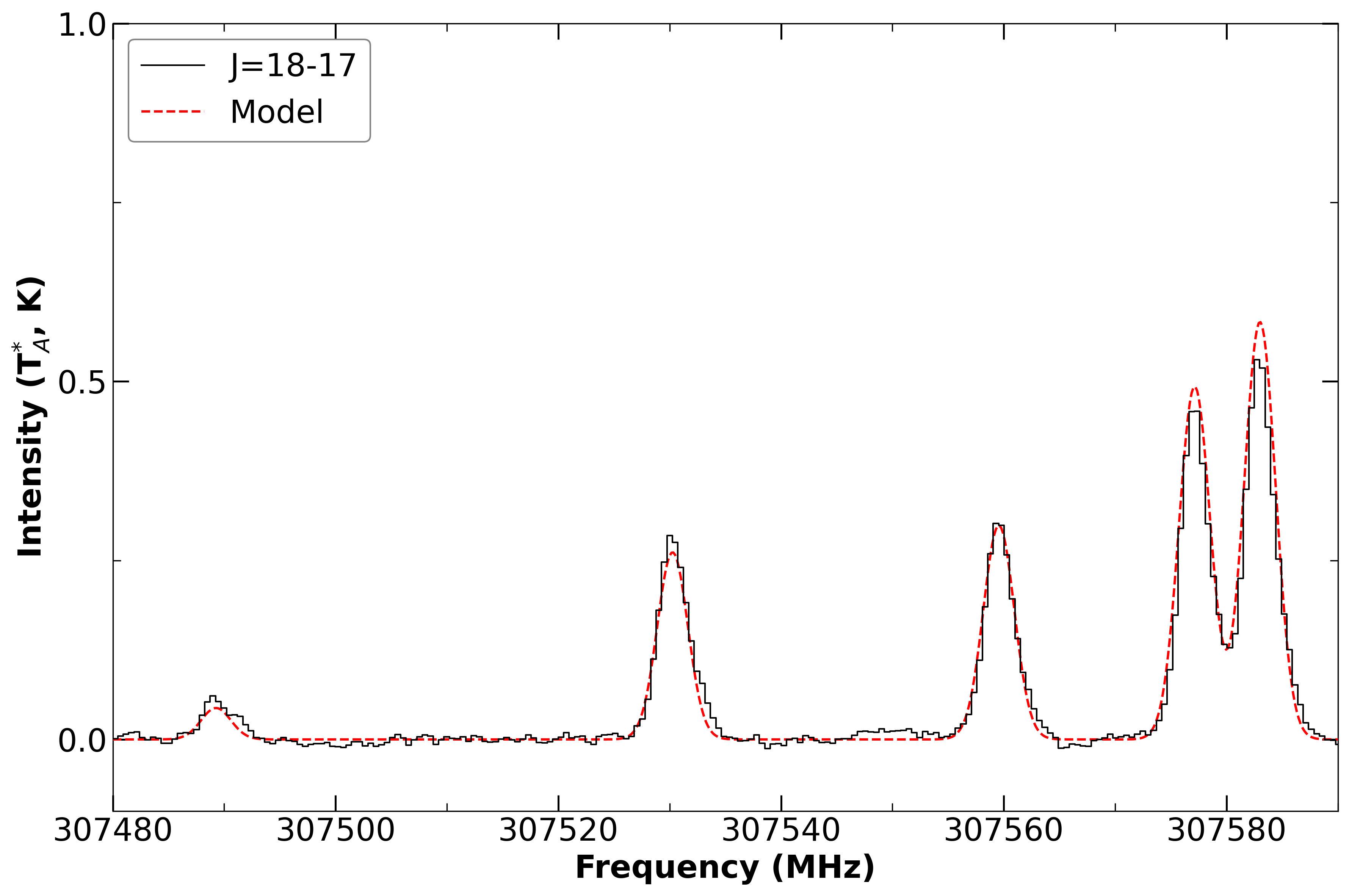}
\includegraphics[width=0.42\linewidth]{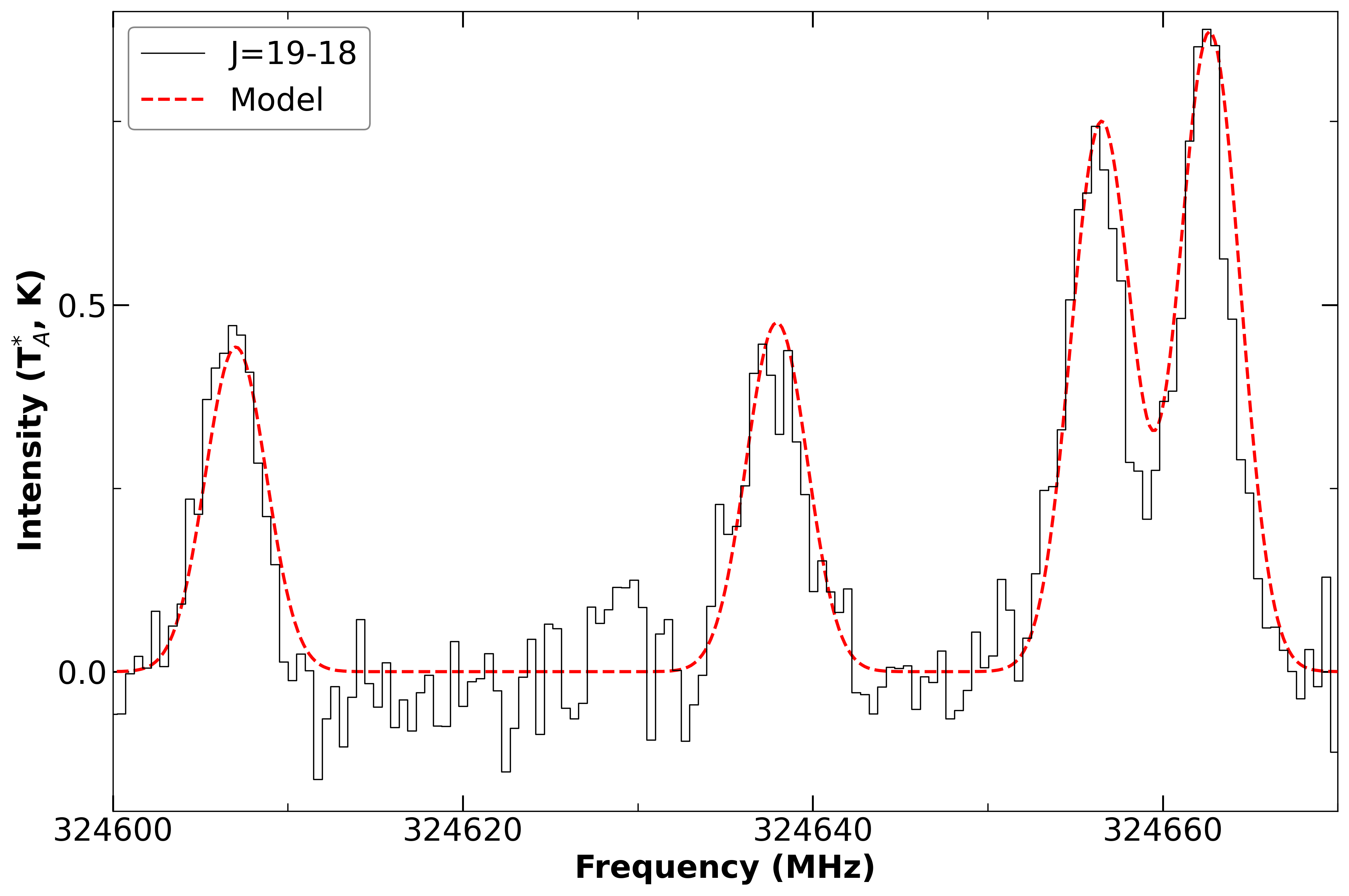}\hspace{0.5cm}
\includegraphics[width=0.42\linewidth]{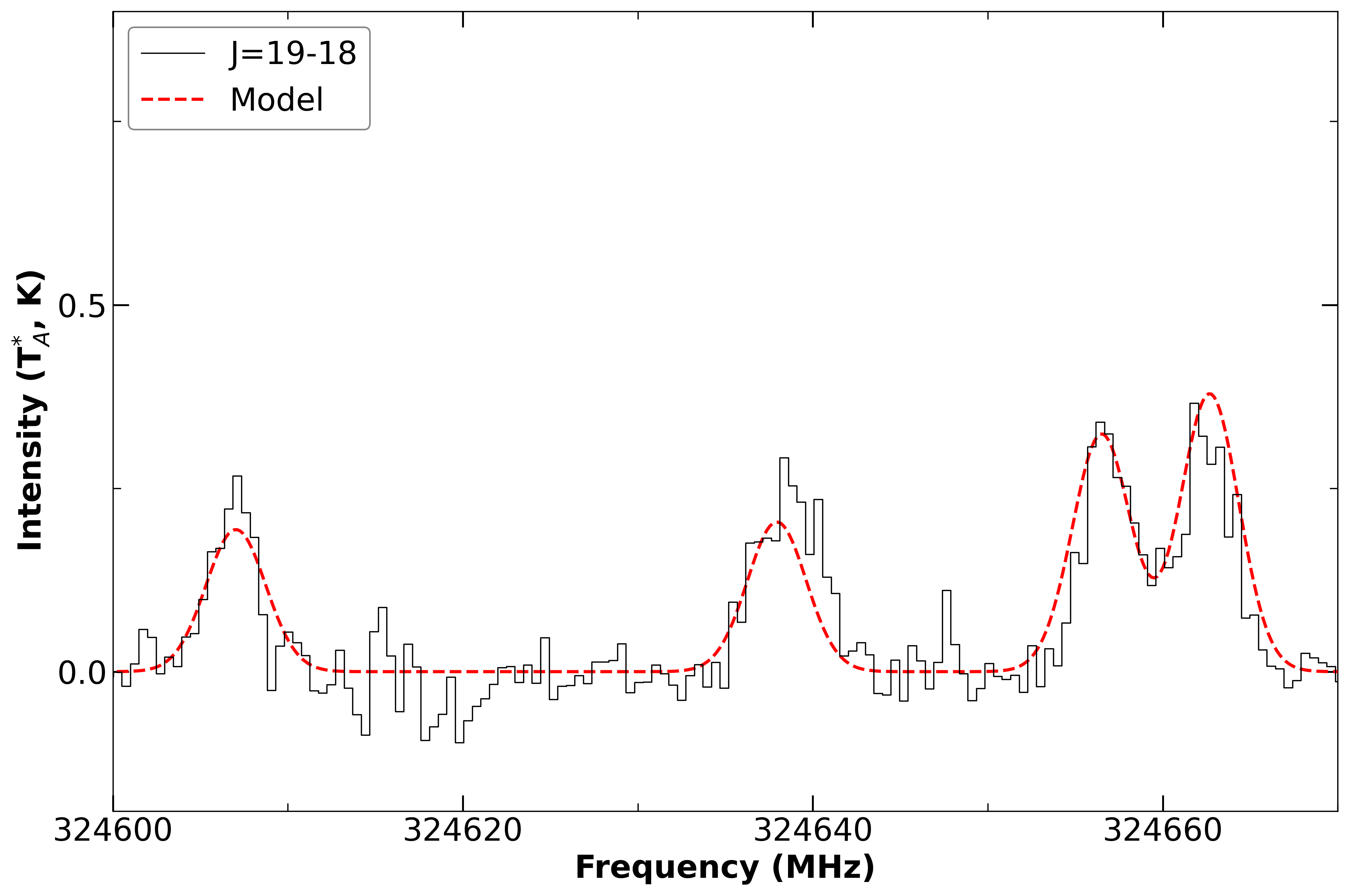}
\includegraphics[width=0.42\linewidth]{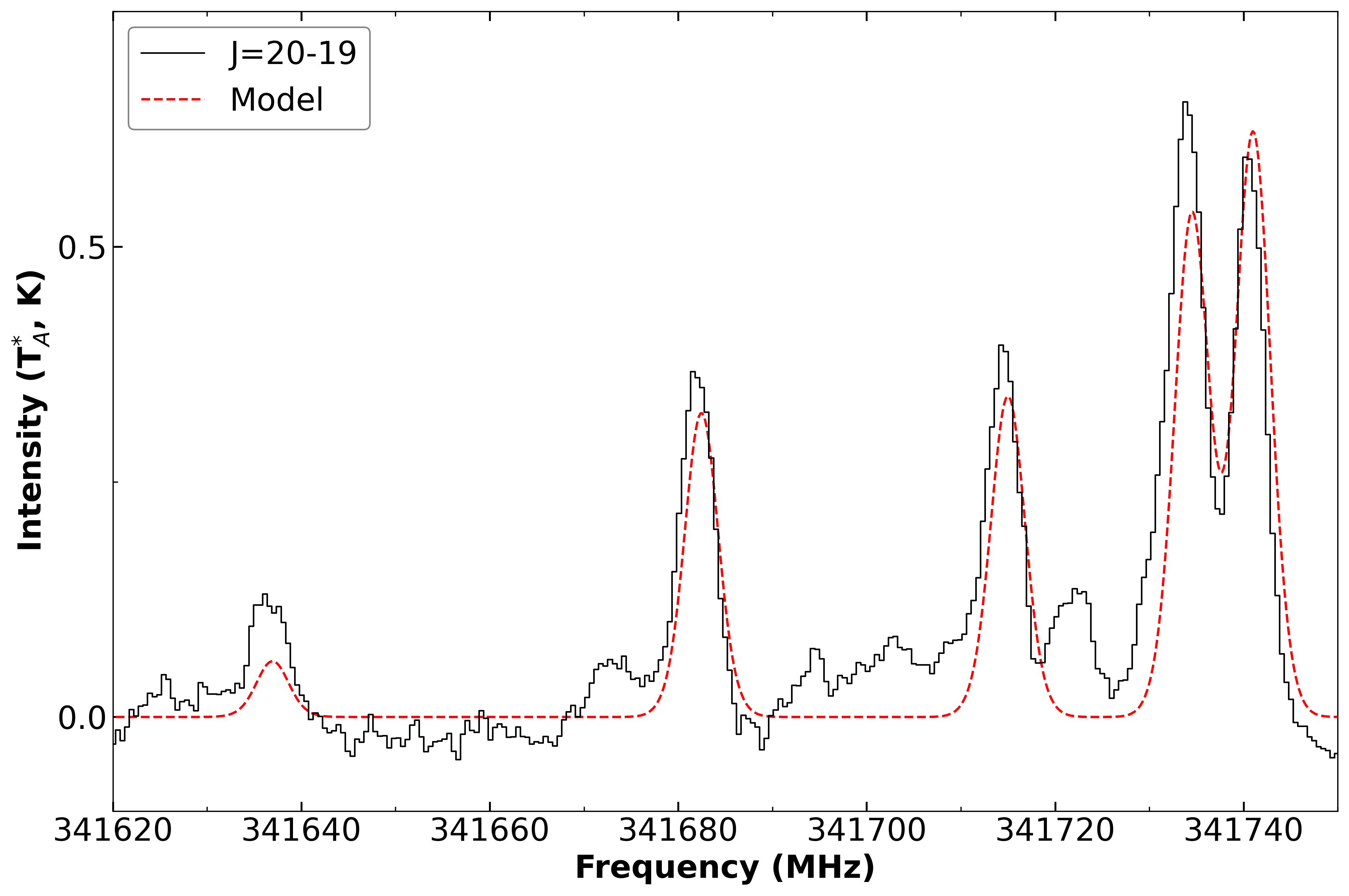}\hspace{0.5cm}
\includegraphics[width=0.42\linewidth]{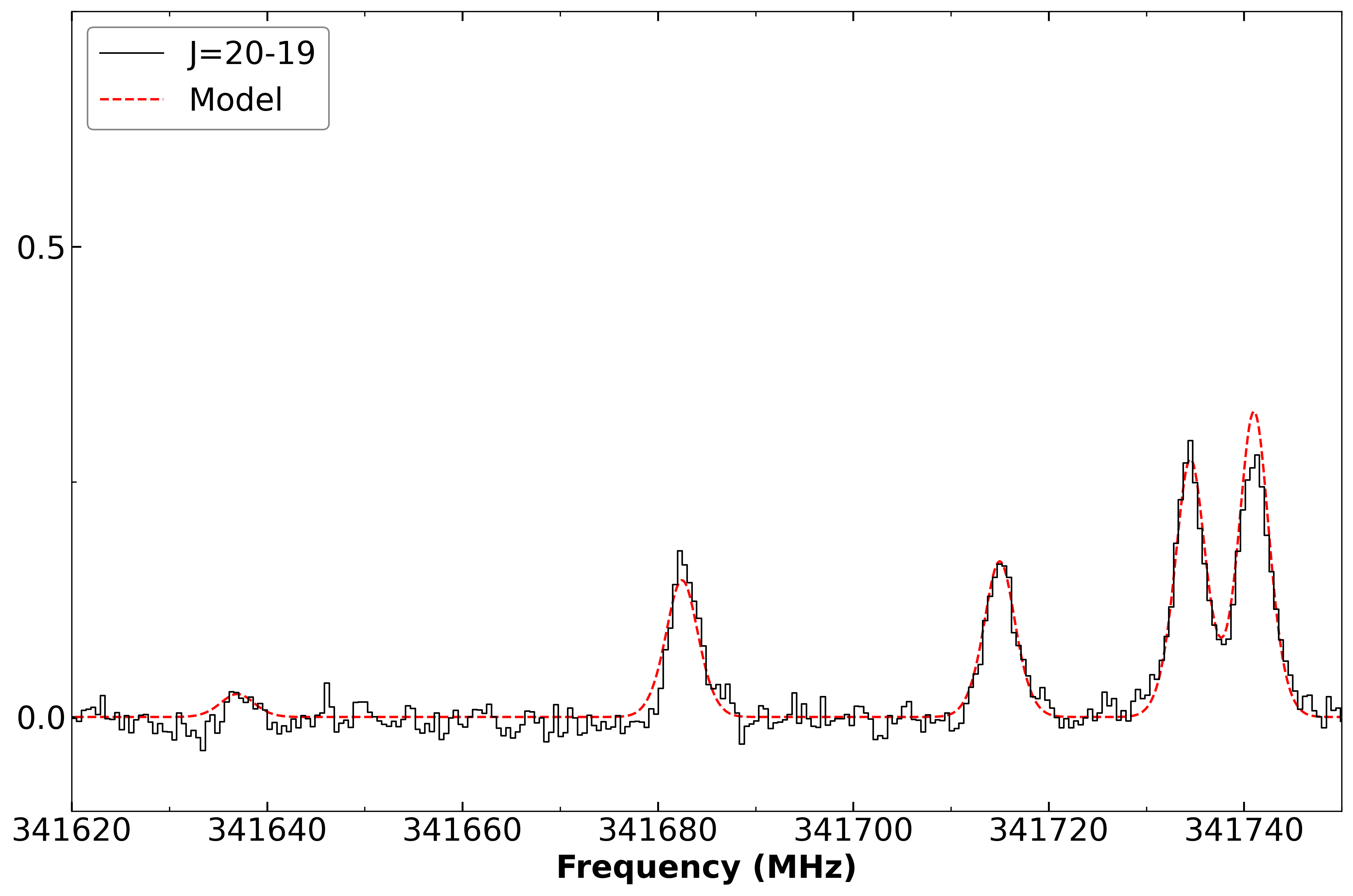}
\caption{Continuation of figure for \MA\ lines from $J$=17--16 to $J$=20--19.}
\label{ma-lines}
\end{figure*}

\begin{figure}
\centering
\includegraphics[width=1.0\linewidth]{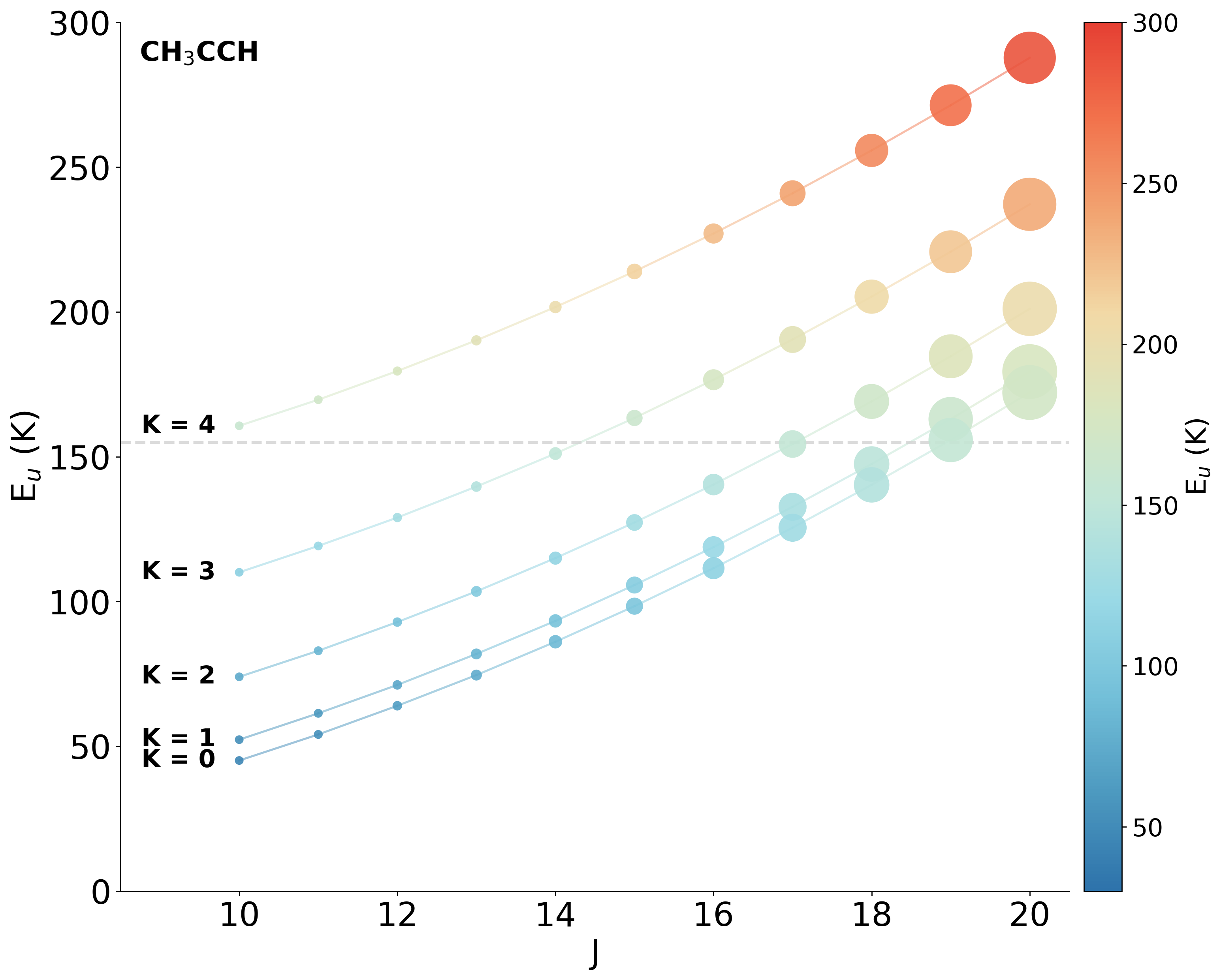}
\caption{Schematic $K$-ladder energy diagram of \MA\ showing the upper-state energy $E_u$  as a function of the rotational quantum number $J$  for the $K=0-4$ ladder. The marker size of each point is scaled by the $A_{\rm ul}$ coefficient of the corresponding transition. The dotted horizontal line marks $E_u \thickapprox$155~K, the boundary between the low- and high-excitation components adopted in Sect~\ref{sec4.1}.}   
\label{e-diagram}
\end{figure}

\subsection{Detection of \CCH\ and \TC\ isotopologues}\label{sec3.2}

\begin{figure*}[!h]
\centering
\includegraphics[width=0.8\linewidth]{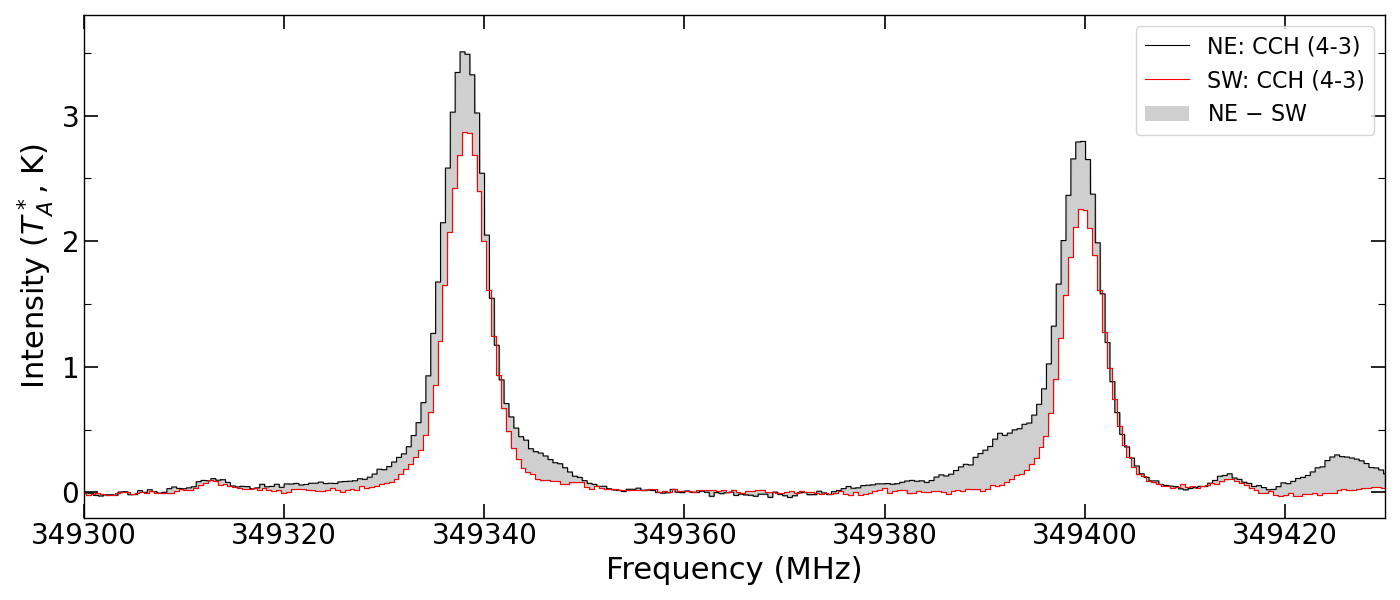}
\caption{Comparison of the strongest hyperfine components of the \CCH\ $N$=4–3 transition toward \NE\ (black) and \SW\ (red). The gray shading marks the intensity excess of NE relative to SW.} \label{cch-lines}
\end{figure*}

\CCH\ was identified via the hyperfine components of the $N$ = 4--3 transition in both the \NE\ and \SW\ regions. 
Four hyperfine components were analysed in each region.  The spectroscopic and line fit parameters of \CCH\ in the NE and SW regions are presented in Tables~\ref{tab:ma-ne} and \ref{tab:ma-sw}, respectively. Likewise, Fig.~\ref{cch-lines} compares the strongest \CCH\ ($N=$4--3) components toward the NE and SW regions. The comparison of the common \CCH\ transitions in NE and SW reveals systematically stronger emission toward NE, as highlighted by the shaded region in Fig.~\ref{cch-lines}.  The line profiles exhibit broader wings than expected for a pure Gaussian, suggesting moderate optical depth effects or multiple velocity components within the beam, both of which point to a departure from LTE in the excitation of this molecule. 

\begin{figure*}[!h]
\centering
\includegraphics[width=8.5cm,keepaspectratio]{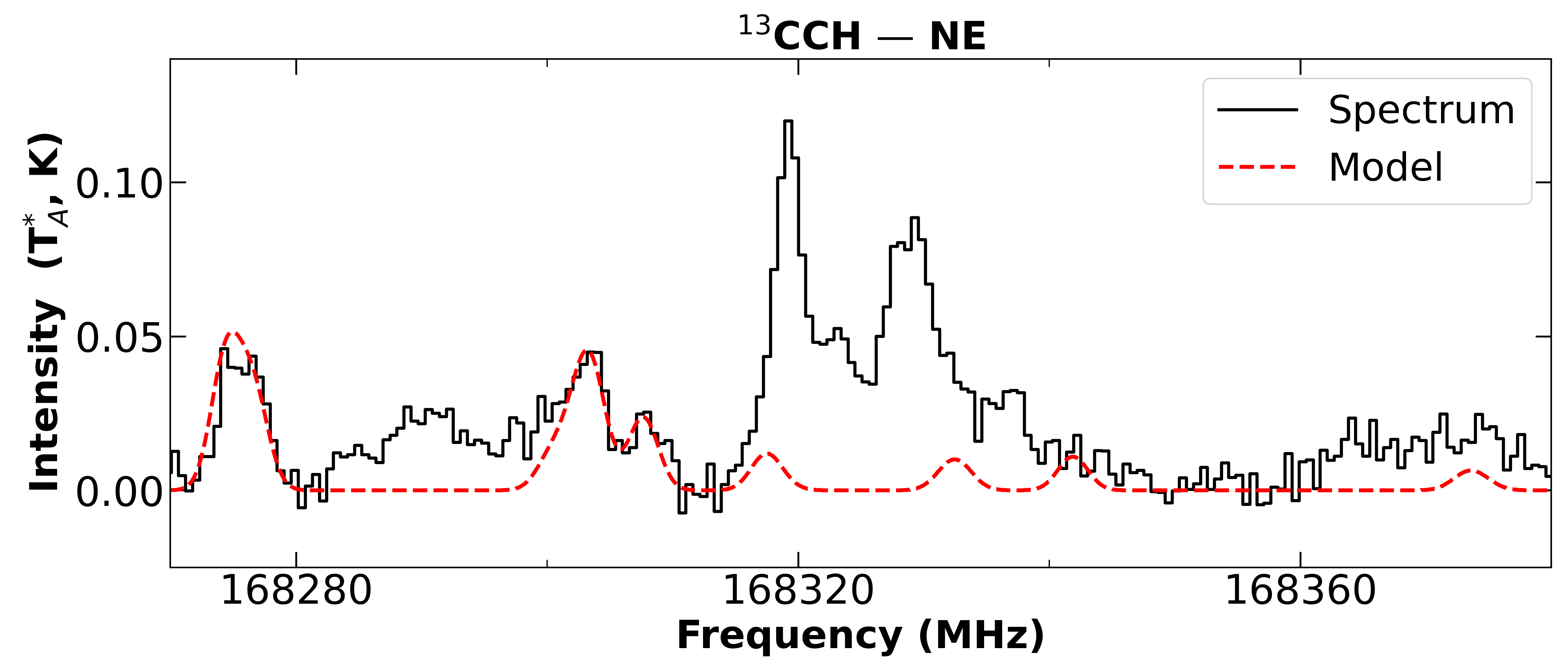}
\includegraphics[width=8.5cm,keepaspectratio]{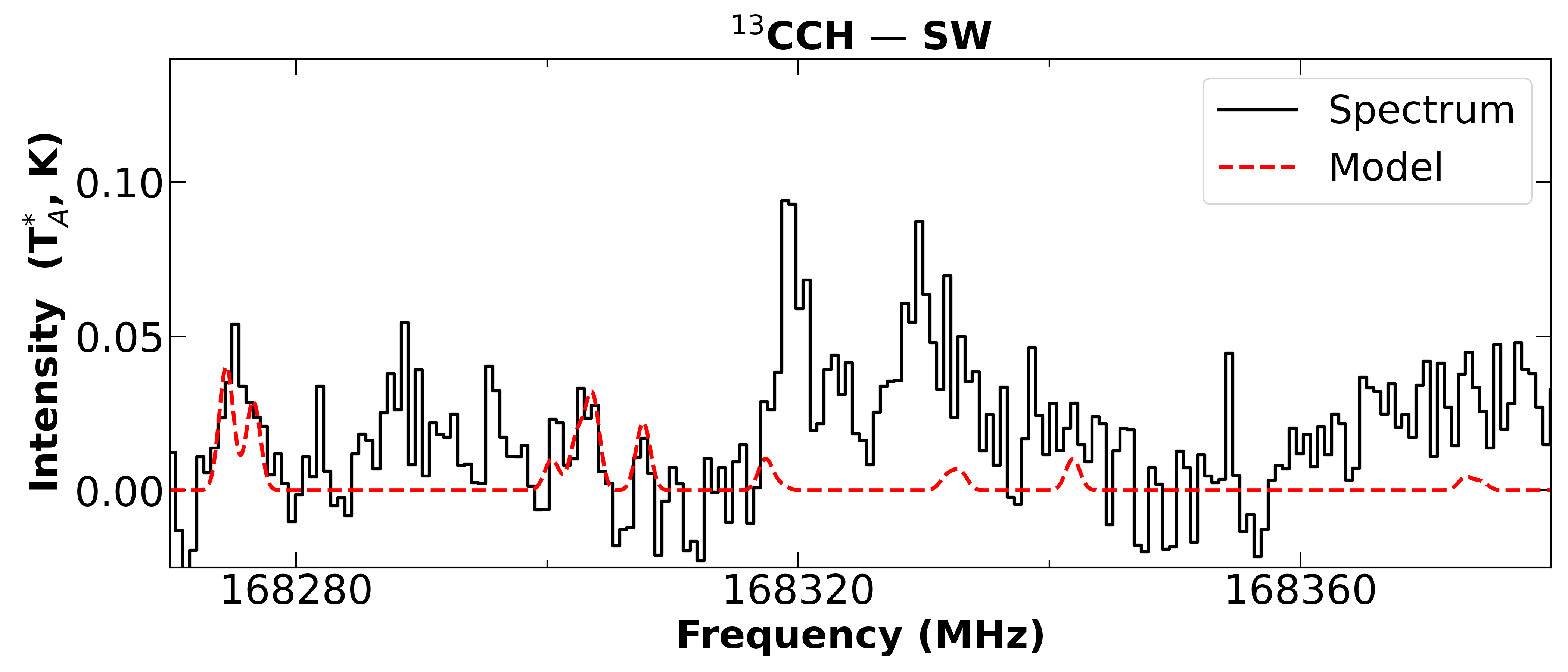}\\
\includegraphics[width=8.8cm,keepaspectratio]{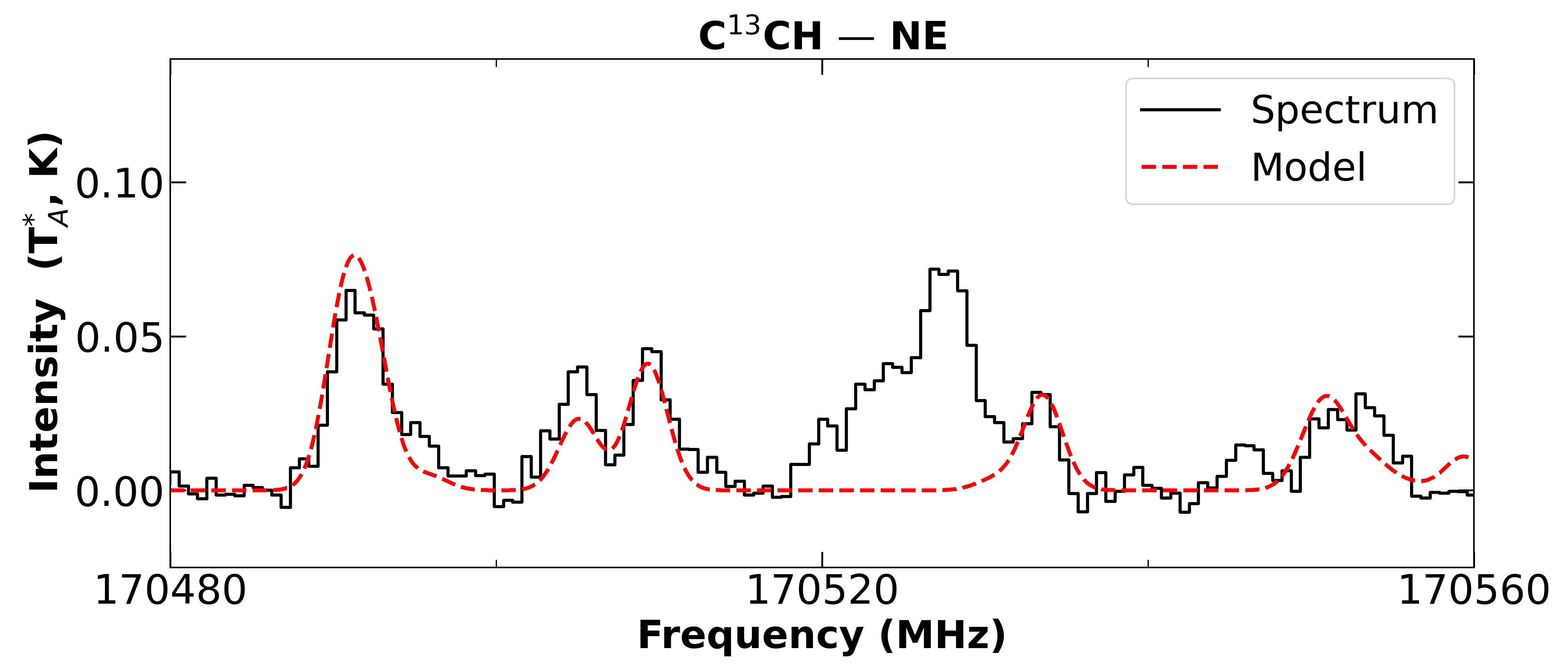}
\includegraphics[width=8.8cm,keepaspectratio]{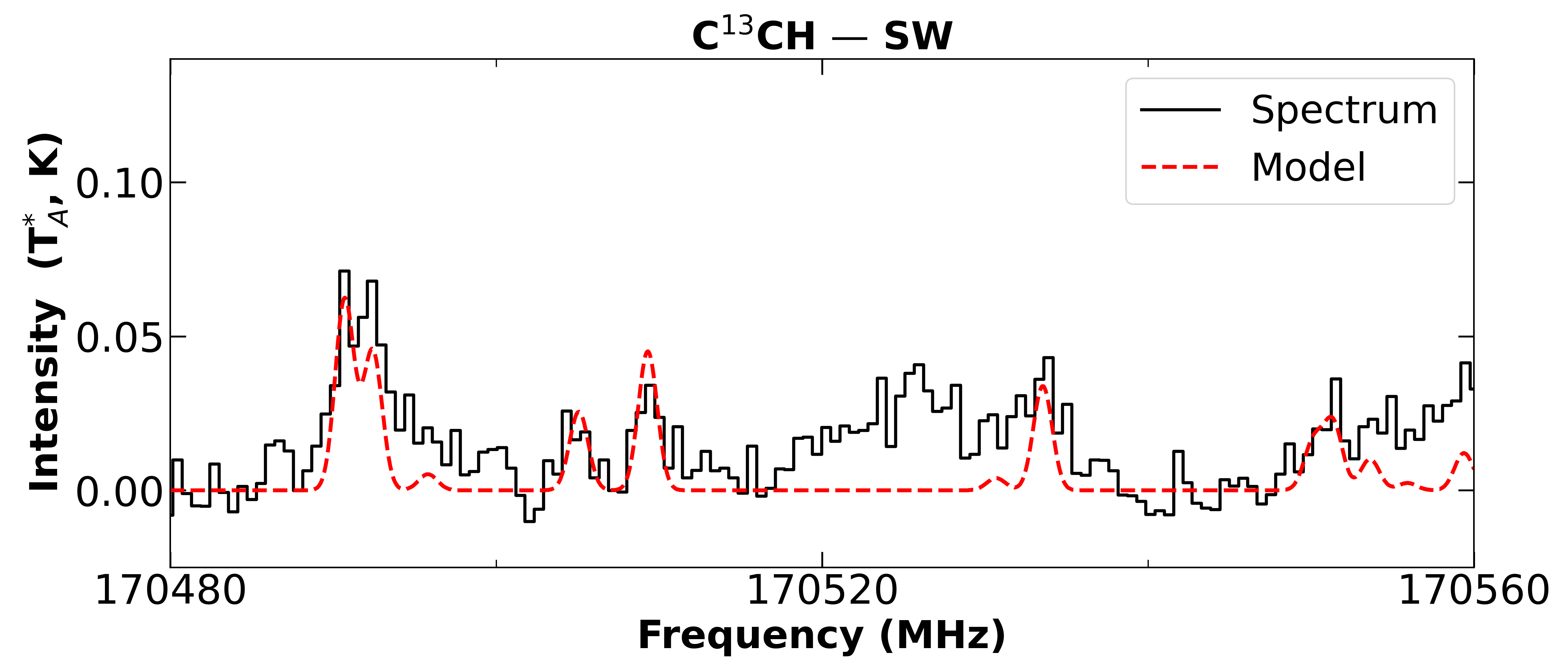}
\caption{Comparison of spectral lines for the $^{13}$C isotopologues of \CCH. The upper and lower panels display \TCC\ and \CTC\ lines, respectively, towards \NE\ and \SW.\ The red dashed lines indicate the LTE model fits, assuming $T_{\rm kin} = \boldsymbol{T_{\rm ex}}$ from the \CCH\ analysis.}
\label{isos-cch}
\end{figure*}

We inspected the spectra for the presence of deuterated and $^{13}$C isotopologues of \CCH\ toward both regions. The \TCC\ and \CTC\ species are detected in NE and SW, whereas no evidence is found for the deuterated counterpart, CCD. Figure~\ref{isos-cch} presents the identifications of \TCC\ and \CTC\ in both regions. The emission of both species is faint, as expected for $^{13}$C-substituted isotopologues, and the emission of \CTC\ is slightly stronger than that of \TCC. The hyperfine structure is not clearly resolved, and additional spectral features overlap with these transitions, preventing the unambiguous isolation of individual components. Qualitative and quantitative analyses were therefore performed by modelling synthetic line profiles of both species in each region (see Fig.~\ref{isos-cch}), an assessment is given in Sect.~\ref{sec4.2}.

\section{Radiative transfer analysis}\label{sec4}

\subsection{Rotational diagram analysis of \MA}\label{sec4.1}

To estimate the excitation temperature (\TE)\ and the total column density (\NC)\ of \MA\ in both \NE\ and \SW,\ we constructed rotational diagrams under the assumption of LTE. In this framework, collisional processes are assumed to couple the level populations to the kinetic temperature. Under the additional assumptions that the emission is optically thin and uniformly fills the telescope beam, the level populations follow a Boltzmann distribution. Consequently, \TE\ and \NC\ can be  derived directly from the slope and intercept of a linear fit to the RD:

\begin{equation}
\centering
\ln\left(\frac{N_u}{g_u}\right)=\ln\left[\frac{N_{\rm col}}{Q(\boldsymbol{T_{\rm ex}})}\right]-\frac{E_u}{k \boldsymbol{T_{\rm ex}}}- \ln C_{\tau}- \ln \left(\frac{\Omega_a}{\Omega_s} \right),
\label{eq:rot_diag}
\end{equation}

\noindent where $Q$(\TE)\ is the partition function evaluated at the excitation temperature; $N_u$, $g_u$, and $E_u$ are the column density, statistical weight, and energy of the upper level of each transition, respectively. The term $C_{\tau}$ is the optical depth correction factor, given by $C_{\tau} = \tau/[1-\exp(-\tau)]$, where $\tau$ is the optical depth of the transition. The last term of Eq.~\ref{eq:rot_diag} accounts for the beam dilution factor, which scales the column density when the emitting source does not fully fill the telescope beam \citep{G&L1999}. 

The spectral dataset for \MA\ toward NE and SW comprises 39 rotational transitions spanning $J=10$--9 to $J=20$--19, with $K$ ladders from 0 to 4.  Figure~\ref{e-diagram} presents a schematic $K$-ladder energy diagram of all the transitions within this $J$ range. Low-$J$, low-$K$ lines (lower-left in the diagram) have the lowest $E_{\rm u}$ and are therefore the most readily excited, particularly in cold gas. In contrast, high-$J$, high-$K$ transitions require substantially higher excitation and have larger  $A_{\rm ul}$ coefficients, implying faster radiative decay. Accordingly, at fixed $J$ the line intensities decrease with increasing $K$, and the overall intensity decreases with $J$; the $J=10$--9 lines are substantially brighter than the $J=20$--19 lines. This decrease is significantly steeper toward SW: at low $J$ the two regions show comparable line intensities, whereas the NE--SW contrast increases with $J$. This behaviour indicates that the observed \MA\ emission cannot be described by a single excitation component.

\begin{figure}
\centering
\includegraphics[width=1.0\linewidth]{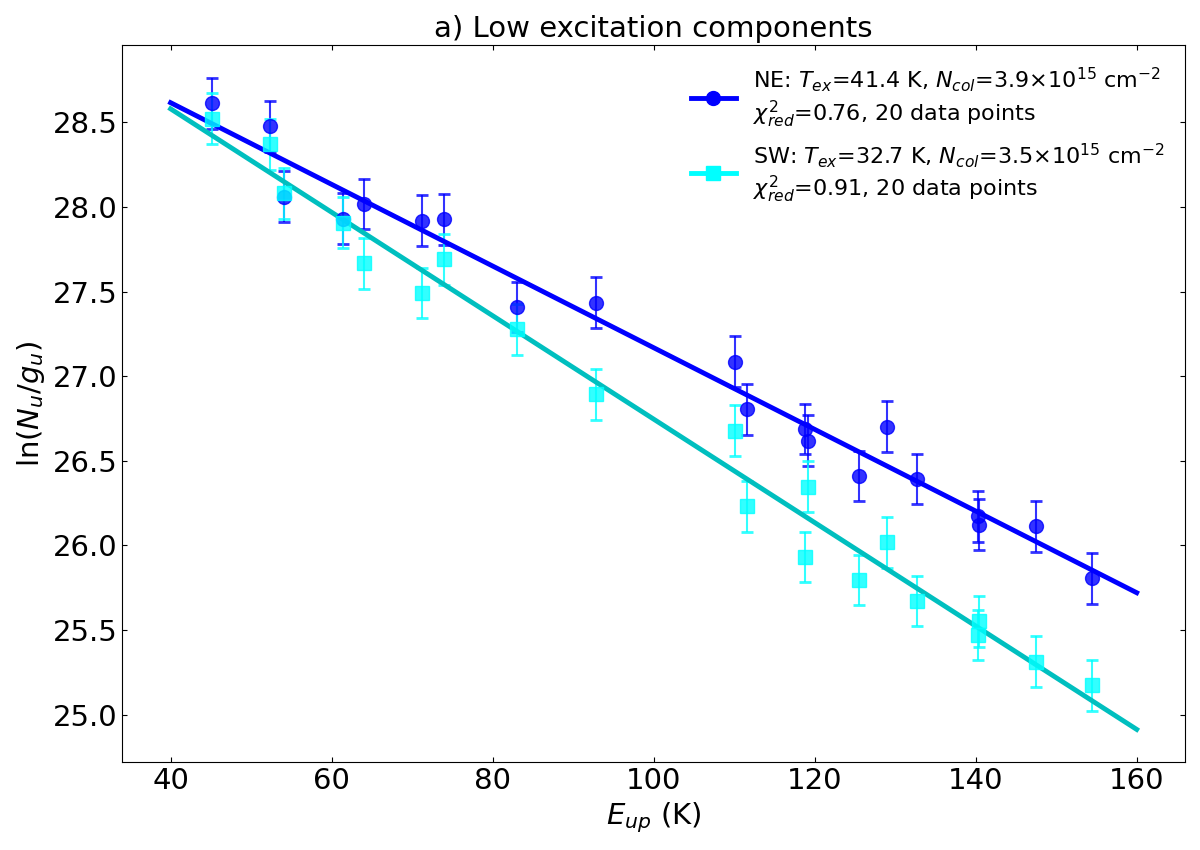}
\includegraphics[width=1.0\linewidth]{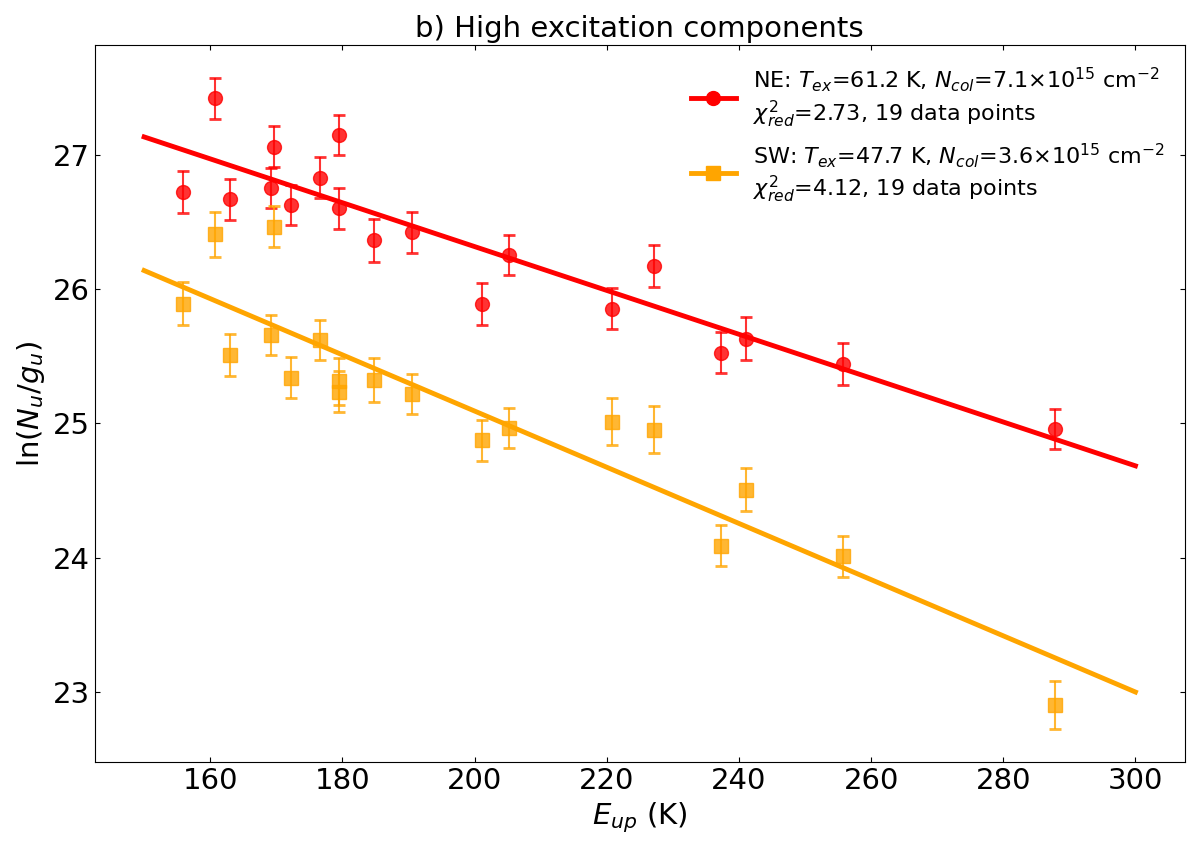}
\caption{Rotational diagrams analyses of \MA: (a) linear regressions fitted to the low excitation components, yielding $\boldsymbol{T_{\rm ex}}=41.4$ and $\boldsymbol{T_{\rm ex}}=32.7$~K for NE and SW, respectively. (b) Same as top, but for the high excitation components, yielding $\boldsymbol{T_{\rm ex}}=61.2$ and $\boldsymbol{T_{\rm ex}}=47.7$~K for NE and SW, respectively.  Error bars include the dominant 15\% calibration uncertainty and the $1\sigma$ noise contribution (Sect.~\ref{sec2.1}).}   \label{rds}
\end{figure}

In both NE and SW, the rotational diagrams exhibit a break in slope at $E_{\rm u}\thickapprox 155$~K, separating a low-$E_{\rm u}$ population with a shallower trend from a high-$E_{\rm u}$ population with a steeper trend. We therefore partitioned the 39 transitions into two excitation components at this knee; the robustness of this decomposition is examined in Appendix~\ref{app2}, together with a comparison against a single-component fit. The low-excitation component includes 20 transitions with $E_u\leq 154.4$~K, while the high-excitation component comprises the remaining 19 transitions with $E_u\geq 155.8$~K. The low-excitation component contains most of the $K=0$ and 1 ladders, whereas the high-excitation component encompasses most of the $K=3$ and 4 ladders. The remaining $K$-ladders are bisected between both components according to their $J$ rotational level. As a result, we present the rotational diagrams in Fig.~\ref{rds}; the low- and high-excitation components are shown in the upper and lower panels, respectively.

Opacity and beam-dilution corrections were applied to each excitation component in both the NE and SW regions. For the high-excitation component, which traces higher-frequency transitions and therefore smaller beamwidths, the emission is expected to be more compact; accordingly, we adopted beam-dilution corrections of $\simeq$8\arcsec\ for the NE and $\simeq$10\arcsec\ for the SW, as inferred from the ALMA maps. For the low-excitation component, which traces lower-frequency transitions and larger beamwidths, the emission is assumed to originate from colder, more extended gas, and a nominal source size of 20\arcsec \ was adopted for both regions. The results are presented in Table~\ref{tab:rd-results}. These solutions were used to simulate the LTE lines of \MA,  which are exhibited in Fig.~\ref{ma-lines}. 

\begin{table*}[h!]
\centering
\caption{Results of the LTE excitation analysis based on rotational diagrams of \MA.\ Parameters are reported for the low- and high-excitation components, corresponding to transitions with upper-level energies $E_u < 155$~K and $E_u > 155$~K, respectively.}
\label{tab:rd-results}
\begin{tabular}{l c c c c c c c  } 
\hline\hline
\noalign{\smallskip}
& \multicolumn{3}{c}{Low-ex. comp. ($E_u \leq 154.38$~K)} & & \multicolumn{3}{c}{High-ex. comp. ($E_u \geq 155.83$~K)}   \\
\cline{2-4} \cline{6-8} \noalign{\smallskip}
Region & $\boldsymbol{T_{\rm ex}}$ & $N$                  & ${\chi^2_{\rm red}}$ & &$\boldsymbol{T_{\rm ex}}$ & $N$                  & ${\chi^2_{\rm red}}$ \\
       & (K)          & (10$^{15}$ cm$^{-2}$)&                      & &(K)          & (10$^{15}$ cm$^{-2}$)&        \\
\noalign{\smallskip}
\hline
\noalign{\smallskip}
NE & $41.4 \pm 1.4$ & 3.9 $\pm$ 0.4 & 0.76 & & $61.2 \pm 6.1$ & 7.1 $\pm$ 2.3 & 2.73  \\
SW & $32.7 \pm 1.0$ & 3.5 $\pm$ 0.3 & 0.91 & & $47.7 \pm 4.9$ & 3.6 $\pm$ 1.5 & 4.12 \\
\noalign{\smallskip}
\hline
\end{tabular}
\tablefoot{Beam-dilution corrections of $\simeq$8\arcsec\ (NE) and $\simeq$10\arcsec \ (SW) were applied to the high-excitation component, while an extended source size of 20\arcsec \ was adopted for the low-excitation component in both regions.}
\end{table*}

\subsection{Excitation analysis of \CCH\ and its $^{13}$C isotopologues}\label{sec4.2}

The excitation parameters of \CCH\ toward the NE and SW were derived using the adaptive random-walk MCMC algorithm implemented for non-LTE calculations using RADEX \citep{van2007}. Both models adopted C$_2$H--He collisional rates scaled to H$_2$ and assumed spherical geometry, the APEX beam response for the 349.108--349.603~GHz window in the rest frame, and $T_\mathrm{bg}=2.73$~K. For each position, the sampler explored the posterior distributions of three free parameters: kinetic temperature ($T_\mathrm{kin}$), column density $N(\mathrm{C}_2\mathrm{H})$, and volume density $n(\mathrm{H}_2)$. The linewidth and $V_\mathrm{LSR}$ were fixed to the values determined from Gaussian profile decomposition. We adopted an extended source size of $\simeq$~20\arcsec,\ consistent with beam-filling conditions on spatial scales comparable to the nominal APEX beam at the analysed \CCH\ rest frequencies.

\begin{figure*}
\centering
\includegraphics[width=0.6\linewidth]{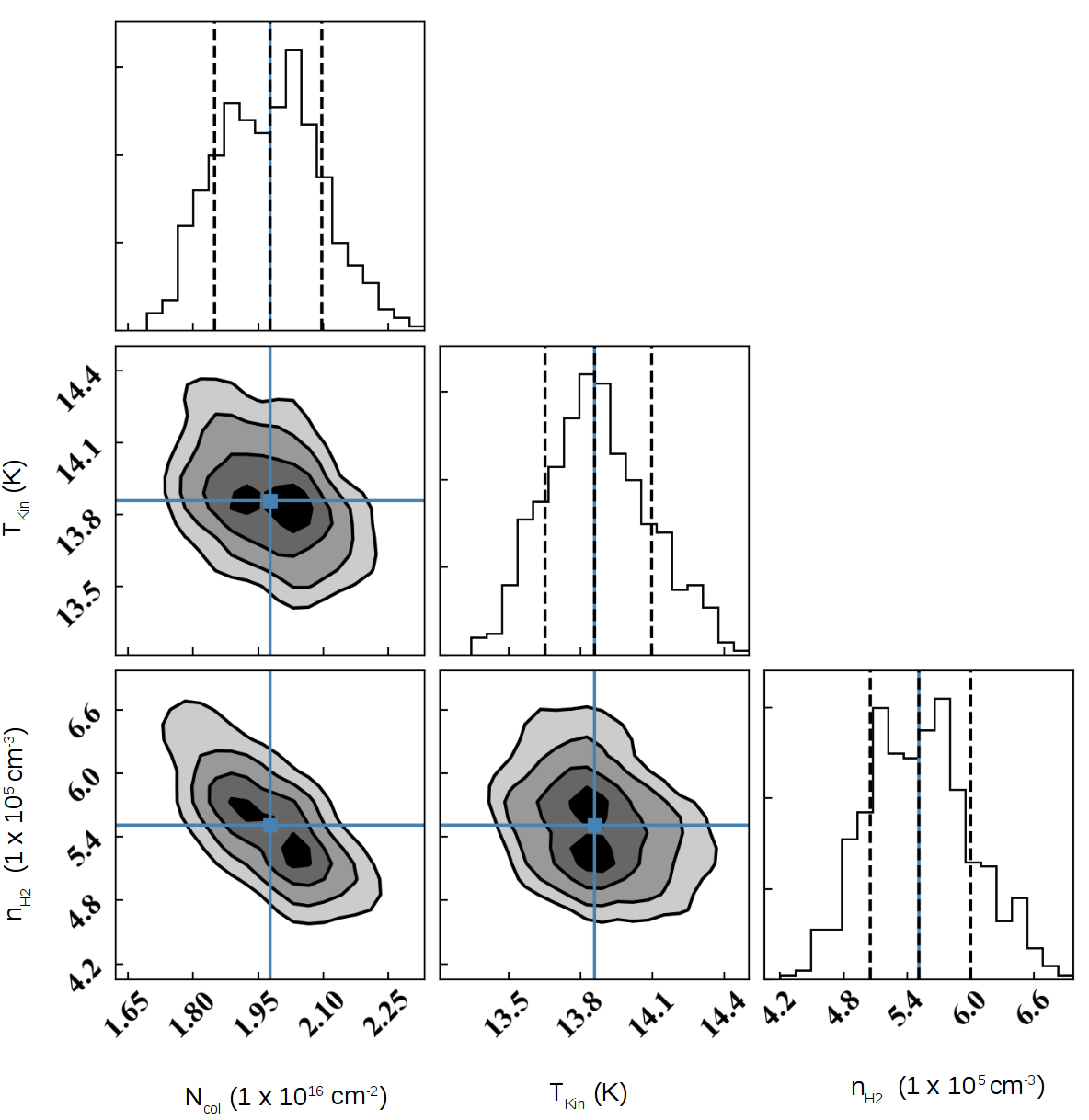}
\includegraphics[width=0.6\linewidth]{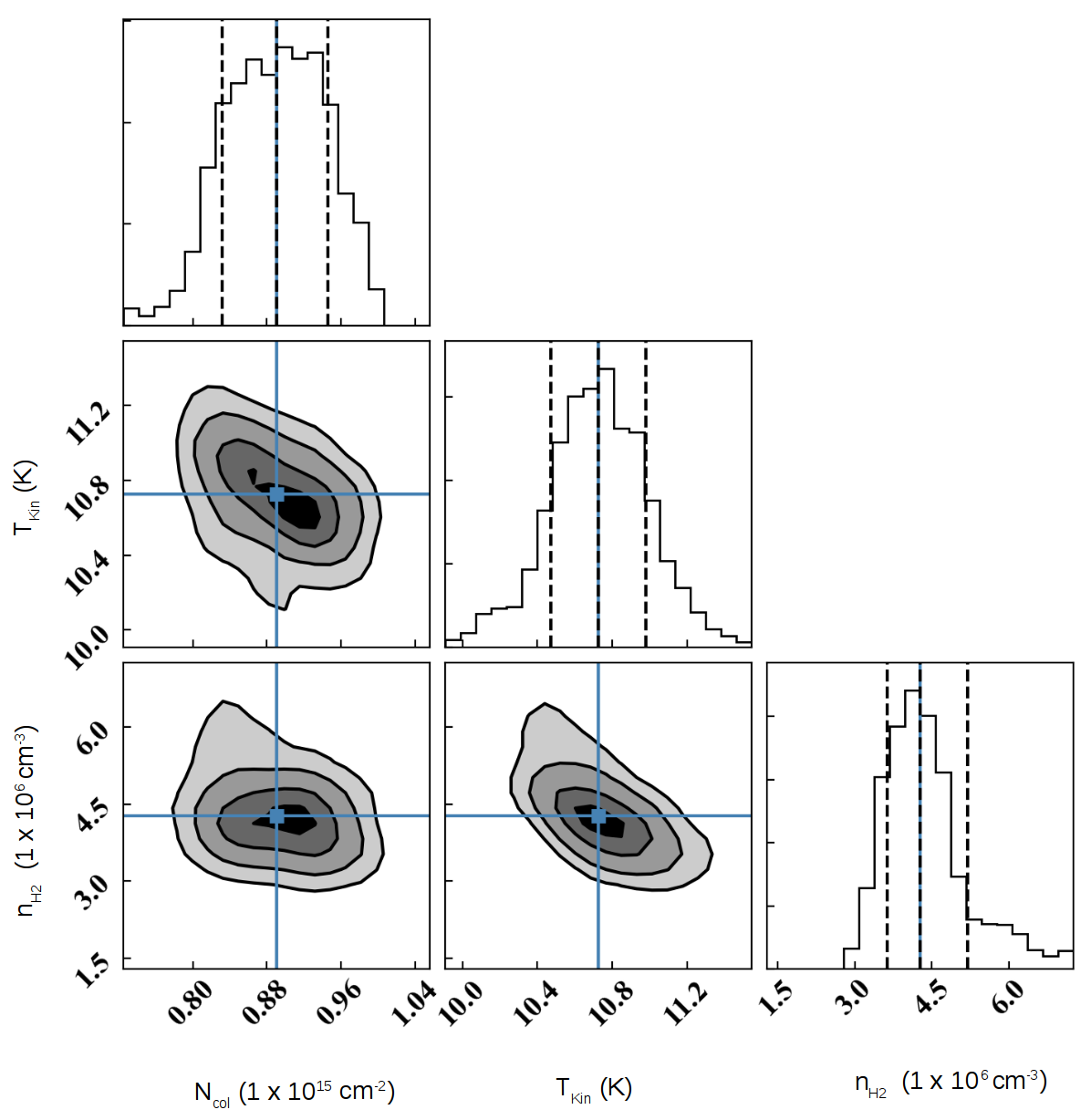}
\caption{Corner plots of the non-LTE-MCMC posterior distributions for \CCH\ in \NE\ (top) and \SW\ (bottom). Diagonal panels show marginalized constraints on  H$_2$ density ($n$(H$_2$), cm$^{-3}$), \CCH\ column density ($N_{\rm col}$, cm$^{-2}$), and kinetic temperature ($T_{\rm kin}$, K) with dashed lines marking the median and $\pm 1\sigma$ uncertainty range. Off-diagonal contours show 2D covariances between parameter pairs; the blue lines indicate the reference values.}
\label{CDs-C2H}
\end{figure*}

The corner diagrams in Fig.~\ref{CDs-C2H} present the posterior probability distributions and covariance contours for NE and SW, following the non-LTE MCMC setup presented in Sect.~\ref{sec2.1}. The diagonal panels show the marginalized one-dimensional histograms; the black dashed vertical lines mark the 16th and 84th percentiles, and the blue vertical lines indicate the posterior medians. The off-diagonal panels display the two-dimensional joint distributions, with shaded contours drawn at the 1$\sigma$, 2$\sigma$, and 3$\sigma$ levels; the blue cross marks the posterior median in each joint plane. For both NE and SW, the off-diagonal confidence contours are compact and well-localized without pronounced elongation, indicating that $N$(C$_2$H), $T_\mathrm{kin}$, and $n(\mathrm{H}_2)$ are independently constrained, with only weak residual covariances. The $n(\mathrm{H}_2)$ posterior for SW shows a mild tail toward higher densities, but the resulting mean--median offset is small. In all panels, the posterior medians lie well within the 1$\sigma$ contours, indicating that the three parameters are well constrained by the data. Table~\ref{tab:c2h-results} summarises the posterior means and standard deviations.

\begin{table}[h!]
\centering
\caption{Beam-averaged \CCH\ physical parameters for the NE and SW regions. Free parameter estimates are posterior means from the converged  non-LTE MCMC calculations; uncertainties are the corresponding posterior standard deviations.}
\label{tab:c2h-results}
\begin{tabular}{l c c c } 
\hline\hline
\noalign{\smallskip}
Region & $T_{\rm kin}$ & $N$(C$_2$H)                   & $n$(H$_2$)                   \\
       & (K)           & (10$^{16}$ cm$^{-2}$) & (10$^5$ cm$^{-3}$)        \\
\noalign{\smallskip}
\hline
\noalign{\smallskip}
NE & $13.87 \pm 0.21$ & $1.97 \pm 0.12$ & $5.53 \pm 0.48$ \\
SW & $10.76 \pm 0.24$ & $0.89 \pm 0.05$ & $42.9 \pm 6.9$ \\
\noalign{\smallskip}
\hline
\end{tabular}
\end{table}

The observed spectral differences between the NE and SW components are consistent with the non-LTE parameters inferred for each source (Table~\ref{tab:c2h-results}). The larger intensity and broader width of the C$_2$H lines in NE can be attributed to a column density $\sim$2.2 times larger and a kinetic temperature $\sim$3~K higher, despite its lower H$_2$ volume density. The broad line profiles in NE may reflect a dynamically complex or more turbulent medium. In SW, the lower kinetic temperature appears to be offset by a higher H$_2$ volume density ($\sim4\times10^{6}$~cm$^{-3}$), likely because more efficient collisional pumping is required to excite the observed transitions in colder gas. Furthermore, the narrower line widths suggest a more quiescent gas, which may correspond to an H$_2$-dense relative to NE.

The \TCC\ and \CTC\ spectra were analysed with LTE radiative-transfer models coupled to an MCMC sampler. For these low-abundance isotopologues, we assumed thermal coupling to the gas by fixing the excitation temperature to the kinetic temperature inferred from the \CCH\ analysis ($\boldsymbol{T_{\rm ex}}=T_{\rm kin}$). Under this assumption, only the column density and velocity dispersion (line width) were treated as free parameters, while the source size was fixed to 20\arcsec \ and the systemic velocity to $V_{\rm LSR}=-11.3$~km~s$^{-1}$ (NE) and $-12.3$~km~s$^{-1}$ (SW). The resulting column densities and the associated $^{12}$C/$^{13}$C ratios are reported in Table~\ref{tab:isotopologues}, and the corresponding best-fit LTE spectra are shown in Fig.~\ref{isos-cch}.

The LTE--MCMC solutions yield systematically larger \CTC\ column densities than \TCC\ in both regions. This offset is plausibly attributable to the cleaner baselines and higher signal-to-noise ratio of the \CTC\ lines, which translate into tighter constraints; nevertheless, the \TCC\ and \CTC\ column densities are statistically consistent within the formal uncertainties. The derived isotopologue ratios, \TCC/\CTC, are $\simeq 0.74$ (NE) and $\simeq 0.68$ (SW), suggesting preferential incorporation of $^{13}$C at the terminal carbon position (the C--H bond). Given \CCH\ formation routes involving C and CH$_2$, a detailed examination of fractionation may help identify which reactant supplies $^{13}$C more effectively \citep{Turner2000,Taniguchi2024}. The observed preference for \CTC\ is consistent with isotopomer-exchange reactions,
\TCC\ + H $\rightleftharpoons$ \CTC\ + H + 8.1~K \citep{Furuya2011},
which favour \CTC\ at the low temperatures characteristic of the cold gas traced by C$_2$H in NE and SW. This trend is consistent with carbon-chain $^{13}$C fractionation driven by zero-point-energy differences, as previously observed in cold, dense clouds \citep[e.g.][]{Furuya2011,Taniguchi2024}.

\begin{table}[t!]
\centering
\caption{Column densities of the $^{13}$C isotopologues of \CCH\ and isotope ratios, derived assuming LTE with $\boldsymbol{T_{\rm ex}} = T_{\rm kin}$ taken from the C$_2$H analysis (14~K for NE, 11~K for SW).}
\label{tab:isotopologues}
\begin{tabular}{l c c } 
\hline\hline
\noalign{\smallskip}
Species &   NE &  SW                \\
\noalign{\smallskip}
\hline
\noalign{\smallskip}
$N$(\TCC)\  & $(1.51\pm0.63)\times10^{14}$ & $(1.46\pm0.65)\times10^{14}$ \\
$N$(\CTC)\  & $(2.03\pm0.02)\times10^{14}$ & $(2.16\pm0.26)\times10^{14}$ \\
CCH/$^{13}$CCH & $131\pm56$ & $61\pm28$ \\
CCH/C$^{13}$CH & $97\pm6$ & $41\pm6$ \\
\noalign{\smallskip}
\hline
\end{tabular}
\tablefoot{LTE-MCMC mean $\pm$ 1$\sigma$ in cm$^{-2}$.}
\end{table}

The isotopic ratios derived for \CCH\ are \CCH/\TCC $\thickapprox$ 131 (NE) and 61 (SW), together with \CCH/\CTC $\thickapprox$ 97 (NE) and 41 (SW). These values fall within the range reported for other Galactic environments. 
In the cold dark cloud TMC-1, \cite{Sakai2010} obtained values larger than 250 and 170, respectively, while the low-mass star-forming region L1527 showed ratios larger than 135 and 80, respectively, both showing moderate $^{13}$C dilution relative to the local interstellar ratio of $\sim$ 60-70 \citep{Milam2005}. Comparable elevated ratios are found in other starless cores \citep{Taniguchi2019}. In the TW Hya protoplanetary disk, \cite{Bergin2024} measured \CCH/\TCC $\thickapprox$ 65, close to the elemental value.  The RCW 122 ratios are therefore consistent with the observed scatter and reflect the combined influence of formation pathways, isotopic fractionation, and local physical conditions.

\section{Astrochemical implications}\label{sec5}

\subsection{Chemical model setup}\label{sec5.1}

\begin{figure}[!t]
\centering
\includegraphics[width=1.0\linewidth]{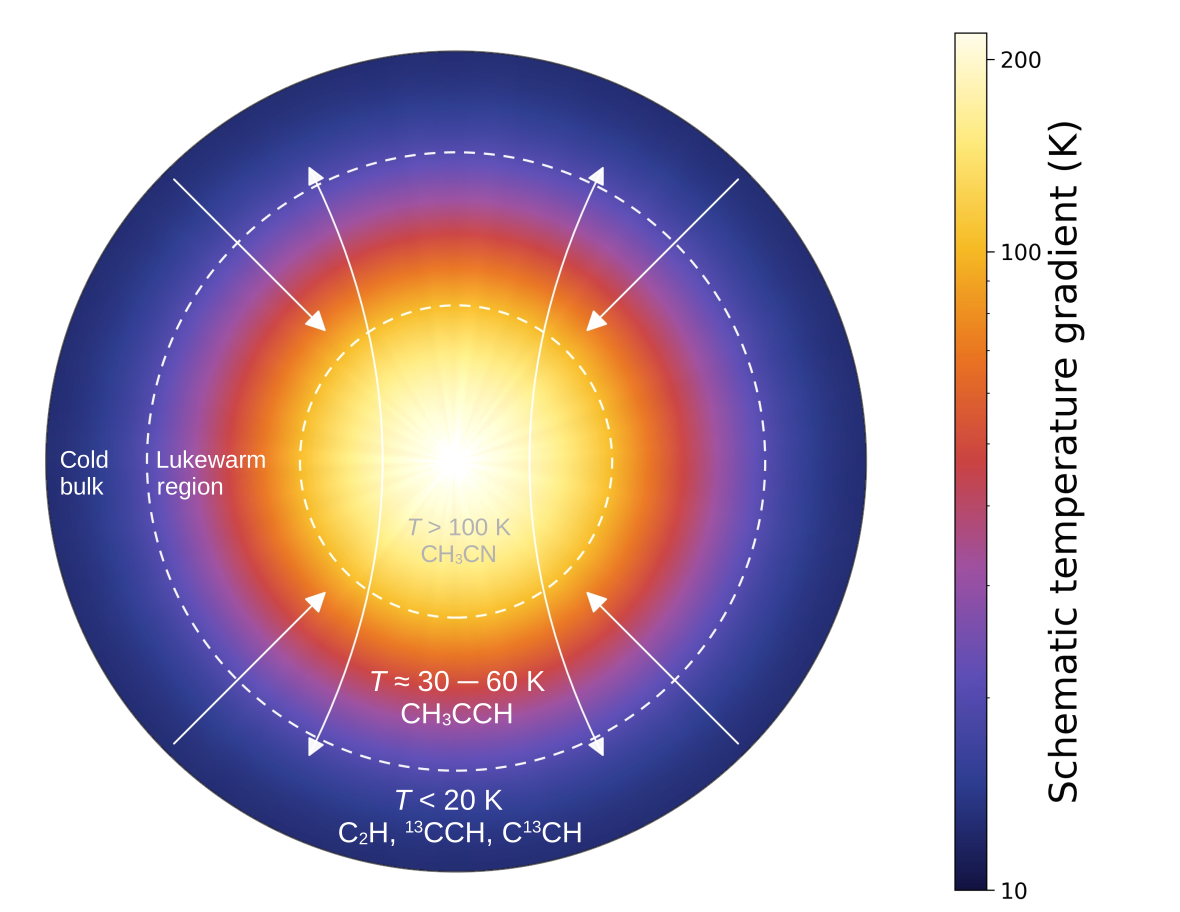}
\caption{Schematic of the thermal and chemical stratification. Molecular tracers delineate distinct thermal zones. \CCH\ and its isotopologues trace the coldest and most pristine gas. \MA\ marks an intermediate lukewarm region of more processed gas surrounding the  high-mass YSO. The innermost region corresponds to $T >$ 100 K domain, typically traced by, e.g., CH$_3$CN.}
\label{fig:SCHEM}
\end{figure}

The formation pathways and abundances of \CCH\ and \MA\ have been investigated using gas--grain chemical models constrained by the physical conditions derived from the LTE and non-LTE analyses of Sect.~\ref{sec4}. The simulations follow a high-mass  YSO evolutionary sequence consisting of the collapse of an initial molecular cloud complex, the formation of a  high-mass YSO, and the subsequent development of a central H\,II region driven by ionizing radiation from the central YSO (see Fig.~\ref{fig:SCHEM} for a schematic illustration).

\begin{table}[!t]
\caption{Initial abundances computed in the chemical models of \NE\ and \SW (extracted from \citealt{sant2022}).}
\label{tab:init-abun}
\centering
\begin{tabular}{llll}
\hline\hline
Species & $n_i/n_{\rm H}$ & Species & $n_i/n_{\rm H}$ \\
\hline
H$_2$ 	& 0.5 	                &  He      & 9.0 $\times 10^{-2}$ \\
N  	    & 6.2 $\times 10^{-5}$   &  O       & 2.4 $\times 10^{-4}$ \\
C$^+$   & 1.7 $\times 10^{-4}$   &  S$^+$   & 1.5 $\times 10^{-5}$ \\
Fe$^+$  & 3.0 $\times 10^{-9}$   &  Si$^+$  & 8.0 $\times 10^{-9}$ \\
Na$^+$  & 2.0 $\times 10^{-9}$   &  Mg$^+$  & 7.0 $\times 10^{-9}$ \\
Cl$^+$  & 1.0 $\times 10^{-9}$   &  P$^+$   & 2.0 $\times 10^{-10}$\\
F       & 6.7 $\times 10^{-9}$   &          &           \\
\hline
\end{tabular}
\end{table}

\begin{figure}[!h]
\centering
\includegraphics[width=1.0\linewidth]{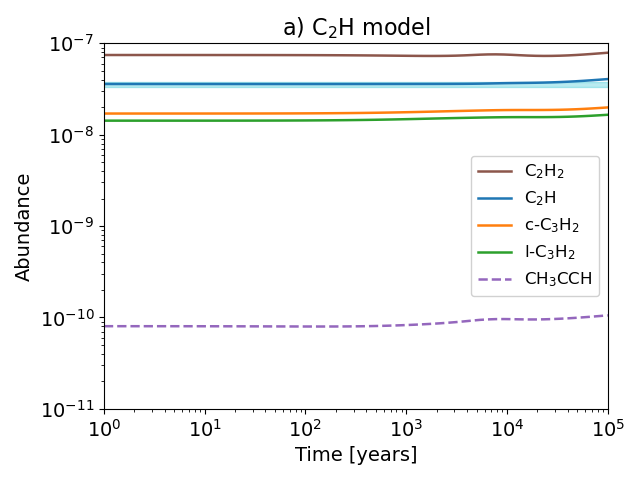}
\includegraphics[width=1.0\linewidth]{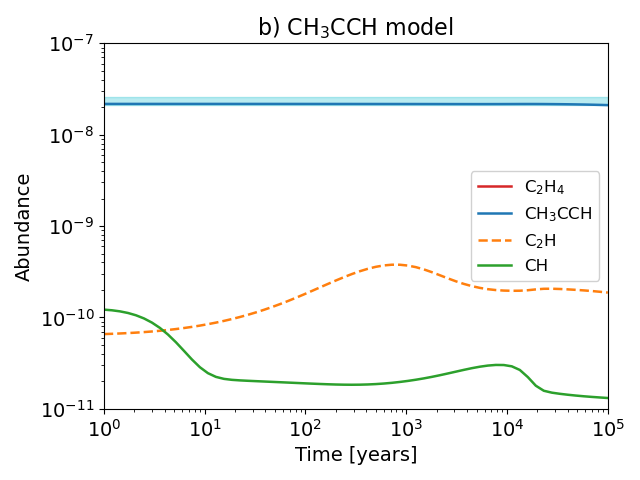}
\caption{Molecular abundances  relative to H computed in the final stage for (a) \CCH\ and related species,  adopting $T = 15$~K, $n = 2\times10^{4}$~cm$^{-3}$,
$\zeta = 10\zeta_0$. (b) \MA\ and related species, adopting
$T =60$~K, $n = 2\times10^{5}$~cm$^{-3}$,
$\zeta = 0.1\zeta_0$.
Dashed lines indicate species whose abundances are underpredicted in each stage
(\MA\ in a, \CCH \ in b).
Horizontal light blue strips show the observed abundance  (relative to H) ranges estimated for NE and SW.}\label{fig:kinetics}
\end{figure}

\CCH\ and \MA\ were modelled following
the sequences described in Sect.~\ref{sec2.2}
(see Table~\ref{tab:sim_params}). For \CCH,\ ion--molecule reactions dominate
its formation under the cold conditions of the second stage, which makes
the cosmic-ray ionisation rate a controlling parameter; we therefore adopted
an enhanced rate of $10\zeta_0$ (with $\zeta_0 = 1.3\times10^{-17}$~s$^{-1}$).
The resulting abundances of \CCH\ and chemically related species, for a
representative computation at $T = 15$~K, are shown in
Fig.~\ref{fig:kinetics}a. In the final stage, the modelled \CCH\ abundance (relative to H)
approaches a quasi-steady value of $\sim$4.1$\times10^{-8}$, equivalent to $\sim$8.2$\times10^{-8}$ relative to H$_2$.
This is within a factor of $\thickapprox$ 1.1--1.2 of the NE and
SW observed abundance (Table~\ref{tab:abundances}).

The formation of \MA\ involves a substantially more complex reaction network and typically requires gas--grain chemistry to reproduce the observed abundances. A significant fraction of \MA\ is produced on grain mantles and subsequently returned to the gas phase through thermal desorption, thereby enhancing its gas-phase abundance \citep{miettinen2006,Hickson2016}. In agreement with this scenario, previous studies have found efficient
grain-surface production of \MA\ in gas--grain models of the low-mass protostar IRAS 16293--2422 \citep{Andron2018}. Thus, using abundances inherited from the molecular-cloud stage as initial conditions, we computed a second, peak-temperature stage in which \MA\ is injected into the gas phase, and then followed its evolution in a third, lukewarm stage consistent with the LTE temperatures (30--60~K). In this final stage we adopted a reduced ionisation rate ($0.1\zeta_0$), which favours \MA\ survival. The resulting abundances are shown in Fig.~\ref{fig:kinetics}b for a representative temperature of 60~K. In this stage, the modelled \MA\ abundance reaches $\sim2.1\times10^{-8}$  relative to H ($\simeq 4.2\times10^{-8}$ relative to H$_2$), which is within a factor $\thickapprox$ 1.6--2.8 of the observed values of \MA\ in both NE and SW  (Table~\ref{tab:abundances}).

\begin{table*}[!h]
\centering
\caption{Physical conditions used at each stage of the chemical simulations for \CCH\ and \MA, starting with a common model (Stage 1) and continuing with separate stages for the chemistry of each species.}
\label{tab:sim_params}
\begin{tabular}{l cc c ccc}
\hline\hline
\noalign{\smallskip}
Parameter                & \multicolumn{2}{c}{\CCH}
                & \multicolumn{4}{c}{CH$_3$CCH} \\
\cline{2-3} \cline{5-7}
       & Stage 1 & Stage 2 & &Stage 1 & Stage 2 & Stage 3 \\
\hline
$n$(H$_2$) (cm$^{-3}$)          & $2 \times 10^{4}$ & $2 \times 10^{4}$ & & $2 \times 10^{4}$ & $2 \times 10^{6}$ & $2 \times 10^{5}$ \\
$T$ (K)                       & 10       & 10--20       & & 10       & 50--200      & 30--60   \\
$\zeta$ (s$^{-1}$) & 10$\zeta_0$       & 10$\zeta_0$       & & 10$\zeta_0$       & 10$\zeta_0$       & 0.1$\zeta_0$      \\
Time (Myr)                     & 1 & 0.1 & & 1 & 0.01 & 0.1 \\
\hline
\end{tabular}
\tablefoot{$\zeta_0=1.3\times10^{-17}$s$^{-1}$, corresponding to the standard cosmic-ray ionisation rate.}
\end{table*}

\subsection{Formation and destruction pathways}\label{sec5.2}

In the first stage, the \CCH\ radical is overwhelmingly formed via dissociative recombination and neutral--neutral reactions

\begin{equation*}
\begin{aligned}
&\mathrm{C_2H_2^+ + e^- \rightarrow H + C_2H} \\
&\mathrm{C_3H^+ + e^- \rightarrow C + C_2H} \\
&\mathrm{C + CH_2 \rightarrow H + C_2H}
\end{aligned}
\end{equation*}

Destruction in this phase is controlled by reactions with ionised and neutral carbon, reflecting the high gas-phase elemental carbon abundance that characterises this stage. In the second stage, the \CCH\ bulk is no longer dominated by a single channel. Instead, it is fed by a broad ensemble of dissociative recombinations of larger hydrocarbon molecular cations

\begin{equation*}
\begin{aligned}
&\mathrm{C_2H_3^+ + e^- \rightarrow C_2H + H + H} \\
&\mathrm{C_5H_3^+ + e^- \rightarrow C_2H + \mathit{c}\text{-}C_3H_2} \\
&\mathrm{C_5H_3^+ + e^- \rightarrow C_2H + \mathit{l}\text{-}C_3H_2} \\
&\mathrm{C_4H_3^+ + e^- \rightarrow C_2H + C_2H_2}
\end{aligned}
\end{equation*}

Destruction in this second phase is dominated by reactions with N, C$_3$, C, and O. Overall, the \CCH\ abundance is governed by a competition between the recombination-driven recycling of hydrocarbon cations and the N-, C-, and O-initiated destruction channels. Under the cold conditions relevant to \CCH,\ dissociative recombination with
electrons is the key process sustaining the chemistry, because neutral--neutral reactions are highly inefficient and the radical is formed almost exclusively through the recombination of hydrocarbon cations. This makes cosmic-ray ionisation an important driver of the \CCH\ chemistry at low temperatures.

Concerning \MA,\ in the first stage, its formation is initially dominated by the dissociative recombination reaction

\begin{equation*}
\mathrm{C_3H_5^+ + e^- \rightarrow H + CH_3CCH}
\end{equation*}

Destruction is governed by reactions with C$^+$, which drive the formation of several molecular cations, along with some reactions with neutral C leading to neutral species. The second stage marks a turning point for the \MA\ abundance, since the dominant production route becomes thermal desorption from grain
mantles, which is responsible for injecting the bulk of the gas-phase \MA. The third and final stage is of particular interest, as it represents the phase in which gas-phase carbon-chain chemistry becomes established, at lukewarm temperatures representative of the conditions derived from the observations. Production is governed by the dissociative recombination of hydrocarbon cations and a neutral--neutral reaction:

\begin{equation*}
\begin{aligned}
&\mathrm{C_3H_5^+ + e^- \rightarrow H + CH_3CCH} \\
&\mathrm{C_4H_5^+ + e^- \rightarrow CH + CH_3CCH} \\
&\mathrm{CH + C_2H_4 \rightarrow H + CH_3CCH}
\end{aligned}
\end{equation*}

An analogous interplay of grain-surface synthesis, thermal desorption, and gas-phase processing was found in low-mass protostar IRAS 16293--2422 \citep{Andron2018}. \MA\ destruction in the final stage proceeds mainly through reactions with C and H$_3$O$^+$, with other neutral species contributing to a lesser extent. Although the third stage is dominated by gas-phase ion
chemistry, the \MA\ abundance retains the memory of the thermal desorption event from the previous stage (e.g., \citealt{Hickson2016}). That process set the initial gas-phase budget from which the late-time network evolved, making thermal desorption the ultimate source of the gas-phase \MA\ observed at late times. It is worth noticing that this last stage
requires a low cosmic-ray ionisation rate to sustain small abundances of molecular cations, which would otherwise reduce the \MA\ abundance through destruction reactions. The late-time \MA\ abundance is therefore sensitive to the cosmic-ray flux --- making it, together with the desorbed mantle reservoir, one of the two key factors setting the final abundance level of \MA\ in the lukewarm regions surrounding the high-mass YSO.

\subsection{Comparison with observations}\label{sec5.3}

The chemical modelling of \MA\ and \CCH\ in both regions allowed us to derive and constrain their abundances. We infer that the chemistry of \CCH\ and that of \MA\ are reasonably independent and that the two molecules trace different gas layers that have undergone distinct chemical processing. \CCH\ traces colder, early-stage, and relatively unprocessed gas, whereas \MA\ requires higher temperatures and more extensive chemical processing, consistent with a later stage of chemical evolution. This interpretation is found to be similar for both sources. In the context of high-mass  YSOs, a single type of chemical model can reproduce the observed abundances (Sect.~\ref{sec5.1}) on evolutionary timescales comparable to those expected for such objects. Our results suggest that, given that NE and SW were formed within the same molecular complex, they share similar physical and chemical properties, although they may differ in age or in the evolutionary stage at which they are observed. This hypothesis requires further investigation, including new observations and a study of species that trace the hot-core region.

Table~\ref{tab:abundances} compares the physical conditions obtained from \MA\ and \CCH\ toward NE and SW with those reported for other Galactic sources. The temperature components derived from \MA\ for NE (41-- 61 K) and SW (33-- 48 K) are found within the range commonly found in high-mass star-forming regions and are consistent with warm molecular gas associated with embedded high-mass YSOs. The most closely comparable temperature regimes are reported for G331.512-0.103 (see e.g., \citealt{sant2022}), G12.89+0.89, G16.86-2.16, and several of the luminous infrared sources studied by \citet{miettinen2006}, which exhibit temperatures of roughly 23-- 50 K and \MA\ abundances comparable to, or only slightly higher than, the values estimated toward NE and SW. This supports the interpretation that the gas probed by \MA\ in RCW122 is warm but not unusually so, placing NE and SW within the broad population of chemically active high-mass YSOs. Other well-studied high-mass regions, such as Orion~KL, Sg~B2, S140, and DR21, show comparable temperatures, typically around 30-- 50 K, but systematically lower \MA\ abundances, generally of the order of a few $\times 10^{-9}$. The most pronounced differences are seen toward Sgr~A and the low-mass protostar IRAS~16293$-$2422, which display \MA\ abundances of order $10^{-7}$ despite temperatures overlapping the range derived toward NE and SW. Since these abundances cannot be explained by thermal excitation alone, they likely reflect differences in chemical histories, density structures, or local environmental conditions.

For \CCH,\ the temperatures obtained for NE (14 K) and SW (11 K) are comparable to those found in the CCH-rich source Mon R2 (12–16 K; \citealt{ginard2012}), the  cold low-mass protostar R CrA IRS7B (15–25 K; \citealt{wata2012}), and  cold dark starless clouds such as L1521B (6.6 K) and L134N (6.4 K) \citep{Taniguchi2019}). These environments share not only similarly low excitation temperatures but also \CCH\ abundances comparable to, or only moderately lower than, those estimated for NE and SW. This behaviour supports the commonly observed picture in which \CCH\ is abundant in cold or weakly processed gas, where efficient carbon-chain chemistry can operate while destruction processes remain relatively limited. A similar combination of low-to-moderate temperature and high \CCH\ abundance is also seen in photon-dominated regions such as the Horsehead Nebula, IC 63 and L1688-W, where \CCH\ abundances reach values of the order of 10$^{-8}$ \citep{tey2004}. In contrast,  classical high-mass star-forming clouds, including Orion-KL, Orion A, M17 SW, W51, S106, and DR 21(OH),  and NGC 2024, display systematically higher temperatures, generally spanning 10–70 K, and \CCH\ abundances that are one to two orders of magnitude lower, mostly in the range 10$^{-10}$–10$^{-9}$ \citep{blake1987, wooten80,watt1988}. This contrast reinforces the scenario in which \CCH\ is tracing a cold and primitive bulk gas surrounding the \NE\ and \SW\ clumps.

\begin{table*}[h!]
\centering
\caption{Molecular abundance, temperature, and column density of CH$_3$CCH and \CCH\ derived for Galactic HMCs and star-forming regions, as reported in the literature.} 
\begin{tabular}{lcccc}
\hline
\hline 

\vspace{0.1cm}
 &    & \large{CH$_3$CCH}     &    &     \\
 \hline
Region & Temperature (K) &  Column density (cm$^{-2}$) & Abundance  & References\\

\hline
 \NE  &  61   &  $7.1 \times 10^{15}$  & $2.7 \times 10^{-8}$ & This work \\
      &  41   &  $3.9 \times 10^{15}$  & $1.5 \times 10^{-8}$  & This work \\
 \SW  &  48   &  $3.6 \times 10^{15}$  & $2.7 \times 10^{-8}$  & This work \\
      &  33   &  $3.5 \times 10^{15}$  & $2.6 \times 10^{-8}$  & This work \\
 Orion-KL (ext.~ridge)  &  52   &  1 $\times$ 10$^{15}$       & 3.3 $\times$ 10$^{-9}$ & \citet{blake1987} \\
Sgr~A                   &  50   & 6.4 $\times$ 10$^{15}$    &   3 $\times$ 10$^{-7}$ & \citet{minh1992} \\
Sgr~B2                  &  45.7 & 3.0  $\times$ 10$^{14}$   &   $\ge$ 1.8 $\times$ 10$^{-9}$  & \citet{kuip1984} \\
S~140                   &  32.1 & 1.7  $\times$ 10$^{14}$   &   2.4 $\times$ 10$^{-9}$  & \citet{kuip1984} \\
Tau~MC1                 &  21.7 & 3.7 $\times$ 10$^{13}$    &  (2.5 - 5) $\times$ 10$^{-9}$  & \citet{kuip1984} \\
DR~21                   &  30.1 & 1.0 $\times$ 10$^{14}$    &   $\sim$ 1 $\times$ 10$^{-9}$  & \citet{kuip1984} \\
G331.512-0.103          &  50   &   7.5 $\times$ 10$^{15}$  & (0.8 - 2.8) $\times$ 10$^{-8}$ & \citet{sant2022}   \\       
G12.89+0.89             &  33   &   1.0 $\times$ 10$^{15}$   &  4.2 $\times$ 10$^{-8}$  &  \citet{Taniguchi2018} \\          
G16.86-2.16             &  29  &  5.4 $\times$ 10$^{14}$   & 3.2 $\times$ 10$^{-8}$   &  \citet{Taniguchi2018} \\
G28.28-0.36             &  23  & 3.7 $\times$ 10$^{14}$    &  7.6 $\times$ 10$^{-8}$  & \citet{Taniguchi2018} \\
G34.3+0.15              &  46.0  & 8.2 $\times$ 10$^{14}$    & 2.8 $\times$ 10$^{-9}$   & \citet{hun-dae2000} \\
IRAS~12326-6245         &  34.0   & 2.0 $\times$ 10$^{14}$   & 3.1 $\times$ 10$^{-9}$   & \citet{miettinen2006} \\
G326.64+0.61            &  29.0   &   3.4 $\times$ 10$^{14}$ & 6.0 $\times$ 10$^{-9}$   & \citet{miettinen2006} \\
OH328.81+0.63           &  30.5   &   4.9 $\times$ 10$^{14}$ &  6.6 $\times$ 10$^{-9}$ & \citet{miettinen2006} \\
G330.95-0.19            &  31.9   &   3.5 $\times$ 10$^{14}$ &  4.4 $\times$ 10$^{-9}$ & \citet{miettinen2006} \\
G345.01+1.8N            &  34.5   &   3.5 $\times$ 10$^{14}$ &  4.8 $\times$ 10$^{-9}$ & \citet{miettinen2006} \\
IRAS~16562-3959         &  35.9   &   1.5 $\times$ 10$^{15}$ &  1.8 $\times$ 10$^{-8}$ &  \citet{miettinen2006} \\
NGC~6334F               &  35.8   &   6.3 $\times$ 10$^{14}$ &  3.9 $\times$ 10$^{-9}$  &  \citet{miettinen2006} \\
G345.00-0.23            &  30.6   &   4.0 $\times$ 10$^{14}$ &  6.7 $\times$ 10$^{-9}$  &  \citet{miettinen2006} \\
IRAS~16293-2422         &  27   &   1.9 $\times$ 10$^{16}$ & 2.6  $\times$ 10$^{-7}$  &  \citet{cazaux2003} \\
\hline
\vspace{0.1cm}
 &    & \large{\CCH}     &    &     \\
 \hline
 \NE  &  14   &  $2 \times 10^{16}$  & $7.5 \times 10^{-8}$ & This work \\
 \SW  &  11   &  $9 \times 10^{15}$  & $6.6 \times 10^{-8}$  & This work \\
Orion-KL (ext. ridge)  &  60   &   1.6 $\times$ 10$^{15}$   &  5.3 $\times$ 10$^{-9}$ & \citet{blake1987} \\
Ori A                  &   70   &  1.9 $\times$ 10$^{15}$   &   2 $\times$ 10$^{-10}$ &  \citet{wooten80} \\
NGC~1333               &  25   &  $<$2.0 $\times$ 10$^{13}$ &  $<$ 2 $\times$ 10$^{-11}$ & \citet{wooten80}\\
NGC~2264               & 25     &  1.0 $\times$ 10$^{14}$   &   2 $\times$ 10$^{-10}$ & \citet{wooten80} \\
Rho Ophiuchi           & 25     &  8.4 $\times$ 10$^{13}$   & 1 $\times$ 10$^{-10}$   & \citet{wooten80} \\
Serpens                & 22     &  1.3 $\times$ 10$^{14}$   &  3 $\times$ 10$^{-10}$  & \citet{wooten80} \\
M17 SW                 &  50   & 1.4 $\times$ 10$^{15}$      &    3 $\times$ 10$^{-10}$ & \citet{wooten80} \\
W51                    & 10 - 50 & (17.7 - 3.1) $\times$ 10$^{14}$ & (8.9 - 1.6) $\times$ 10$^{-10}$ & \citet{watt1988} \\
S106                   & 10 - 50 & (11.9 - 2.1) $\times$ 10$^{14}$ & (2.0 - 0.4) $\times$ 10$^{-9}$ & \citet{watt1988} \\
DR21 (OH)              & 10 - 50 & (5.2 - 0.9) $\times$ 10$^{14}$ & (8.7 - 1.5) $\times$ 10$^{-10}$ & \citet{watt1988} \\        
NGC 2024               & 10 - 50 & (6.4 - 1.1) $\times$ 10$^{14}$ & (2.1 - 0.4) $\times$ 10$^{-9}$ & \citet{watt1988} \\ 
Mon R2                 &  12 - 16     & (4.1 - 3.7) $\times$ 10$^{14}$  & (5.4 - 8.6) $\times$ 10$^{-9}$   &  \citet{ginard2012} \\
RCrA IRS7B             & 15 - 25 & (9.8 - 4.1) $\times$ 10$^{14}$ &  (6.3 - 4.8) $\times$ 10$^{-9}$ & \citet{wata2012} \\
Horsehead              &  - - -  & (17.0 - 22.7) $\times$ 10$^{13}$ & (2.3 - 2.2) $\times$ 10$^{-8}$ & \citet{tey2004} \\
IC 63                  &  - - - & (5.3 - 8.6) $\times$ 10$^{13}$ & (1.1 - 2.1)  $\times$ 10$^{-8}$&  \citet{tey2004} \\     
L1688-W                &  - - - & (3.0 - 2.0) $\times$ 10$^{13}$ &  (0.6 - 0.8)  $\times$ 10$^{-8}$ &   \citet{tey2004} \\
L1521B                 & 6.6    & 1.66 $\times$ 10$^{14}$  & 1.75 $\times$ 10$^{-8}$   & \citet{Taniguchi2019} \\
L134N                  & 6.4    &  7.3 $\times$ 10$^{13}$  & 6.1  $\times$ 10$^{-9}$ &  \citet{Taniguchi2019} \\
\hline
\end{tabular}
\label{tab:abundances}
\end{table*}

\section{Conclusions and final remarks}\label{sec6}

\CCH\ and \MA\ in the \NE\ and \SW\ regions trace a cold-to-warm gradient, consistent with scenarios depicting molecular stratification around high-mass YSOs. The models support distinct chemistries: \CCH\ is a low-temperature, cosmic-ray--driven product of primitive gas, supported by detections of \TCC\ and \CTC; \MA\ is a later-time species whose abundance is explained by thermal desorption from grains, with no secure isotopologue detections. The inferred evolution is compatible with high-mass YSO timescales ($\sim$10$^5$~yr), following the transition from the cloud stage. The main remarks are as follows:

\begin{itemize}
    \item This work reports the robust detection of \MA\ in both NE and SW across 170.8--341.8~GHz, covering rotational levels from $J=10$--$9$ to $J=20$--$19$ and $K$-ladders from $K=0$ to $K=4$. The line intensities follow the expected excitation structure of \MA:\ for a fixed $J$, intensities decrease with increasing $K$, and the overall intensity decreases with increasing $J$, with the strongest emission arising from the lower rotational levels. The most important difference between the two regions is the marked decrease of high-$J$ emission toward SW relative to NE. This behaviour is consistent with a warmer excitation reservoir in NE, while the high-$J$ emission in SW indicates a more strongly sub-thermal excitation regime. 
    
    \item \CCH\ was detected via the hyperfine components of its $N =4-3$ transition in both the NE and SW regions, with systematically stronger emission toward NE. The line profiles show broader wings than expected for a pure Gaussian, indicating moderate optical depth effects and/or multiple velocity components and a departure from LTE. $^{13}$C isotopologues \TCC\ and \CTC\ were detected in both regions, while no deuterated counterpart CCD was found. The \CTC\ emission is stronger than \TCC, indicating that the two carbon atoms in \CCH\ are chemically non-equivalent and that the \CTC\ preference is consistent with isotopomer exchange favoured at the low temperatures traced by \CCH\ in both regions.
 
    \item We constructed LTE rotational diagrams for \MA\ towards the NE and SW regions; the 39 detected transitions were partitioned into low- and high-excitation components, separated at $E_u \thickapprox $155 K, where the clear break in the rotational diagram indicates distinct excitation regimes that a single-component fit could not adequately reproduce. The fits yield excitation temperatures of 41.4 $\pm$ 1.4 K and 61.2 $\pm$ 6.1 K for the low- and high-excitation components in the NE, and 32.7 $\pm$ 1.0 K and 47.7 $\pm$ 4.9 K in the SW, with corresponding column densities of 3.9 $\pm$ 0.4 and 7.1 $\pm$ 2.3 ($\times$10$^{15}$ cm$^{-2}$) in  NE and 3.5 $\pm$ 0.3 and 3.6 $\pm$ 1.5 ($\times$10$^{15}$ cm$^{-2}$) in  SW.
    
    \item The excitation conditions of \CCH\ toward the NE and SW regions were derived from non-LTE MCMC calculations. NE exhibits a higher column density $N$(\CCH)\ $= (1.97 \pm 0.12) \times 10^{16}$ cm$^{-2}$ and a warmer kinetic temperature $T_{\rm kin} = 13.87 \pm 0.21$ K, whereas SW is colder $T_{\rm kin} = 10.76 \pm 0.24$ K and $N$(\CCH)\ $= (8.9 \pm 0.5) \times 10^{15}$ cm$^{-2}$.  The LTE-MCMC analysis shows that \CTC\ is slightly more abundant than \TCC\ toward both the NE and SW regions, indicating preferential incorporation of $^{13}$C at the terminal carbon position. The resulting $^{12}$C/$^{13}$C ratios are consistent with values found in other Galactic environments, reflecting the combined effects of chemistry, isotopomer fractionation, and local physical conditions.
    
    \item Gas–grain chemical models constrained by the derived physical conditions successfully reproduce the observed abundances of \CCH\ and \MA,\ confirming their utility as tracers for distinct evolutionary phases. The analysis demonstrates that \CCH\ formation is driven by ion–molecule reactions in cold ($10$–20 K), minimally processed gas regulated by a high cosmic-ray ionization rate (10$\zeta$). In contrast, \MA\ requires a complex thermal history involving a warm-up phase ($200$ K) followed by gas-phase enrichment via thermal desorption and cooling. These results indicate that while both species coexist within the high-mass  YSO sequence, they chemically segregate the environment: \CCH\ traces the cold, early-stage envelope, whereas \MA\ traces warmer, chemically evolved gas associated with later high-mass YSO evolution.
    
    \item Our analysis of the NE and SW regions reveals that while they share parallel excitation behaviours, \CCH\ and \MA\ trace chemically distinct gas layers, effectively serving as markers for different evolutionary stages. \CCH\ characterizes a cold ($\sim$11–14 K), primitive environment with abundances comparable to dark clouds, indicating early-stage, unprocessed gas. In contrast, \MA\ probes warmer ($\sim$33–61 K), chemically evolved gas associated with embedded high-mass YSOs. Notably, the \MA\ abundances derived here are systematically higher than those in classical high-mass regions like Orion-KL, suggesting a specific chemical richness in these sources. Ultimately, the decoupling of these species confirms that \CCH\ traces the cold envelope while \MA\ traces the warm interior, placing the NE and SW clumps within the population of chemically active high-mass YSOs that may differ in age or evolutionary phase.

\end{itemize}

\begin{acknowledgements}
       We thank the anonymous referee for the constructive revision, which has substantially improved the manuscript. LB and RF gratefully acknowledges support from the ANID Basal projects 210003 and FB210003. XHL is funded by the Xinjiang Talent Development Fund (XJRC-2025-KJ-PY-KJLJ-104), the National Science Foundation of China (No. 12473025), and the Natural Science Foundation of Xinjiang Uygur Autonomous Region (No. 2024D01E37). ZX acknowledges the Xinjiang Talent Development Fund (XJRC-2025-KJ-YJ-CXPT- 067). ND and MO are members of the Carrera del Investigador Cient\'ifico of CONICET, Argentina.
       This work is based on the following ALMA data: ADS/JAO.ALMA $\#$2015.1.01209 and $\#$2019.1.00195 . ALMA is a partnership of ESO (representing its member states), NSF (USA) and NINS (Japan), together with NRC (Canada), MOST and ASIAA (Taiwan), and KASI (Republic of Korea), in cooperation with the Republic of Chile. The Joint ALMA Observatory is operated by ESO, AUI/NRAO and NAOJ.
\end{acknowledgements}

%
\bibliographystyle{aa} 
\bibliography{bibliography_fixed.bib} 
\appendix

\section{Parametrization of the chemical models}
\label{app1}

This appendix details the chemical models presented in Sects.~\ref{sec2.2} and \ref{sec5.1}. We first examine how the adopted cosmic-ray ionisation rate and warm-up peak temperature affect the modelled \CCH\ and \MA\ abundances, and then show the full time evolution of the dominant species in the two-stage \CCH\ and three-stage \MA\ models.

\subsection{Sensitivity of \CCH \ and \MA\ abundances to the cosmic‑ray ionization rate}

\begin{figure}[!htbp]
\centering
\includegraphics[width=0.8\linewidth]{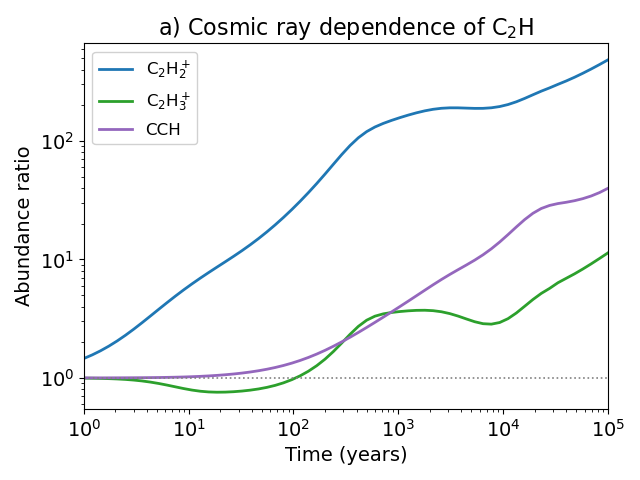}\\
\includegraphics[width=0.8\linewidth]{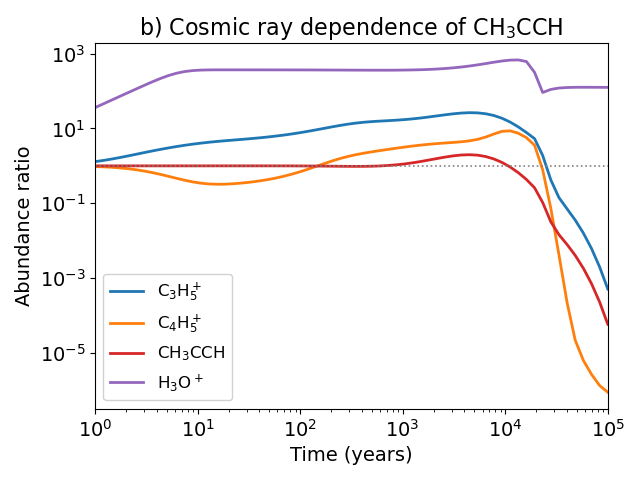}
\caption{Sensitivity of the modelled fractional abundances ($X_i$) to the cosmic-ray ionization rate, shown for (a) \CCH \ and (b) \MA \ and their associated species. The dependence is quantified by the abundance ratio $X_i(\zeta=1.3\times10^{-15}\,\mathrm{s}^{-1})/X_i(\zeta=1.3\times10^{-18}\,\mathrm{s}^{-1})$.}
\label{fig:crs-app}
\end{figure}

The cosmic-ray ionisation rate ($\zeta$) controls the abundances of species such as H$_3^+$ and He$^+$  and thus the ion--molecule chemistry \citep{Gaches2026} that initiates the formation of both \CCH\ and \MA\ at low temperatures (Sect.\ref{sec5.2}). To assess the impact of $\zeta$ on the chemistry of these species, models were tested for $\zeta = (0.1$--$100)\zeta_0$ ($\zeta_0=1.3\times10^{-17}$s$^{-1}$).  Figure~\ref{fig:crs-app} reports the sensitivity as $X_i(\zeta=1.3\times10^{-15})/X_i(\zeta=1.3\times10^{-18})$, where $X_i$ is the fractional abundance relative to H. For \CCH\ (Fig.\ref{fig:crs-app}a), the ratio is $>1$ at all times and increases monotonically, reaching $\sim$10--100 at late times; the effect is larger for the precursors C$_2$H$_2^+$ and C$_2$H$_3^+$ ($\sim$10 to a few times $10^2$), which enhance the dissociative-recombination channels forming \CCH.\ By contrast, \MA\ shows a non-monotonic response (Fig.\ref{fig:crs-app}b): the ratio is near unity at early times but decreases to $\sim10^{-3}$ at late times because higher $\zeta$ increases H$_3$O$^+$ (by up to $\sim$2 dex), and H$_3$O$^+$ together with atomic C dominates \MA\ destruction in the third stage. This motivates adopting a low $\zeta$ during the third stage to limit cation-driven destruction and preserve the post-desorption \MA\ abundance through late times.

\subsection{Thermal effect on \MA\ desorption and destruction}

\begin{figure}
\centering
\includegraphics[width=0.8\linewidth]{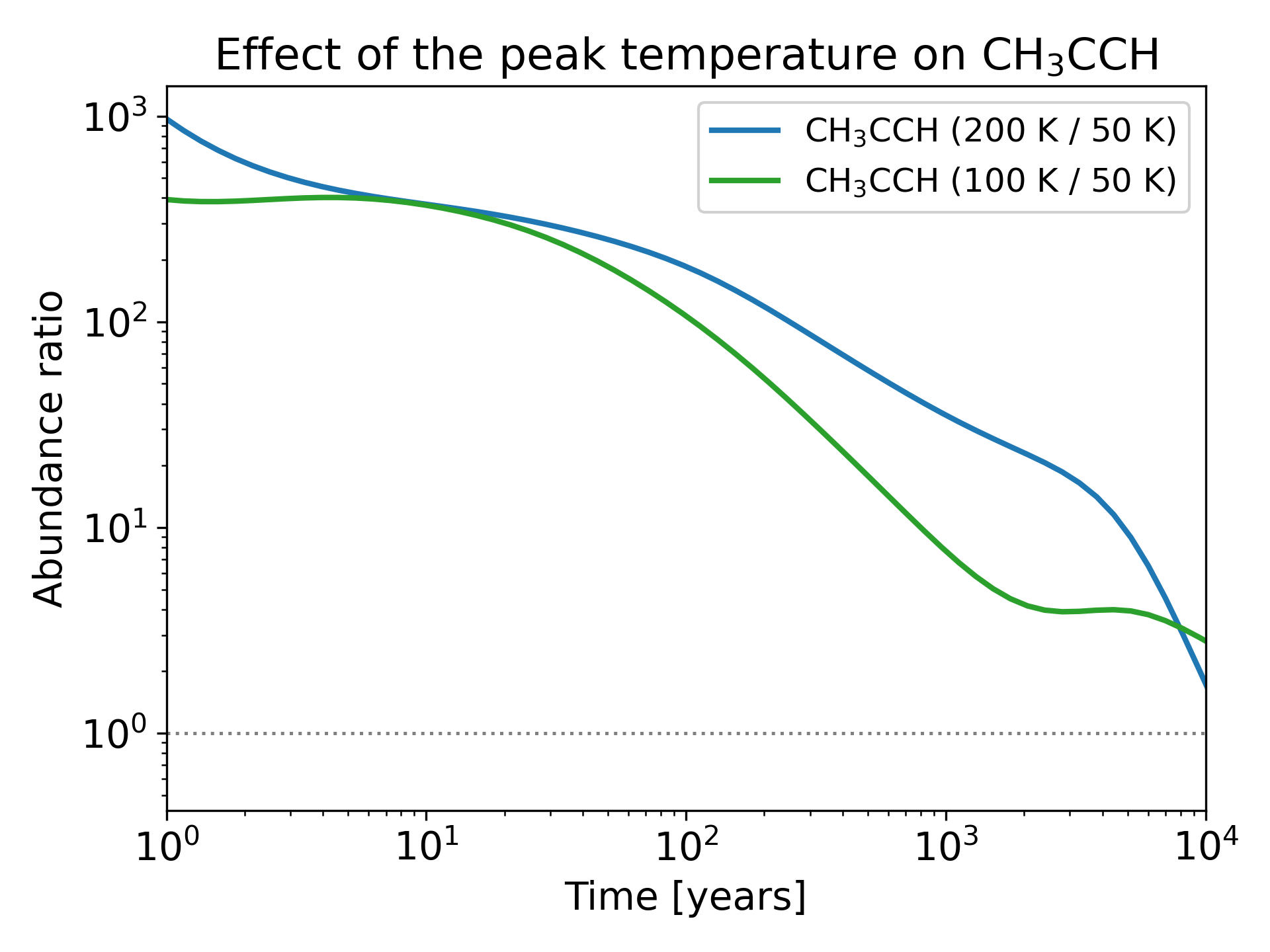}
\caption{Effect of the peak temperature reached during the warm-up stage, $T_{\rm peak}$, on the gas-phase abundance of \MA.\ The curves show the abundance ratios $X({\rm CH}_3{\rm CCH}; T_{\rm peak})/X({\rm CH}_3{\rm CCH}; 50~{\rm K})$ for $T_{\rm peak} = 100$ and 200~K.}
\label{fig:temp-app}
\end{figure}

A key parameter of the \MA\ model is the peak temperature of the warm-up phase, $T_{\rm peak}\leq200$~K, which favours thermal desorption of \MA\ from grain mantles. Figure~\ref{fig:temp-app} shows the abundance ratios $X({\rm CH}_3{\rm CCH};T_{\rm peak})/X({\rm CH}_3{\rm CCH};50\,{\rm K})$ for $T_{\rm peak}=100$ and 200~K. At early times ($t\lesssim10^2$~yr) both ratios exceed unity by 1--2 dex, indicating earlier and more complete mantle release at higher $T_{\rm peak}$. At late times the ratios approach unity. The response is therefore non-monotonic, reflecting competition between desorption-driven replenishment of gas-phase \MA\ and temperature-enhanced destruction by channels involving atomic C and H$_3$O$^+$, which becomes efficient above $\sim$100~K. This warm-up stage, with $T_{\rm peak} \leq$~200~K,  is consistent with the observed gas-phase abundance of \MA\ and earlier findings in the literature \citep{miettinen2006,Hickson2016,sant2022}.

\subsection{Abundance evolution in the two-stage \CCH\ model}

\begin{figure}[!htbp]
\centering
\includegraphics[width=0.8\linewidth]{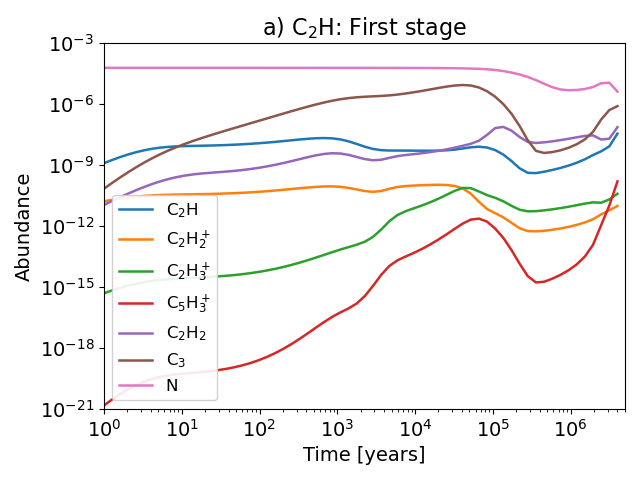}
\includegraphics[width=0.8\linewidth]{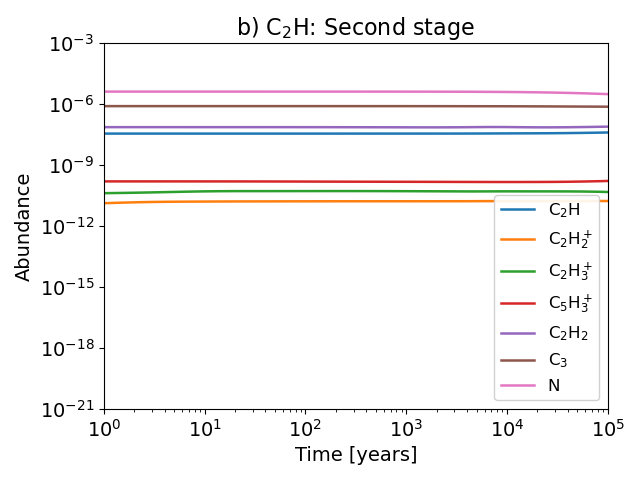}
\caption{Time evolution of the abundances of \CCH\  relative to H and of the species dominating its formation and destruction. Panel (a), stage 1: $n({\rm H}_2) = 2\times10^{4}$ cm$^{-3}$, $T = 10$ K, $\zeta = 10\zeta_0$. Panel (b), stage 2: $n({\rm H}_2) = 2\times10^{4}$ cm$^{-3}$, $T = 15$ K, $\zeta = 10\zeta_0$.}
\label{fig:cch-tracers}
\end{figure}

Figure~\ref{fig:cch-tracers} presents the temporal evolution of \CCH\ and the main species governing its formation and destruction in the two-stage \CCH\ model (Table~\ref{tab:sim_params}). During the first stage (Fig.~\ref{fig:cch-tracers}a), the hydrocarbon ions C$_2$H$_2^+$ and C$_2$H$_3^+$ build up within the C$^+$-initiated hydrocarbon network; their dissociative recombination produces C$_2$H, which approaches a quasi-steady abundance of $\sim10^{-8}$ after $\sim$~0.1~Myr. During the second stage (Fig.~\ref{fig:cch-tracers}b), 
the abundance profiles remain close to quasi-steady state: the C$_2$H abundance varies by only $\sim$15\% between the beginning and end of the stage (based on the model initial and final abundances). The relative ordering of the dominant species is preserved, indicating that the modest temperature increase does not qualitatively modify the cold-gas chemistry.

\subsection{Abundance evolution in the three-stage \MA\ model}

\begin{figure}[!htbp]
\centering
\includegraphics[width=0.8\linewidth]{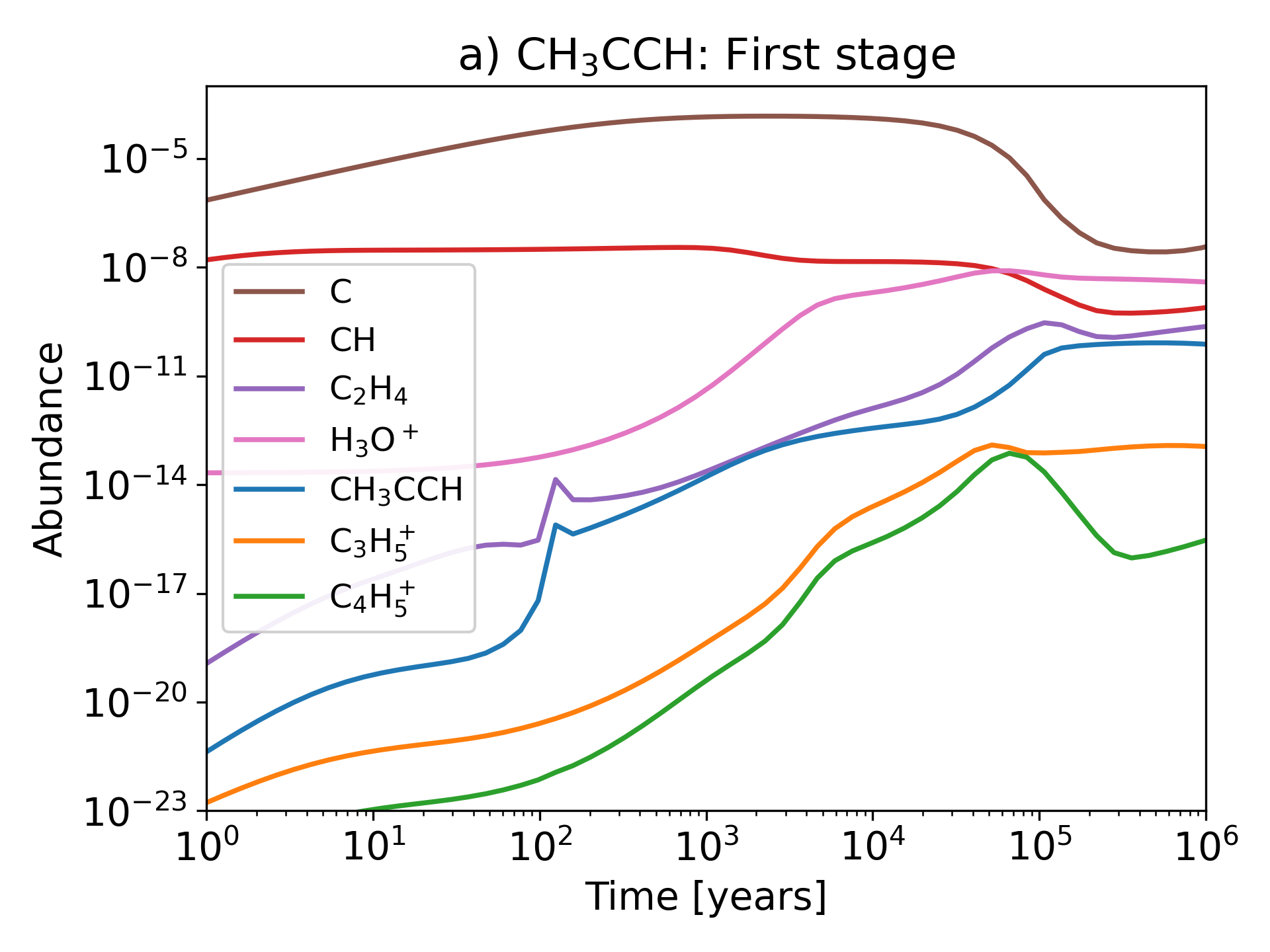}
\includegraphics[width=0.8\linewidth]{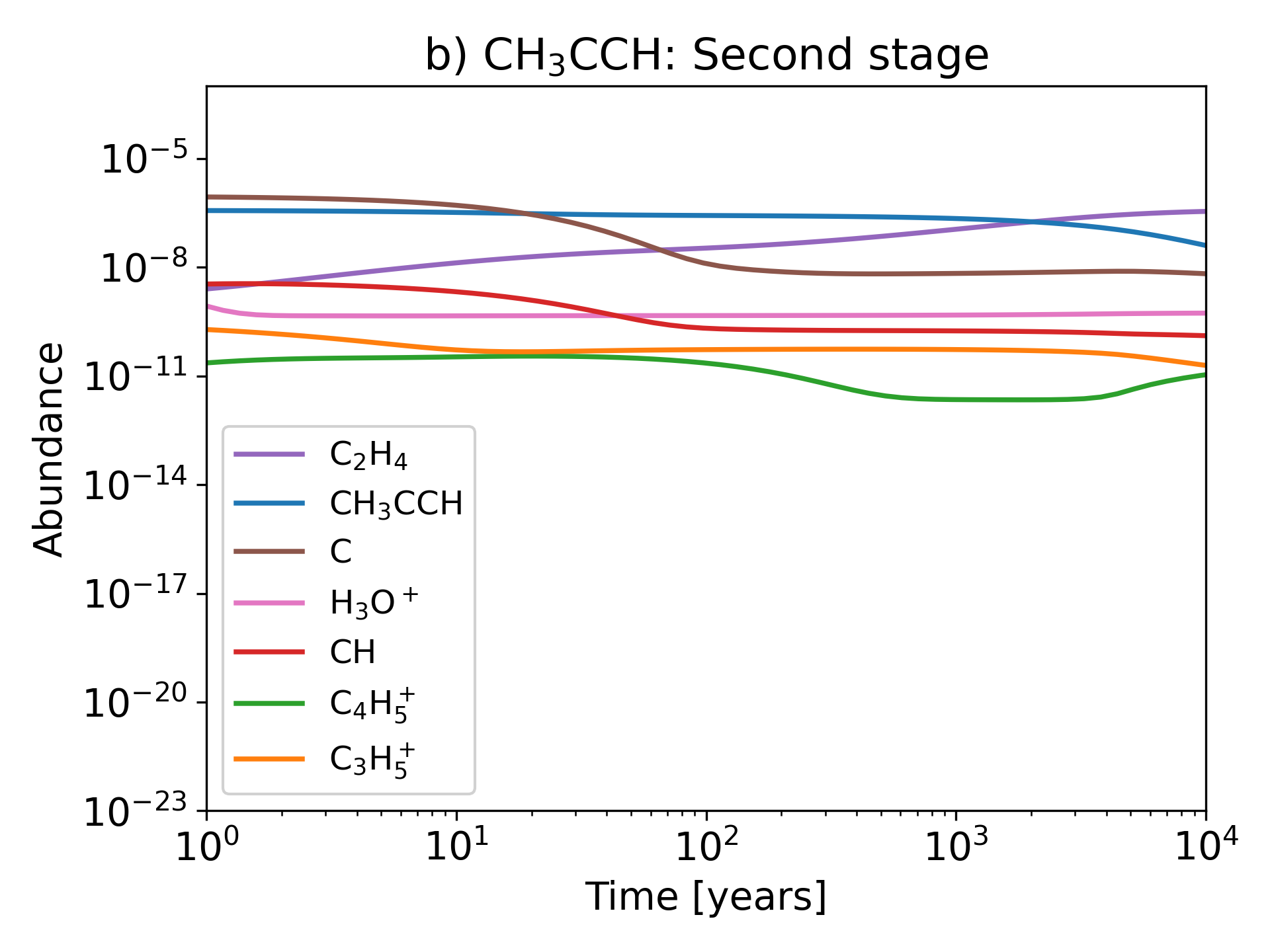}
\includegraphics[width=0.8\linewidth]{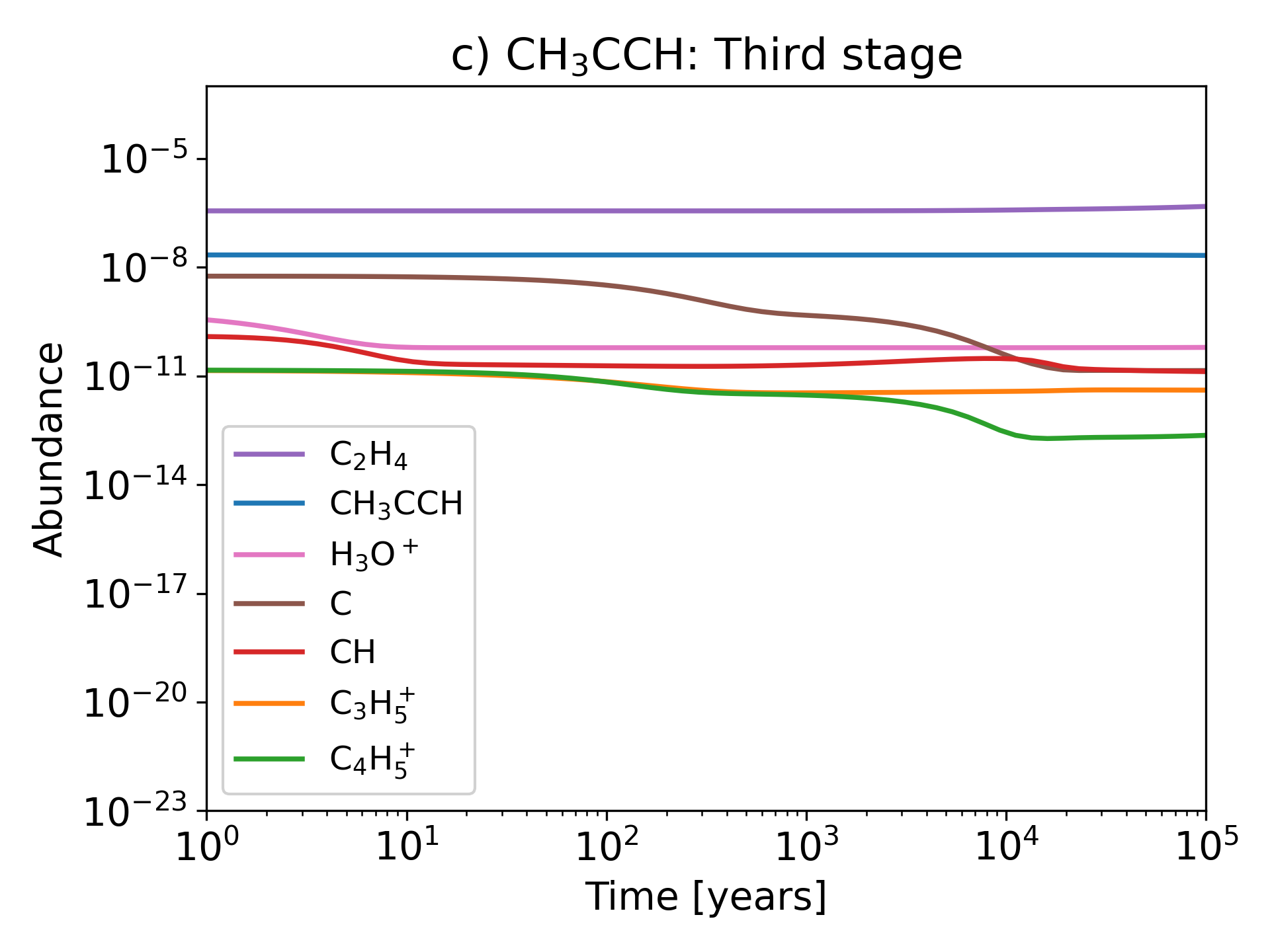}
\caption{Time evolution of the abundances of \MA\  relative to H and of the species dominating its formation and destruction. Panel a, stage 1: $n({\rm H}_2) = 2\times10^{4}$ cm$^{-3}$, $T = 10$ K, $\zeta = 10\zeta_0$. Panel b, stage 2: $T=200$~K, $n({\rm H}_2) = 2\times10^{6}$ cm$^{-3}$,  $\zeta = 10\zeta_0$. Panel c, stage 3: $T = $60 K, $n({\rm H}_2) = 2\times10^{5}$ cm$^{-3}$, and $\zeta = 0.1\zeta_0$.}
\label{fig:ma-tracers}
\end{figure}

Figure~\ref{fig:ma-tracers} summarises the three-stage evolution of \MA\ and the dominant related species. In the cold phase (Fig.~\ref{fig:ma-tracers}a), \MA\ forms efficiently on grain surfaces, but remains largely frozen out and the gas-phase abundance stays low ($X\lesssim10^{-10}$). The gas phase is instead dominated by atomic C and CH, with C$_2$H$_4$ and the ions C$_3$H$_5^+$ and C$_4$H$_5^+$ at the $10^{-10}$--$10^{-14}$ level.

During the second stage (Fig.~\ref{fig:ma-tracers}b), thermal desorption of \MA\ increases the gas-phase abundance by $>2$ dex to $\sim2\times10^{-8}$, consistent with the observed range (1.5--2.7$\times10^{-8}$). In parallel, C$_2$H$_4$ becomes the most abundant species of the set ($\sim4\times10^{-7}$), whereas C$_3$H$_5^+$ and C$_4$H$_5^+$ remain low ($\sim10^{-11}$), indicating that ion--molecule pathways contribute but are not the dominant source of gas-phase \MA.\ In the third stage (Fig.~\ref{fig:ma-tracers}c), the \MA \ abundance stays nearly constant at 2.1$\times10^{-8}$, varying by only $\sim3\%$ between the start and end of the stage, which indicates that it reaches a quasi-steady state. Meanwhile atomic C, CH, and H$_3$O$^+$ decrease by 1--3 dex as the ionisation level drops. The late-time \MA\ abundance is therefore set during the warm-up and is preserved during cooling provided $\zeta$ remains low.

\section{Low- and High-Excitation Components in the \MA \ Emission}
\label{app2}

This appendix justifies the partition of the \MA\ transitions into low- and high-excitation components adopted in Sect.~\ref{sec4.1}. We show that a single-temperature fit to the full rotational diagram systematically underestimates the highest-$E_{\rm u}$ transitions, and we quantify the improvement provided by the two-component fit.

\begin{figure}[!htbp]
\centering
\includegraphics[width=0.8\linewidth]{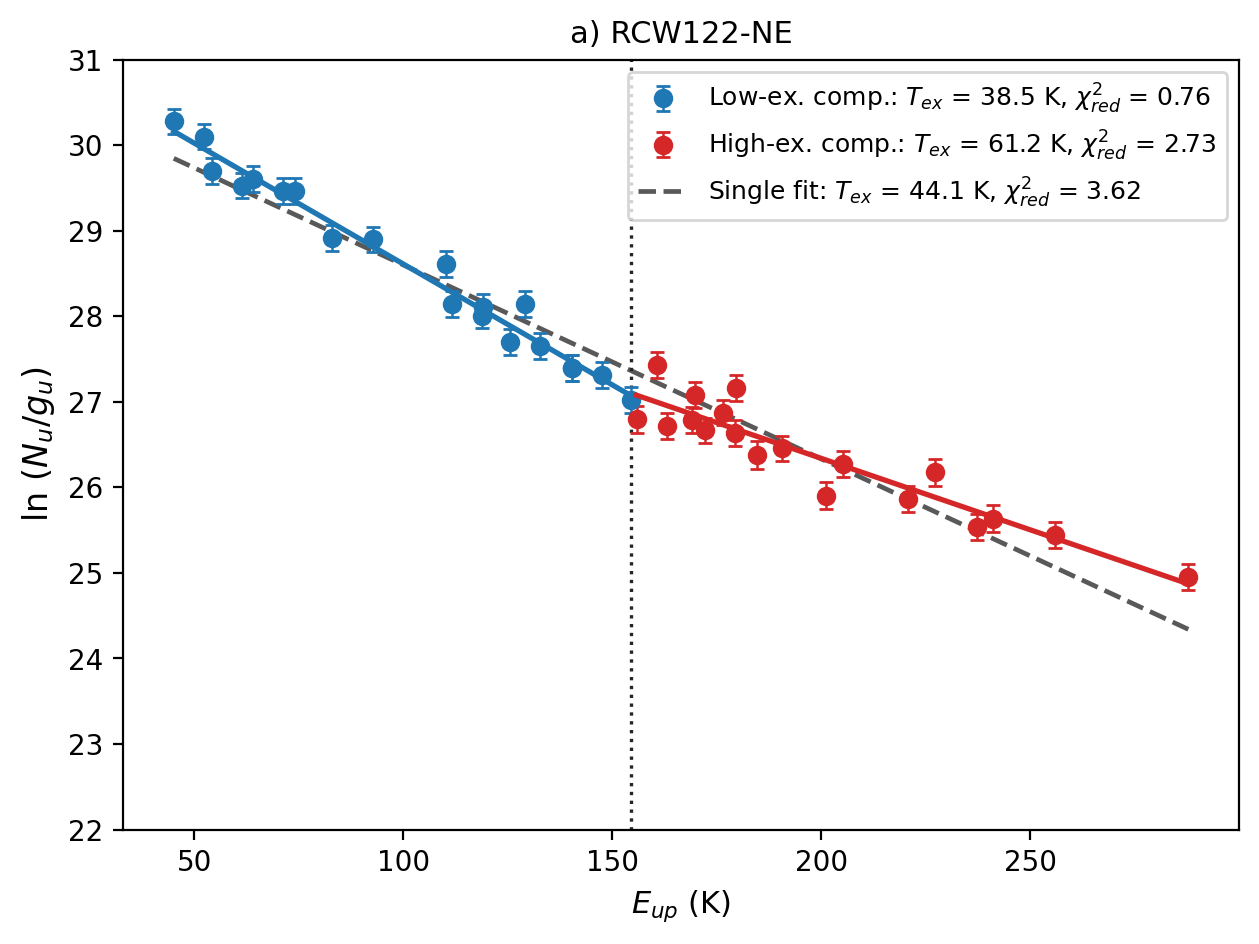}
\includegraphics[width=0.8\linewidth]{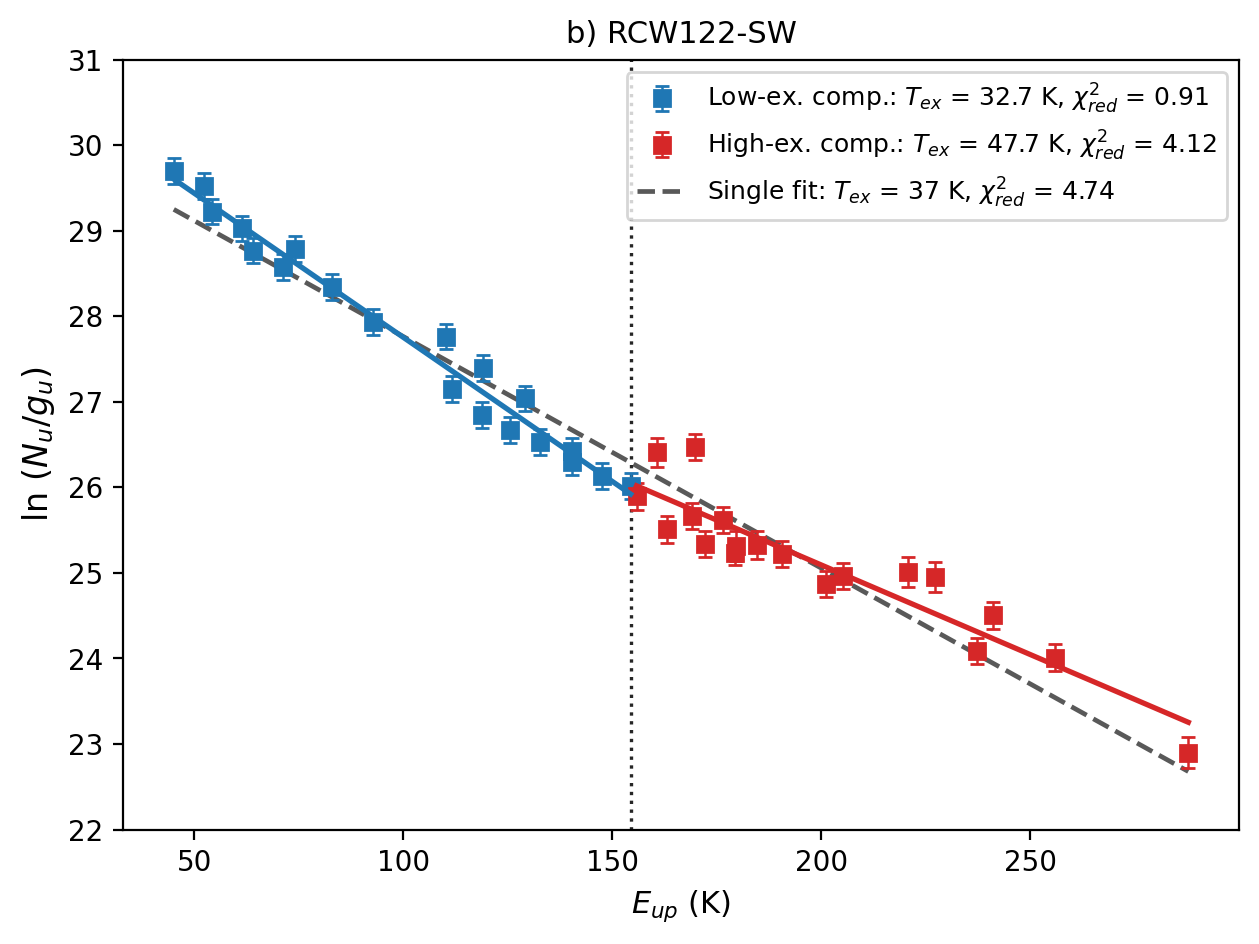}
\caption{Rotational diagrams of \MA\ toward (a) \NE\ and (b) \SW,\ corrected for beam dilution assuming compact emitting regions ($\simeq 8\arcsec$ and $\simeq 10\arcsec$, respectively). The vertical dotted line indicates the break in slope at $E_{\rm u}\approx 155\,$K. Blue (red) symbols denote transitions included in the low- (high-) excitation fit; solid lines show the corresponding weighted least-squares regressions. The dashed line shows a single-temperature fit to the full $K$-ladder data set.  Error bars include the dominant 15\% calibration uncertainty and the $1\sigma$ noise contribution (Sect.~\ref{sec2.1}).}
\label{rds-app}
\end{figure}

Under LTE, the rotational level populations follow a Boltzmann distribution. Consequently, a rotational diagram of $\ln(N_u/g_u)$ versus upper-level energy $E_{\rm u}$ is expected to be linear, with slope $-1/\boldsymbol{T_{\rm ex}}$. For \MA, the asymmetric-top and internal-rotation structure produces multiple $K$-resolved $J$ ladders; however, in LTE the transitions within the ground vibrational state should still be consistent with a single straight line. In both NE and SW, the observed \MA\ rotational diagrams (Fig.~\ref{rds-app}) deviate from a single-temperature description. When beam dilution is corrected assuming compact emitting regions ($\simeq 8\arcsec$ for NE and $\simeq 10\arcsec$ for SW), the diagrams exhibit a clear change in slope, with a knee at $E_{\rm u}\approx 155\,$K. The feature is less evident for more extended source-size assumptions ($\gtrsim 20\arcsec$). The knee lies between the $J{=}17_2{-}16_2$ line ($E_{\rm u}\approx 154.4\,$K) and the $J{=}19_0{-}18_0$ line ($E_{\rm u}\approx 155.8\,$K), which we adopt as an empirical boundary between two excitation regimes.

Table~\ref{tab:app1} summarizes the parameters obtained for each component and for the single-component fit. The two-component approach provides improved descriptions of the data, with reduced chi-squared values of 0.76 and 2.73 for the low- and high-excitation components in NE, and 0.91 and 4.12 in SW, respectively. We interpret the two excitation regimes as evidence for a temperature gradient within the \MA-emitting gas; adopting a single temperature would bias the solution toward low excitation and underestimate the highest-$E_{\rm u}$ transitions.

\begin{table*}[!htbp]
\caption{Rotational-diagram parameters for \MA\ toward NE and SW. We report results for a two-component excitation model and, for comparison, a single-temperature fit to the full data set. The high-excitation component (and the single-component fit) is corrected for beam dilution assuming compact emitting regions ($\simeq 8\arcsec$ for NE; $\simeq 10\arcsec$ for SW), whereas the low-excitation component is corrected assuming an extended $\sim 20\arcsec$ source.}
\label{tab:app1}
\centering
\begin{tabular}{c c c c c c c}
\hline\hline
Region & Exc. Component  & No. of lines & $E_{\rm u}$ (K) & $\boldsymbol{T_{\rm ex}}$ (K) & $N$ (cm$^{-2}$)   & $\chi^2_{\rm red}$  \\
\hline
\NE &Low exc. & 20& $\leq$154.38   & 41.4 $\pm$ 1.4 & $ (3.9 \pm 0.4) \times 10^{15}$ & 0.76\\
&High exc.    & 19& $\geq$155.83   & 61.2 $\pm$ 6.1 & $ (7.1 \pm 2.3) \times 10^{15}$ & 2.73\\
&Single fit   & 39& 45.11--287.78  & 44.1 $\pm$ 1.5 & $ (1.6 \pm 0.2) \times 10^{16}$ & 3.62\\
\hline
\SW &Low exc.& 20 & $\leq$154.38   & 32.7 $\pm$ 1.0 & $ (3.5 \pm 0.3) \times 10^{15}$ & 0.91\\
&High exc.   & 19 & $\geq$155.83   & 47.7 $\pm$ 4.9 & $ (3.6 \pm 1.5) \times 10^{15}$ & 4.12\\
&Single fit  & 39& 45.11--287.78  & 37.0 $\pm$ 1.2 & $ (8.1 \pm 1.2) \times 10^{15}$ & 4.74\\
\hline
\end{tabular}
\tablefoot{$E_{\rm u}$ is the upper-level energy. $\boldsymbol{T_{\rm ex}}$ and $N$ denote the rotational (Boltzmann) temperature and total column density derived from weighted linear fits to the rotational diagrams. The low- and high-excitation subsets are defined by $E_{\rm u}\leq 154.38\,$K and $E_{\rm u}\geq 155.83\,$K, respectively.}
\end{table*}

\end{document}